\documentclass[epsfig,12pt,onecolumn]{article}
\usepackage{amsfonts}
\usepackage{amssymb}
\usepackage{multicol}
\usepackage{graphicx}
\usepackage{float}
\usepackage{caption}
\usepackage{xcolor}
\usepackage[utf8]{inputenc}
\usepackage{amsmath}

\makeatletter \@addtoreset{equation}{section}
\renewcommand{\theequation}{\thesection.\arabic{equation}}
\def \be{\begin{equation}}
\def \ee{\end{equation}}
\def \bea{\begin{eqnarray}}
\def \eea{\end{eqnarray}}

\newcommand{\nc}{\newcommand}
\nc{\al}{\alpha} \nc{\bib}{\bibitem} \nc{\la}{\lambda}
\nc{\C}{\mbox{\hspace{1.24mm}\rule{0.2mm}{2.5mm}\hspace{-2.7mm} C}}
\nc{\R}{\mbox{\hspace{.04mm}\rule{0.2mm}{2.8mm}\hspace{-1.5mm} R}}
\input{tcilatex}
\begin{document}

\title{\textbf{Black hole solutions of three dimensional E}$_{6}$\textbf{%
-gravity }}
\author{R. Sammani, E.H Saidi, R. Ahl Laamara \\
{\small 1. LPHE-MS, Science Faculty}, {\small Mohammed V University in
Rabat, Morocco}\\
{\small 2. Centre of Physics and Mathematics, CPM- Morocco}}
\maketitle

\begin{abstract}
This paper aims to construct exceptional Bañados-Teitelboim-Zanelli (BTZ) black holes carrying E$_{6}$
charges as solutions to the 3D higher spin Anti-de Sitter (AdS) gravity with E$_{6}$
boundary conditions. Guided by Tits-Satake graphs of real forms of the e$%
_{6} $ Lie algebra, we build three remarkable E$_{6}$-higher spin black hole
models: the linear-exceptional and the ortho-exceptional BTZ solutions
result from splitting the extremal nodes in the E$_{6\left( 6\right) }$
Tits-Satake diagram while the pure exceptional-exceptional model follows
from the folding down to F$_{4\left( 4\right) }$. And with the help of Hasse
diagram visualizations, we study the ensuing higher spin spectrums to
develop the corresponding metrics using two types of gauge transformations.
For completeness, we examine the thermodynamics of the standard BTZ coupled
to E$_{6}$ higher spin gravity fields by computing the partition function
exploiting a one to one correspondence between the factors of the vacuum
characters and the roots of the E$_{6}$ root system.

\textbf{Keywords: }\emph{BTZ black hole, Landscape of HS-BTZ black holes, E}$%
_{\mathbf{6}}$\ \emph{gravity, Real forms of E}$_{\mathbf{6}}$\emph{\ Lie
algebra, Tits-Satake graphs, Hasse diagrams, Partition function.}
\end{abstract}

\tableofcontents

\section{Introduction}

\qquad In this study, we are interested in the Landscape of Ba\~{n}%
ados-Teitelboim-Zanelli (BTZ) black holes \cite{BTZ} solutions coupled to
higher spins (HS) \cite{spin3}-\cite{hb1} within the Chern-Simons (CS)
formulation. We impose the most general AdS$_{3}$ boundary conditions for
higher spin AdS gravity first considered by Grumiller and Riegler (GR) \cite%
{GR}. We aim to shed more light on the gray areas of the Landscape regarding
the corner of HS exceptional BTZ solutions while focussing on building
models for HS- BTZ black holes carrying E$_{6}$ charges.

The choice of GR boundary conditions of AdS$_{3}$ gravity is essential as it
is based on several compelling considerations, among which we cite: $\left(
\mathbf{i}\right) $\textrm{\ }a well-defined variational principle since a
gauge variation of the bulk CS action gives rise to an anomalous term
remedied by treating the charges and the chemical potentials functions on
un-equal footing. One can restore a healthy variational principle on-shell
by only allowing the charges to vary as required by GR boundary conditions. $%
\left( \mathbf{ii}\right) $ The emerging metrics given by the generalised
Fefferman-Graham form, where all the charges and chemical potential
functions in the CS description, are also captured in the canonical metric
formulation. As we will see subsequently, the usual BTZ metric form lacks
intersection terms between the radial and the boundary coordinates \cite{BTZ}%
. However with the use of the GR gauge transformation ($e^{\rho
L_{0}}e^{L_{+}})$ instead of the diagonal gauge change $e^{\rho L_{0}},$ we
can recuperate these metric components to get the generalised
Fefferman-Graham class$.$ $\left( \mathbf{iii}\right) $ Upon imposing the GR
boundary conditions, the asymptotic symmetry is given by two copies of the
affine $sl(2)_{k}$ algebra for spin 2 AdS gravity and bigger ranks
asymptotic affine symmetries for higher spins. This is an explicit
manifestation of the prominent duality between Chern-Simons theories and
Wess-Zumino-Witten models \cite{wzw1}-\cite{wzw4}; and conveys how the
theory's physical states are arranged in representations of the boundary
loop algebra \cite{GR}. $\left( \mathbf{iv}\right) $\ The possibility of
recovering all formerly discovered boundary conditions after imposing
certain restrictions on the GR constraints \cite{rec1}-\cite{rec5}. For
example by imposing the Drienfield-Sokolov (DS) highest weight reduction on
the GR spin 2 boundary conditions, we can retrieve Brown \& Henneaux's
boundary constraints and the associated two copies of the conformal Virasoro
algebra.

Depending on the higher spin symmetry, one can extend the GR boundary
conditions for spin 2 gravity to construct various HS-AdS$_{3}$ models \cite%
{2spin3}-\cite{ozil3} via the principal $sl(2,\mathbb{R})$ embedding \cite%
{hs4, hs41} in the higher gauge symmetry. In \cite{son}, using Tits-Satake
diagrams of real forms of Lie algebras \cite{SPIN}, a graphical description
of the principal embedding was proposed for several split real forms, namely
$sl(N,\mathbb{R}),$ $so(N,N+1)$, $sp(2N)$ and $so(N,N)$. With this
description, one can work out different HS BTZ solutions for a given higher
spin symmetry. For instance, AdS$_{3}$ gravity with orthogonal split real
form $so(N,N+1)$ as gauge symmetry has two remarkable solutions: $\left(
\mathrm{\alpha }\right) $ A first solution termed as vectorial results from
cutting the vectorial node in the Tits-Satake diagram giving therefore a $%
so(N-1,N)$ subsymmetry$.$ $(\mathrm{\beta })$ A second solution termed as
spinorial giving a $sl(N,\mathbb{R})$ subalgebra follows from the cutting
the spinorial node in the Tits-Satake diagram.

Given these varieties of solutions, one can define a Landscape of higher
spin BTZ black holes in AdS$_{3}$ gravity with higher ranks gauge symmetries
sitting in the usual Cartan classification of the standard finite
dimensional gauge groups ABCD and the exceptional EFG. However, a close
inspection of the diverse HS-BTZ solutions in literature reveals that only
partial results were derived for the exceptional corner with HS-BTZ black
holes having EFG conserved charges. In fact, apart from the G$_{2}$ higher
spin BTZ black hole \cite{trunc}, the AdS$_{3}$ gravity solutions with the
other exceptional EF gauge symmetries have not been rigorously investigated.

In this paper, we aim to fill this gap by constructing three solutions for
higher spin BTZ black holes with conserved exceptional E$_{6}$ charges. With
a top-down approach, we adopt the E$_{6}$ as the great higher spin gauge
symmetry; then we explore the various subsymmetries to classify the
different possible embeddings of the core $sl(2,\mathbb{R})$ by appropiately
decomposing the adjoint dimension. With this method, we were able to
identify two possible embeddings corresponding to the Levi decompositions
giving therefore two solutions: a linear-exceptional BTZ black hole with a $%
sl(6,\mathbb{R})$ subsymmetry and an ortho-exceptional BTZ black hole with a
$so(5,5)$ subalgebra. We also used a bottom-up approach to construct a third
model using only exceptional symmetries. The key point is to use the $F_{4}$
algebra as a transitional stop by first implanting the known G$_{2}$ in F$%
_{4}$ then both in E$_{6}.$

To carry out this study, we first review the real forms of the exceptional E$%
_{6}$ gauge algebra and the associated Tits-Satake diagrams. Then, with the
help of graphical algebraic tools and Levi-decomposition of gauge
symmetries, we construct the two exceptional higher spin E$_{6}$ BTZ
solutions (linear and orthogonal) by using Hasse diagrams. We also give a
third exceptional model based solely on exceptional symmetries using the
embedding $G_{2}\subset F_{4}\subset E_{6}$. As a consistency check, we
compute the partition functions to show the accordance between the models.

The organisation of this paper is as follows: In section 2, we introduce the
Landscape of 3D higher spin black holes as a means to identify the unrefined
exceptional corner. In section 3, we diligently construct the exceptional E$%
_{6}$ BTZ black holes. We overview the algebraic framework of the E$_{6}$
Lie algebra as a higher spin symmetry in preparation for the derivation of
the exceptional BTZ black holes solutions via three patterns distinguished
by the different subsymmetries and computational methods. And before
concluding, we study the coupling of spin 2 BTZ black hole with the E$_{6}$
higher spin fields and compute the ensuing partition functions in section 4.
Section 5 is reserved for summative comments and remarks. \textrm{We
accompany our study with two appendices, where we report technical details
and explicit algebraic computations. In appendix A, we exploit the sl(N,R)
canonical basis to construct a higher spin E}$_{6}$\textrm{\ algebra based
on the associated root system. In appendix B, we present the higher spin
spectrums upon decomposition towards both orthogonal and linear subalgebras.}

\section{Landscape of higher spin black holes}

In this section, we aim to define a Landscape of higher spin 3D black holes
as the set of HS gauge symmetries giving rise to 3D HS black hole solutions.
First, we review the BTZ black hole solution with $sl(2,\mathbb{R})\times
sl(2,\mathbb{R})$'s most general boundary conditions. Then, we describe
various HS extensions using Tits-Satake diagrams of higher spin symmetries
to outline the different corners of the HS-BTZ Landscape.

\subsection{Most general spin 2 BTZ solution}

In this inaugural subsection, we explore a somewhat new class of asymptotic
AdS$_{3}$ solutions known as the generalized Fefferman-Graham metrics.
Particularly, we study the generalised form of the BTZ black hole metric
using Grumiller \& Riegler (GR)'s gauge and boundary conditions.

In pursuit of this goal, we first consider the three dimensional
gravitational action with a negative cosmological constant $\Lambda
=-1/l_{AdS}^{2}$ as follows%
\begin{equation}
\mathcal{S}_{{\small 3D}}^{{\small grav}}=-\frac{1}{16\pi G_{N}}\int d^{3}x%
\sqrt{g}\left( \mathbf{R}-2\Lambda \right)  \label{action1}
\end{equation}%
with BTZ black hole solution written as \cite{BTZ}%
\begin{equation}
ds^{2}=-f\left( r\right) dt^{2}+\frac{1}{f\left( r\right) }%
dr^{2}+r^{2}\left( d\phi -\frac{J}{2r^{2}}dt\right) ^{2}  \label{btz}
\end{equation}%
with
\begin{equation}
f\left( r\right) =\frac{r^{2}}{l_{{\small AdS}}^{2}}+\frac{J^{2}}{4r^{2}}-M
\end{equation}%
where $M$ and $J$\ are respectively the mass and the angular momentum of the
black hole. The expression $f\left( r\right) $ is singular for two radii,
the event horizon $r_{+}$ and the inner horizon $r_{-}$ given by%
\begin{equation}
r_{\pm }=l_{{\small AdS}}\left[ \frac{M}{2}\left( 1\pm \sqrt{1-\frac{J^{2}}{%
M^{2}l_{{\small AdS}}^{2}}}\right) \right] ^{\frac{1}{2}}
\end{equation}

The gauge description of (\ref{action1}) can be formulated as a difference
of two Chern-Simons (CS) actions, each valued in a $SL(2,\mathbb{R})$
algebra:%
\begin{equation}
\mathcal{S}_{{\small 3D}}^{{\small grav}}=\frac{k}{4\pi }\int tr(AdA+\frac{2%
}{3}A^{3})-\frac{\tilde{k}}{4\pi }\int tr(\tilde{A}d\tilde{A}+\frac{2}{3}%
\tilde{A}^{3})  \label{equi}
\end{equation}%
where $k,$ $\tilde{k}$ are the CS levels related to the 3D Newton constant $%
G_{N}$ as $k=\tilde{k}=l_{AdS}/(4G_{N}).$ In order to recast the BTZ black
hole solution (\ref{btz}) in the gauge reformulation, we $(i)$ impose
boundary conditions to develop, asymptotically, the left and the right gauge
connections $A$ and $\tilde{A}$ and then $(ii)$ insert into $G_{\mu \nu }=%
\frac{1}{2}tr\left[ (A_{\mu }-\tilde{A}_{\mu })(A_{\nu }-\tilde{A}_{\nu })%
\right] $ to compute the metric$.$\newline
On the boundary of AdS$_{3}$ manifold, we parameterise the surface with
light cone coordinates $\xi ^{\pm }=(\varphi \pm t)/\sqrt{2}$ and rewrite
the bulk 1-form gauge potentials $A=A_{\mu }dx^{\mu }$ and $\tilde{A}=\tilde{%
A}_{\mu }dx^{\mu }$ as follows
\begin{equation}
\begin{tabular}{lll}
$A$ & $=$ & $\mathrm{g}^{-1}\mathfrak{A}{\small g}+\mathrm{g}^{-1}d\mathrm{g}
$ \\
$\tilde{A}$ & $=$ & $\mathrm{\tilde{g}}^{-1}\widetilde{\mathfrak{A}}\mathrm{%
\tilde{g}}+\mathrm{\tilde{g}}^{-1}d\mathrm{\tilde{g}}$%
\end{tabular}
\label{radial}
\end{equation}%
The SL$\left( 2,\mathbb{R}\right) $ group elements $\mathrm{g}$ and $\mathrm{%
\tilde{g}}$ of the gauge transformation (\ref{radial}) are set in the GR
gauge as \cite{GR}
\begin{equation}
\mathrm{g}=e^{\rho K_{0}}e^{K_{-}},\qquad \mathrm{g}^{-1}\frac{\partial
\mathrm{g}}{\partial \rho }=K_{0},\qquad \mathrm{\tilde{g}=g}^{-1}
\label{radial2}
\end{equation}%
with $SL\left( 2,\mathbb{R}\right) $ generators $K_{0},$ $K_{\pm }$ obeying%
\begin{equation}
\left[ K_{n},K_{m}\right] =\left( n-m\right) K_{n+m},\qquad n,m=0,\pm
\label{k}
\end{equation}%
and Killing form matrix given by%
\begin{equation}
tr\left( K_{a}K_{b}\right) =\left(
\begin{array}{ccc}
0 & 0 & -1 \\
0 & 1/2 & 0 \\
-1 & 0 & 0%
\end{array}%
\right)
\end{equation}%
The 1-forms $\mathfrak{A}=\mathfrak{A}_{\alpha }d\xi ^{\alpha }$\ and $%
\widetilde{\mathfrak{A}}=\widetilde{\mathfrak{A}}_{\alpha }d\xi ^{\alpha }$\
designate the Chern-Simons boundary gauge fields with charges $(\mathfrak{A}%
_{-}^{0,\pm },\widetilde{\mathfrak{A}}_{+}^{0,\pm })$ and chemical
potentials $(\mathfrak{A}_{+}^{0,\pm },\widetilde{\mathfrak{A}}_{-}^{0,\pm })
$ expanding as follows \cite{GR}%
\begin{equation}
\begin{tabular}{lll}
$\mathfrak{A}_{-}$ & $=$ & $-\frac{2\pi }{\mathrm{k}}\left[ \mathfrak{A}%
_{-}^{+}K_{+}-2\mathfrak{A}_{-}^{0}K_{0}+\mathfrak{A}_{-}^{-}K_{-}\right] $
\\
$\mathfrak{A}_{+}$ & $=$ & $\mathfrak{A}_{+}^{+}K_{+}+\mathfrak{A}%
_{+}^{0}K_{0}+\mathfrak{A}_{+}^{-}K_{-}$%
\end{tabular}%
\end{equation}%
and
\begin{equation}
\begin{tabular}{lll}
$\widetilde{\mathfrak{A}}_{+}$ & $=$ & $\frac{2\pi }{\mathrm{k}}\left[
\widetilde{\mathfrak{A}}_{-}^{+}K_{+}-2\widetilde{\mathfrak{A}}_{-}^{0}K_{0}+%
\widetilde{\mathfrak{A}}_{-}^{-}K_{-}\right] $ \\
$\widetilde{\mathfrak{A}}_{-}$ & $=$ & $\widetilde{\mathfrak{A}}%
_{+}^{+}K_{+}+\widetilde{\mathfrak{A}}_{+}^{0}K_{0}+\widetilde{\mathfrak{A}}%
_{+}^{-}K_{-}$%
\end{tabular}%
\end{equation}%
Using both (\ref{radial}) and (\ref{radial2}), the resulting metric is of
the form%
\begin{equation}
ds^{2}=d\rho ^{2}+2G_{\rho \alpha }d\rho dx^{\alpha }+G_{\alpha \beta
}dx^{\alpha }dx^{\beta }  \label{GFG}
\end{equation}%
known as the generalised Fefferman-Graham metric \textrm{in lieu} of the
standard one
\begin{equation}
ds^{2}=d\rho ^{2}+G_{\alpha \beta }dx^{\alpha }dx^{\beta }
\end{equation}%
computed with the gauge element $\mathrm{g}=e^{\rho K_{0}}.$

\subsection{Higher spin BTZ black holes}

A higher spin BTZ black hole proceeds straightforwardly from the principal
embedding of the core $sl(2,\mathbb{R})$ algebra in a higher dimensional
symmetry following the Gutperle-Kraus algorithm \cite{spin3}. After
uncovering the higher spin content of the higher spin symmetry, one can then
study the higher spin BTZ black hole physics by imposing suitable boundary
conditions on the boundary gauge potentials. For the general case of the
higher spin linear $SL_{N}$ symmetry, the higher spin content corresponds to
the splitting of the adjoint representation dimension into 2j+1
isomultiplets as follows%
\begin{equation}
N^{2}-1=3+5+...+2N-1
\end{equation}%
with isospin j=1,2,..,N-1. The higher spin analogue of (\ref{k}) for the sl$%
_{N}$ generators $W_{n_{\sigma }}^{(\sigma )}$ can be written as
\begin{equation}
\left[ W_{m_{s}}^{(\tau )},W_{n_{\sigma }}^{(\sigma )}\right]
=\dsum\limits_{s=2}^{N-1}c_{n_{\sigma },m_{\tau }|s}^{\tau ,\sigma
}W_{m_{\tau }+n_{\sigma }}^{(s)}  \label{sln}
\end{equation}%
where the $c_{n_{\sigma },m_{\tau }|s}^{\tau ,\sigma }$ are \textrm{%
structure constants}. For the first higher spin case with N=3, the above
relation reads as \cite{hs4}
\begin{equation}
\begin{tabular}{lll}
$\left[ K_{i},K_{j}\right] $ & $=$ & $\left( i-j\right) K_{i+j}$ \\
$\left[ K_{i},\mathtt{W}_{m}\right] $ & $=$ & $\left( 2i-m\right) \mathtt{W}%
_{i+m}$ \\
$\left[ \mathtt{W}_{m},\mathtt{W}_{n}\right] $ & $=$ & $-\frac{1}{3}\left(
m-n\right) \left( 2m^{2}+2n^{2}-mn-8\right) K_{m+n}$%
\end{tabular}%
\end{equation}%
For convenience, one can set $L_{i}=K_{-i}$ and $W_{m}=\mathtt{W}_{-m}$ in
order to rewrite the commutation relations like%
\begin{equation}
\begin{tabular}{lll}
$\left[ L_{i},L_{j}\right] $ & $=$ & $\left( j-i\right) L_{i+j}$ \\
$\left[ L_{i},W_{m}\right] $ & $=$ & $\left( m-2i\right) W_{i+m}$ \\
$\left[ W_{n},W_{m}\right] $ & $=$ & $\frac{1}{3}\left( n-m\right) \left(
2m^{2}+2n^{2}-mn-8\right) L_{n+m}$%
\end{tabular}
\label{w3}
\end{equation}%
with the useful property%
\begin{equation}
\begin{tabular}{lll}
$\left[ L_{0},L_{j}\right] $ & $=$ & $jL_{j}$ \\
$\left[ L_{0},W_{m}\right] $ & $=$ & $mW_{m}$%
\end{tabular}%
\end{equation}%
indicating that the labels $j$ and $m$ are just the charges of the operator $%
L_{0}.$

\begin{figure}[tbph]
\begin{center}
\includegraphics[width=12cm]{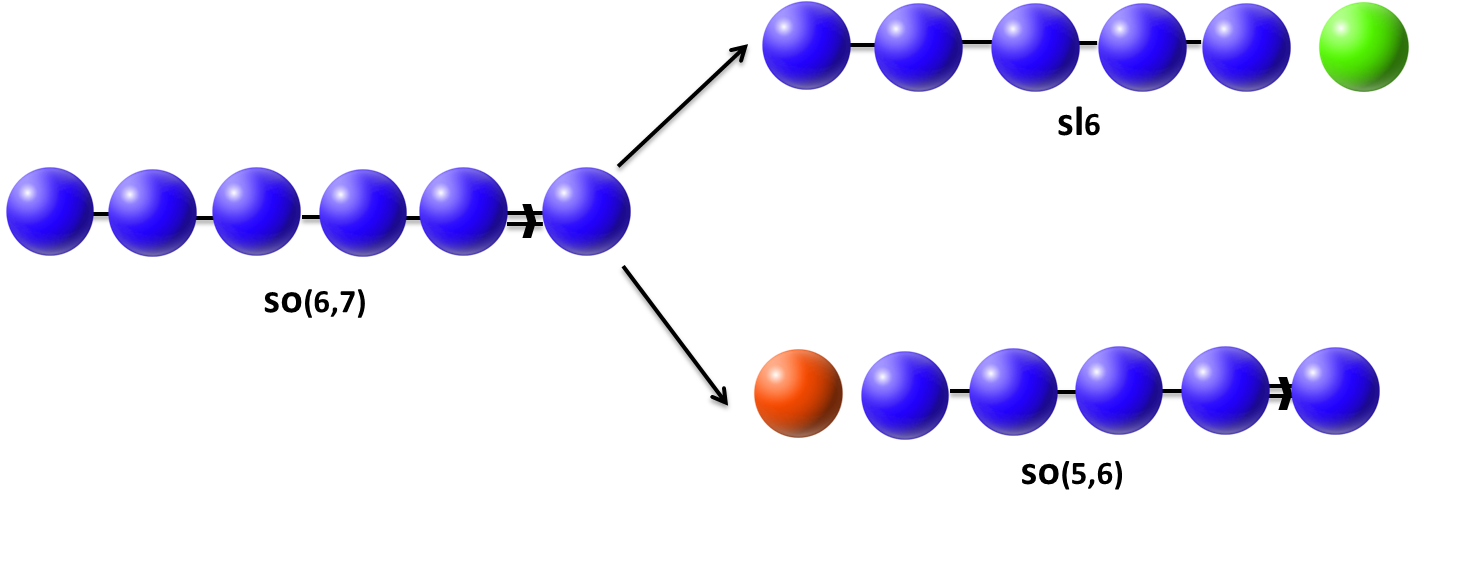}
\end{center}
\par
\vspace{-1.5cm}
\caption{Tits-Satake diagram of $so(6,7)$. The red node (resp. green)
corresponds to $so(1,2)\sim sl(2,\mathbb{R})$ whose cutting leaves $so(5,6)$
(resp. $sl(6))$.}
\label{B6}
\end{figure}

Depending on the desired choice of higher spin symmetry, one can therefore
construct various HS-AdS$_{3}$ models. In \cite{son}, using Tits-Satake
diagrams of real forms of Lie algebras \cite{SPIN}, a graphical description
of the principal embedding was proposed for several split real forms, namely
$sl(N,\mathbb{R}),$ $so(N,N+1)$, $sp(2N)$ and $so(N,N)$. With this
description, one can explore different nuances of a HS-AdS$_{3}$ model with
a single higher spin gauge symmetry. For instance, the higher spin theory
based on the orthogonal split real form $so(N,N+1)$ has two different
conformal spin spectrums, each corresponding to a distinct $sl(2,\mathbb{R})$
embedding, see \textbf{Figure} \textbf{\ref{B6}}.
These subtleties are masked in the linear $sl(N,\mathbb{R})$ case due to the
overlap of both extremal node decompositions.

\begin{figure}[tbph]
\begin{center}
\includegraphics[width=12cm]{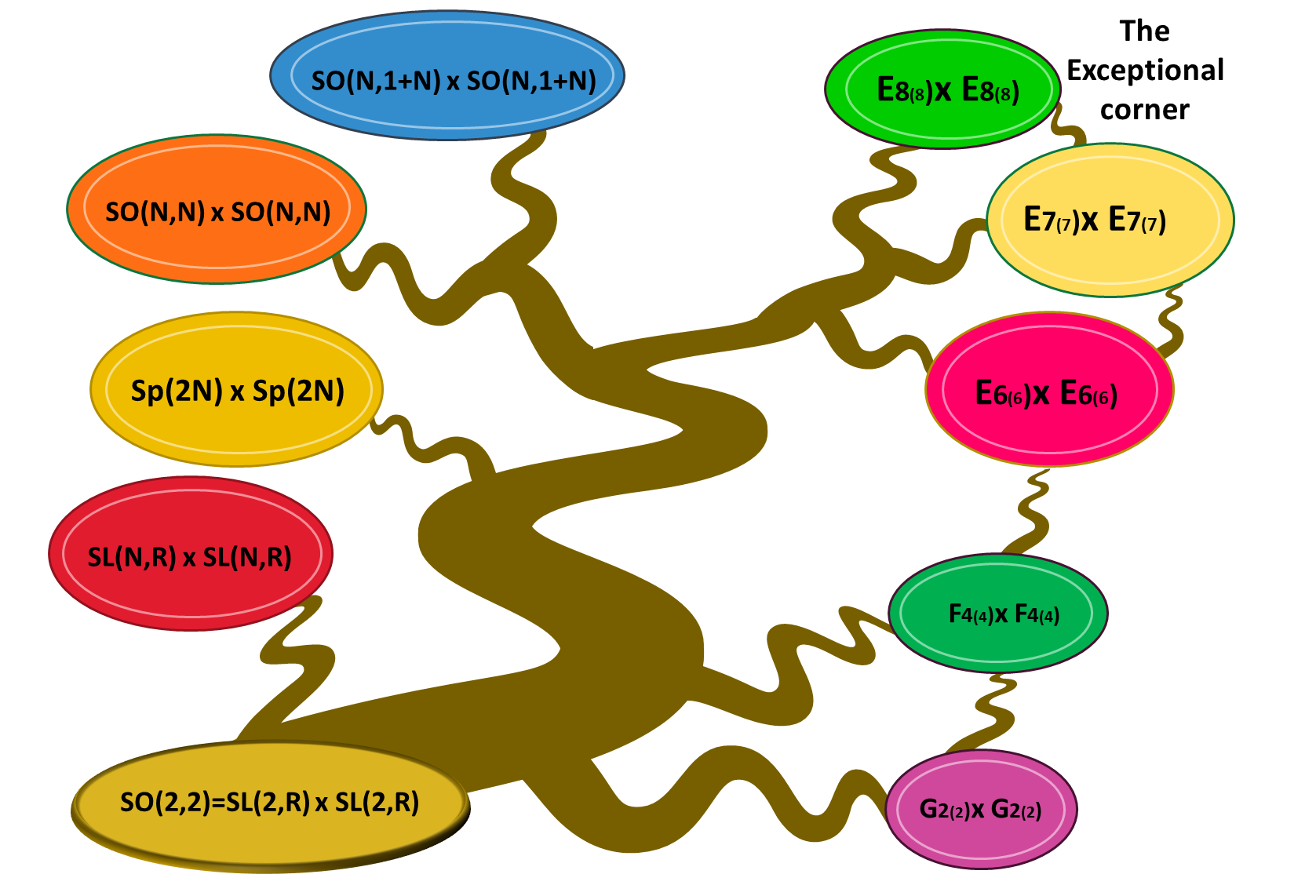}
\end{center}
\par
\vspace{-0.5cm}
\caption{Landscape of higher spin BTZ black holes}
\label{L}
\end{figure}

One can therefore define a Landscape of higher spin BTZ black hole solutions
by designating the various higher spin symmetries giving rise to higher spin
BTZ black hole solutions from the different possible embeddings of the
principal $sl(2,\mathbb{R})$. For instance, starting from the infinite
higher spin algebra $hs\left[ \lambda \right] ,$ an extension of the usual $%
sl(N,\mathbb{R})$ with an infinite tower of higher spins $s=2,...\infty ,$
one can generate different types of higher spin gauge symmetries like $%
sp(2N),so(2N+1)$ and eventually G$_{2}$ via truncations \cite{trunc}, or by
directly adopting the desired higher spin symmetry as we will opt to do
subsequently. In this next \textbf{Figure} \textbf{\ref{L}}, we represent
the different higher spin extensions of AdS$_{3}$ gravity in the
Chern-Simons formulation with higher spin BTZ black hole solutions.
Apart from the G$_{2}$ higher spin symmetry \cite{trunc}, exceptional BTZ
holes have not been thoroughly considered in higher spin theory and concrete
models were not built in higher spin Literature. Possessing such framework
with valuable tools to build interesting higher spin gravitational theories
is a promising opportunity to investigate exotic higher spin black hole
models and the ensuing thermodynamical properties. These exceptional BTZ
black hole solutions are investigated in the next section\textrm{\ }with
emphasis on E$_{6}$.

\section{Exceptional HS-BTZ black holes}

\qquad The main objective of this section is to study the higher spin BTZ
black hole with the exceptional $E_{6}$ higher spin symmetry. For this
purpose, we build the E$_{6}$ higher spin basis in prepartion for the
implementation of the principal embedding of $sl(2,\mathbb{R})$ into the E$%
_{6}$ Lie algebra \cite{Y4, Y5, y} using Tits-Satake graphs and Hasse
diagrams. Then, we investigate three BTZ black hole scenarios: the
linear-exceptional, the ortho-exceptional and the exceptional-exceptional
BTZ solutions. We explicitely derive the higher spin content and the
associated gauge connections to compute the E$_{6}$-BTZ black hole metrics
using different gauge transformations and computational methods.

\subsection{Graphical algebraic tools of $E_{6}$ higher spin symmetry}

As its name suggests, the E$_{6}$ symmetry is an exceptional structure;
however it is nicely described and easily maneuvered by its root system $%
\Phi _{e_{6}}$ in addition to its graphical representations like the Dynkin
and Tits Satake diagrams.

\begin{figure}[ph]
\begin{center}
\includegraphics[width=17cm]{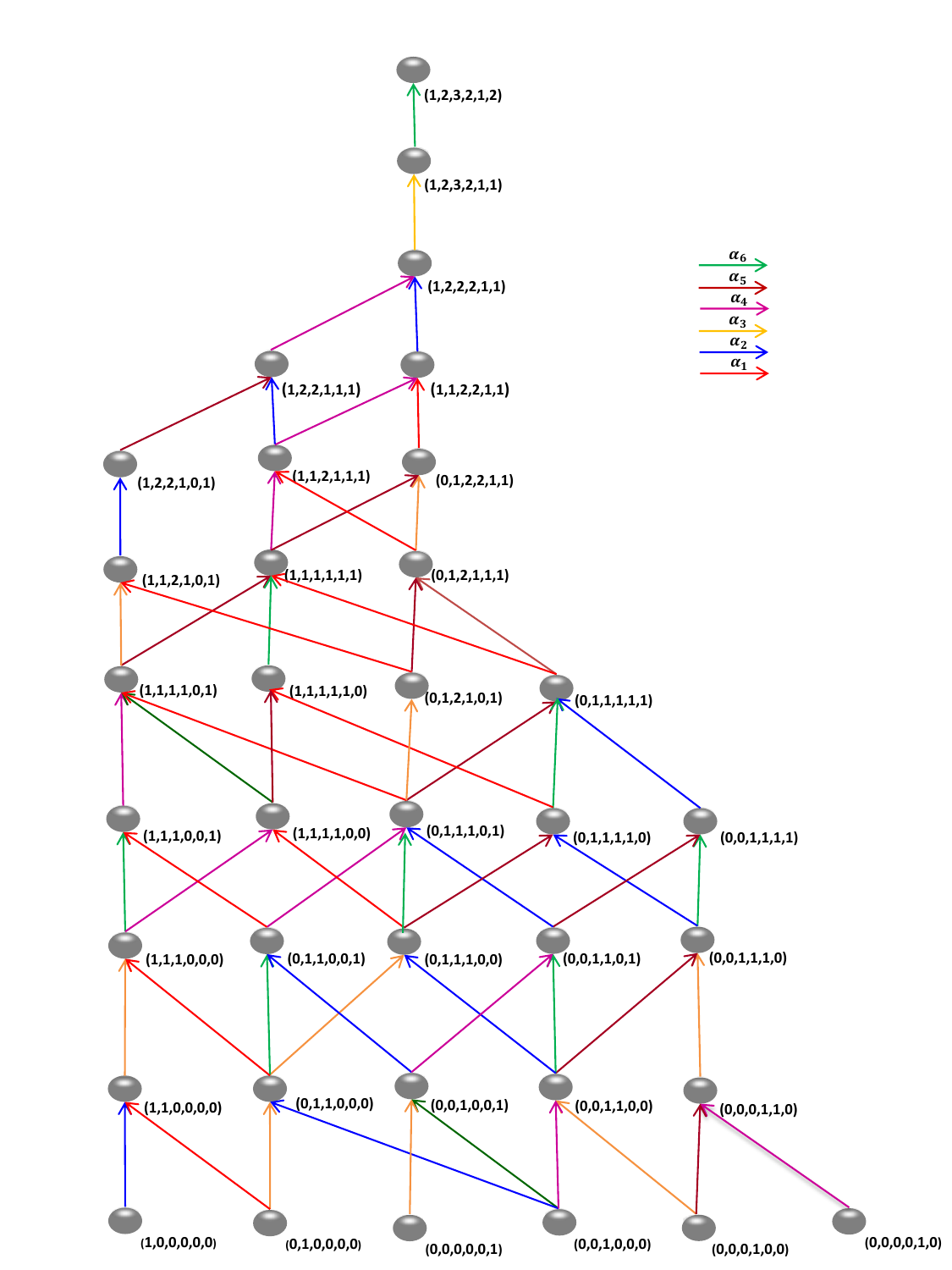}
\end{center}
\par
\vspace{-0.5cm}
\caption{Hasse diagram of the positive root system of the exceptional
algebra $e_{6}$ where each positive root is given by a linear combination $%
\sum_{i=1}^{6}n_{i}\protect\alpha _{i}$ labeled as $\left(
n_{1},n_{2},n_{3},n_{4},n_{5},n_{6}\right) $ with positive n$_{i}$ integers
as indicated in the Figure.}
\label{rs}
\end{figure}

We start by recalling that the simple Lie algebra ${\Large e}_{6}$ has rank
6 and dimension 78; its generators are given by the usual 6 commuting
Cartans
\begin{equation}
H_{\alpha _{1}},\quad H_{\alpha _{2}},\quad H_{\alpha _{3}},\quad H_{\alpha
_{4}},\quad H_{\alpha _{5}},\quad H_{\alpha _{6}}
\end{equation}%
and 36+36 step operators
\begin{equation}
E_{\pm \alpha }\qquad ,\qquad \alpha \in \Phi _{e_{6}}^{+}
\end{equation}%
with $\alpha $ standing for positive roots sitting in the root system $\Phi
_{e_{6}}=\Phi _{e_{6}}^{+}\cup \Phi _{e_{6}}^{-}$. The $\Phi _{e_{6}}^{\pm }$
are generated by six simple roots $\alpha _{1},\alpha _{2},\alpha
_{3},\alpha _{4},\alpha _{5},\alpha _{6}$ with intersection $\alpha
_{i}.\alpha _{j}$ given by the Cartan matrix $K_{ij}\left( e_{6}\right) $
directly read from the Dynkin diagram $\mathcal{D}\left( e_{6}\right) $. A
visual representation of the root subsystem $\Phi _{e_{6}}^{+}$ is given by
the so called Hasse diagram depicted by \textbf{Figure} \textbf{\ref{rs}}
where the explicit realisations of the 36 positive roots namely $\alpha
=\sum n_{i}\alpha _{i}$ are also reported.

\begin{figure}[h]
\begin{center}
\includegraphics[width=9cm]{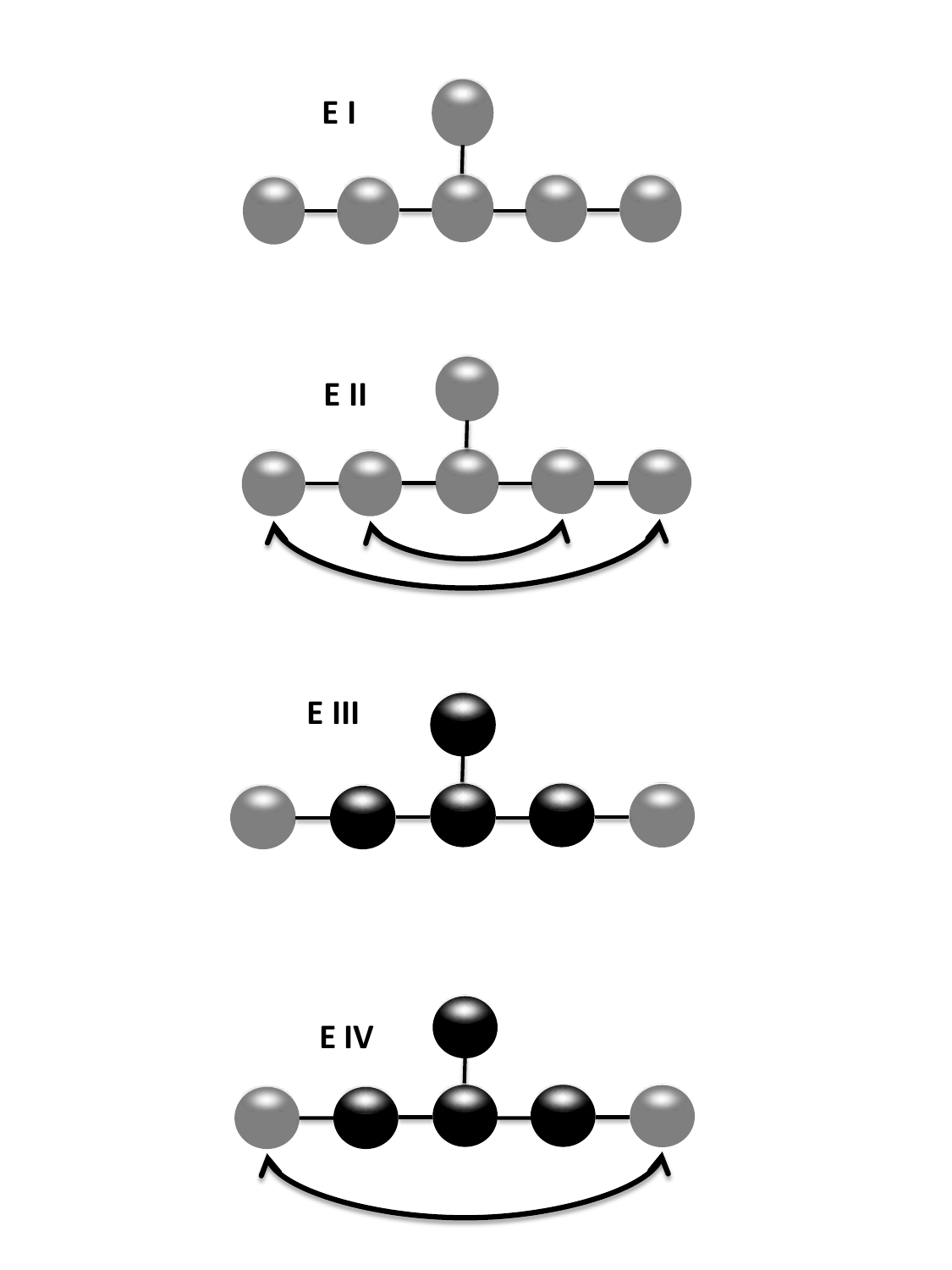}
\end{center}
\par
\vspace{-0.5cm}
\caption{The four Tits-Satake diagrams of real forms of the exceptional
algebra ${\protect\Large e}_{6}$.}
\label{F1}
\end{figure}

\begin{figure}[h]
\begin{center}
\includegraphics[width=13cm]{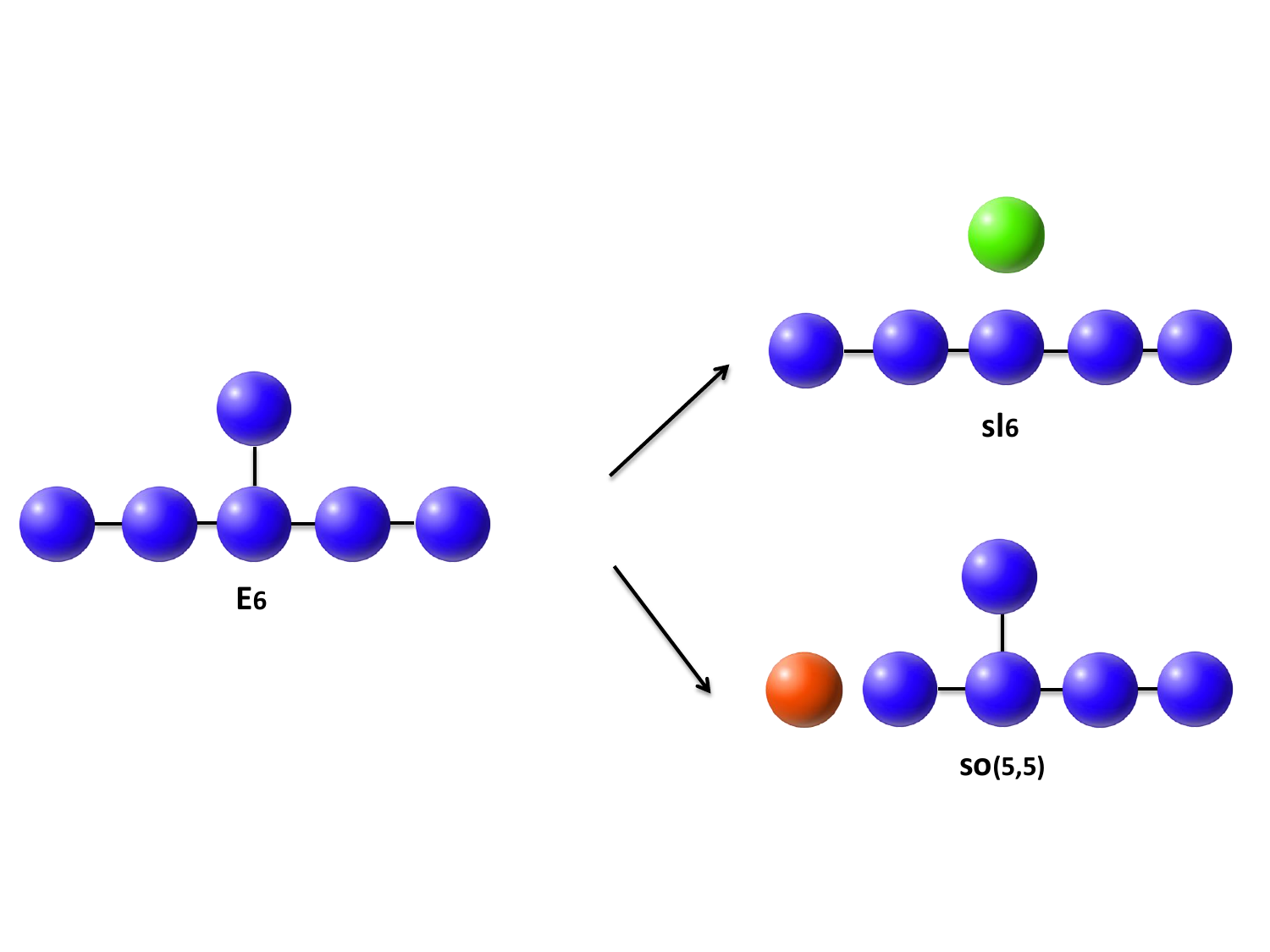}
\end{center}
\par
\vspace{-2.1cm}
\caption{Graphical description of the two different Levi- decompositions of
the exceptional ${\protect\Large E}_{6}$ model. In the top right, the cutted
node is given by $\protect\alpha _{6}.$ In the bottom, the cutted node is $%
\protect\alpha _{1}$ or equivalently $\protect\alpha _{5}$.}
\label{F2}
\end{figure}

Below, we will put these 78 dimensions \{$H_{\alpha _{i}},E_{\pm \alpha }$\}
into (2j+1) HS representations of a principal A$_{1}$ algebra generated by $%
H_{\mathrm{\alpha }},$ $E_{\pm \mathrm{\alpha }}$ associated with some given
root $\mathrm{\alpha ;}$ \textrm{We invite the reader to check out the full
construction of the E}$_{6}$\textrm{\ higher spin (HS) basis and the
associated generators \{}$W_{m_{s}}^{(s)}$\textrm{\} in appendix A}$.$ The
root $\mathrm{\alpha }$ defines the fundamental spin s=2 modes and it is
characterised by the commutation relations
\begin{subequations}
\begin{equation}
\left[ H_{\mathrm{\alpha }},E_{\pm \mathrm{\alpha }}\right] =\pm E_{\pm
\mathrm{\alpha }}\qquad ,\qquad \left[ E_{+\mathrm{\alpha }},E_{-\mathrm{%
\alpha }}\right] =2H_{\mathrm{\alpha }}  \label{cb}
\end{equation}%
To make contact between these commutation relations of complex A$_{1}$ and
those defining its real form $sl(2,\mathbb{R}),$ we compare them with the $%
sl(2,\mathbb{R})$ commutation relations given by eq(\ref{w3}) namely $\left[
L_{n},L_{m}\right] =\left( m-n\right) L_{n+m}.$ A bridge transformation can
be therefore deduced by identifying the realisation of the $L_{n}$'s with
Cartan-Weyl generators as follows
\end{subequations}
\begin{equation}
L_{0}=H_{\alpha }\qquad ,\qquad L_{\pm }=\pm E_{\mp \alpha }
\end{equation}%
Moreover to pick out the root $\alpha $ that defines the principal $sl(2,%
\mathbb{R})$, we use Levi-decomposition of Lie algebras which for E$_{6}$ is
known to be of two types: either $\left( \mathbf{i}\right) $ A decomposition
of E$_{6}$ towards to the $A_{5}$ Lie algebra ($E_{6}\rightarrow A_{5}$); or
$\left( ii\right) $ A decomposition towards $D_{5}$ ($E_{6}\rightarrow
D_{5}).$ In these Levi- decompositions, the 78 generators of E$_{6}$ are
first split into several multiplets of $A_{5}$ (resp. $D_{5}$) as follows
\begin{equation}
\begin{tabular}{lllll}
$E_{6}\rightarrow A_{5}$ & : & $78$ & $=$ & $1+35+21+21$ \\
$E_{6}\rightarrow D_{5}$ & : & $78$ & $=$ & $1+45+16+16$%
\end{tabular}
\label{d1}
\end{equation}%
and then into multiplets of the principal $sl(2,\mathbb{R})$ like%
\begin{equation}
\begin{tabular}{lllll}
$A_{5}$ & : & $35$ & $=$ & $3+5+7+9+11$ \\
$D_{5}$ & : & $45$ & $=$ & $3+5+7+9+21$%
\end{tabular}
\label{d2}
\end{equation}%
In this regard, recall that the root system $\Phi _{e_{6}}$ has 36+36 roots
generated by six simple roots $\alpha _{1},\alpha _{2},\alpha _{3},\alpha
_{4},\alpha _{5},\alpha _{6}$ that can be realised in terms of unit vectors $%
\{\epsilon _{i}\}$ in $\mathbb{R}^{8}$ as follows \textrm{\cite{y}}
\begin{equation}
\begin{tabular}{lll}
$\alpha _{1}$ & $=$ & $\frac{1}{2}\left( \epsilon _{1}-\epsilon
_{2}-\epsilon _{3}-\epsilon _{4}-\epsilon _{5}-\epsilon _{6}-\epsilon
_{7}+\epsilon _{8}\right) $ \\
$\alpha _{i}$ & $=$ & $\epsilon _{i}-\epsilon _{i-1}$ \\
$\alpha _{6}$ & $=$ & $\epsilon _{1}+\epsilon _{2}$%
\end{tabular}
\label{al1}
\end{equation}%
The 36 positive roots $\beta $ of the system $\Phi _{e_{6}}^{+}$ are given
by particular linear combinations $\beta =\sum n_{i}\alpha _{i}$ as depicted
in \textbf{Figure} \textbf{\ref{rs}}; they can be organised into $\left(
2j+1\right) $ dimensional HS multiplets as%
\begin{equation}
\begin{tabular}{lllll}
$A_{5}$ & : & $78$ & $=$ & $\dsum\limits_{j\in I_{{\small sl}_{{\small 6}%
}}}\left( 2j+1\right) $ \\
$D_{5}$ & : & $78$ & $=$ & $\dsum\limits_{j\in I_{{\small so}_{{\small 10}%
}}}\left( 2j+1\right) $%
\end{tabular}
\label{2j}
\end{equation}%
where the sets $I_{{\small sl}_{{\small 6}}}$ and $I_{{\small so}_{{\small 10%
}}}$ are given by%
\begin{equation}
\begin{tabular}{lllllll}
$j\in I_{{\small sl}_{{\small 6}}}$ & $=$ & $\{1,2,3,4,5,21\}$ & $\qquad
,\qquad $ & $j\in I_{{\small so}_{{\small 10}}}$ & $=$ & $\{1,2,3,4,10,16\}$
\\
$s\in J_{{\small sl}_{{\small 6}}}$ & $=$ & $\{2,3,4,5,6,22\}$ & $\qquad
,\qquad $ & $s\in J_{{\small so}_{{\small 10}}}$ & $=$ & $\{2,3,4,5,11,17\}$%
\end{tabular}
\label{414}
\end{equation}%
These roots $\beta $ label the 36+36 step operators $E_{\pm \alpha }$ of the
exceptional Lie algebra ${\Large e}_{6}.$ Moreover, following \cite{RF}, the
${\Large e}_{6}$ algebra possesses four different types of real forms as
illustrated in the following \textbf{Figure \ref{F1}}.
By analogy with the split real form $SL(N,\mathbb{R})$, we direct our
attention here also to the\textrm{\ }split real form of the ${\Large e}_{6}$
algebra. In this case, the corresponding Tits-Satake diagram correspond to (%
\textbf{EI}), an all white node Dynkin graph-like as given in \textbf{Figure}
\textbf{\ref{F1}}. As for the Dynkin graphs of finite Lie algebras, to each
node $\mathfrak{N}_{i}$ ($1\leq i\leq 6$) of the \textbf{EI} Tits-Satake
diagram, we associate a simple root $\alpha _{i}$ and a $sl(2,\mathbb{R})$
Lie algebra. One considers therefore $sl(2,\mathbb{R})$ as the elemental
symmetry of the ${\Large e}_{6}$ algebra allowing to decompose the 78
dimensions of ${\Large e}_{6}$ into several $sl_{2}$ multiplets as in (\ref%
{2j}). With this perspective, the higher spin theory with an ${\Large e}_{6}$
symmetry is constructed by \textrm{thinking about} one extremal $sl(2,%
\mathbb{R})$ node as a fundamental symmetry; while considering the rest of
the generators of the coset space ${\Large e}_{6}/sl_{2}$ as higher
dimensional representations of $sl_{2}.$ Depending on the choice of the
extremal node of the Tits-Satake diagram, say the red node $\mathfrak{N}_{1}$
(equivalently $\mathfrak{N}_{5}$) or the green node $\mathfrak{N}_{6}$, we
distinguish two types of HS (Levi-) decompositions: the red node cutting
exposes the orthogonal subsymmetry while the green node splitting reveals
the linear subalgebra; see \textbf{Figure} \textbf{\ref{F2}}.

Below, we study the ${\Large e}_{6}$ HS theory for both choices of the
extremal nodes. We first deconstruct the algebra to reveal the hidden
sub-symmetries, explore the higher spin conformal content for each case,
derive the boundary equations of motion along with the gauge connections,
and compute the metrics.

\subsection{BTZ black hole solution I: pattern $E_{6}\rightarrow SL_{6}$}

Here, we study the exceptional $E_{6}$ higher spin BTZ black hole solution
labelled by the branching pattern $E_{6}\rightarrow SL_{6}$. \textrm{%
Exploiting the higher spin content of }$E_{6}$\textrm{\ with }$SL_{6}$%
\textrm{\ subsymmetry, explicitely derived in appendix B, we construct the
induced linear-exceptional BTZ metric}%
\begin{equation}
\begin{tabular}{lll}
$G_{\mu \nu }$ & $=$ & $\frac{1}{2}tr\left[ \boldsymbol{E}_{\mu }\boldsymbol{%
E}_{\nu }\right] $ \\
$\boldsymbol{E}_{\mu }$ & $=$ & $A_{\mu }-\tilde{A}_{\mu }$%
\end{tabular}%
\end{equation}%
in terms of 2D boundary fields $\mathfrak{A}_{\pm }$ and $\widetilde{%
\mathfrak{A}}_{\pm }$\ through gauge transformations of $E_{6}$ as follows%
\begin{equation}
\begin{tabular}{lllllll}
$A_{\rho }$ & $=$ & $L_{0}$ & $\qquad ,\qquad $ & $\tilde{A}_{\rho }$ & $=$
& $-L_{0}$ \\
$A_{\pm }$ & $=$ & $\mathrm{g}^{-1}\mathfrak{A}_{\pm }\mathrm{g}$ & $\qquad
,\qquad $ & $\tilde{A}_{\pm }$ & $=$ & $\mathrm{g}\widetilde{\mathfrak{A}}%
_{\pm }\mathrm{g}^{-1}$%
\end{tabular}%
\end{equation}

\paragraph{\textbf{The linear-exceptional BTZ metric:}}

By the label "linear-exceptional" BTZ metric we characterise the metric
associated with the cutting pattern $E_{6}\rightarrow SL_{6}$ of the
Tits-Satake diagram \textbf{EI} of the exceptional $E_{6}.$ This designation
aids in distinguishing it from the other possible cutting pattern E$%
_{6}\rightarrow SO_{5,5}$ which will be labelled the "ortho-exceptional" BTZ
metric.

To determine the linear-exceptional higher spin BTZ metric $G_{\mu \nu }=%
\frac{1}{2}tr\left( \boldsymbol{E}_{\mu }\boldsymbol{E}_{\nu }\right) $
expressed in terms of the bulk vector potentials as $\boldsymbol{E}_{\mu
}=A_{\mu }-\tilde{A}_{\mu }$ with $A_{\pm }=\mathrm{g}^{-1}\mathfrak{A}_{\pm
}\mathrm{g}$ and $\tilde{A}_{\pm }=\mathrm{g}\widetilde{\mathfrak{A}}_{\pm }%
\mathrm{g}^{-1},$ we have to first compute the Dreibein $\boldsymbol{E}_{\mu
}$ and then perform the trace. Being an element of\ the exceptional Lie
algebra of the gauge symmetry E$_{6}$, we can expand it in the higher spin
basis \{$W_{m_{s}}^{(s)}$\} as follows,%
\begin{equation}
\boldsymbol{E}_{\mu }=\dsum\limits_{s\in J_{sl_{{\small 6}%
}}}\dsum\limits_{m_{s}=1-s}^{s-1}E_{\mu {\small (s)}}^{m_{s}}W_{m_{s}}^{(s)}
\end{equation}%
Putting into $G_{\mu \nu }=tr\left[ \boldsymbol{E}_{\mu }\boldsymbol{E}_{\nu
}\right] /2$, we obtain the following expression of the metric%
\begin{equation}
G_{\mu \nu }=\frac{1}{2}\dsum\limits_{s,\sigma \in J_{sl_{{\small 6}}}\times
J_{sl_{{\small 6}}}}\dsum\limits_{m_{s}={\small 1-s}}^{s-1}\dsum\limits_{n_{%
\sigma }=1-\sigma }^{\sigma -1}E_{\mu {\small (s)}}^{m_{s}}\mathcal{K}%
_{m_{s}n_{\sigma }}^{s,\sigma }E_{\nu {\small (\sigma )}}^{n_{\sigma }}
\end{equation}%
with linear-exceptional Killing metric $\mathcal{K}_{m_{s}n_{\sigma
}}^{s,\sigma }$ given by%
\begin{equation}
\mathcal{K}_{m_{s}n_{\sigma }}^{s,\sigma }=tr\left(
W_{m_{s}}^{(s)}W_{n_{\sigma }}^{(\sigma )}\right)
\end{equation}%
Moreover, using the expansions of $A_{\mu }$ and $\tilde{A}_{\mu },$ we get
\begin{equation}
E_{\mu {\small (s)}}^{m_{s}}=A_{\mu {\small (s)}}^{m_{s}}-\tilde{A}_{\mu
{\small (s)}}^{m_{s}}
\end{equation}%
To get the explicit expression of $\boldsymbol{E}_{\mu }$ in terms of the
boundary potential components $\mathfrak{A}_{\pm {\small (s)}}^{m_{s}}$ and $%
\widetilde{\mathfrak{A}}_{\pm {\small (s)}}^{m_{s}},$ we use the gauge
transformations $A_{\pm }=\mathrm{g}^{-1}\mathfrak{A}_{\pm }\mathrm{g}$ and $%
\tilde{A}_{\pm }=\mathrm{g}\widetilde{\mathfrak{A}}_{\pm }\mathrm{g}^{-1}$.
Depending on the nature of the gauge transformation $\mathrm{g}$, we get an
associated BTZ metric $G_{\mu \nu }^{(\mathrm{g})}$. Notice that because any
two gauge transformations $\mathrm{g}_{1}$ and $\mathrm{g}_{2}$ of the E$_{6}
$ symmetry are related to one another via a gauge transformation $\mathrm{g}%
_{\text{\textsc{bridge}}}$, each two distinct choices of the gauge
potentials say $A_{\pm }^{\mathrm{g}_{1}}=\mathrm{g}_{1}^{-1}\mathfrak{A}%
_{\pm }\mathrm{g}_{1}$ and $A_{\pm }^{\mathrm{g}_{2}}=\mathrm{g}_{2}^{-1}%
\mathfrak{A}_{\pm }\mathrm{g}_{2}$ (similarly for the twild ones), are also
related to one another as follows%
\begin{equation}
A_{\pm }^{\mathrm{g}_{2}}=\mathrm{g}_{\text{\textsc{bridge}}}^{-1}A_{\pm }^{%
\mathrm{g}_{1}}\mathrm{g}_{\text{\textsc{bridge}}}
\end{equation}%
with bridge gauge transformation given by $\mathrm{g}_{\text{\textsc{bridge}}%
}=\mathrm{g}_{1}^{-1}\mathrm{g}_{2}.$

Below, we investigate the Gutperle-Kraus (GK) and the generalised
Fefferman-Graham (GFG) frames for the higher spins in the exceptional $%
E_{6}\rightarrow SL_{6}$ gauge theory.

\paragraph{A. Gutperle-Kraus higher spin metric\newline
}

Here, the GK gauge transformation is given by the diagonal choice $\mathrm{g}%
_{{\small GK}}=\mathrm{e}^{\rho L_{0}}$ with $L_{0}$ is the charge operator
of $sl_{2}.$ The expression of the boundary field components $\mathfrak{A}%
_{\pm {\small (s)}}^{m_{s}}$ and $\widetilde{\mathfrak{A}}_{\pm {\small (s)}%
}^{m_{s}}$ namely
\begin{equation}
\begin{tabular}{lll}
$\mathfrak{A}_{\pm }$ & $=$ & $\dsum\limits_{s\in I_{sl_{{\small 6}%
}}}\dsum\limits_{m_{s}=1-s}^{s-1}\mathfrak{A}_{\pm {\small (s)}%
}^{m_{s}}W_{m_{s}}^{(s)}$ \\
$\widetilde{\mathfrak{A}}_{\pm }$ & $=$ & $\dsum\limits_{s\in I_{sl_{{\small %
6}}}}\dsum\limits_{m_{s}=1-s}^{s-1}\widetilde{\mathfrak{A}}_{\pm {\small (s)}%
}^{m_{s}}W_{m_{s}}^{(s)}$%
\end{tabular}%
\end{equation}%
\textrm{are explicitely computed in appendix B.} Using the gauge
transformation property $e^{-\rho L_{0}}W_{m_{s}}^{(s)}e^{\rho
L_{0}}=e^{-m_{s}\rho }W_{m_{s}}^{(s)}$, then putting into $A_{\pm }^{{\small %
GK}}=\mathrm{g}_{{\small GK}}^{-1}\mathfrak{A}_{\pm }\mathrm{g}_{{\small GK}%
} $, we obtain the expression of the bulk gauge potential%
\begin{equation}
\begin{tabular}{lll}
$A_{\pm }^{{\small GK}}$ & $=$ & $\dsum\limits_{s\in I_{sl_{{\small 6}%
}}}\dsum\limits_{m_{s}=1-s}^{s-1}e^{-m_{s}\rho }\mathfrak{A}_{\pm {\small (s)%
}}^{m_{s}}W_{m_{s}}^{(s)}$ \\
$\tilde{A}_{\pm }^{{\small GK}}$ & $=$ & $\dsum\limits_{s\in I_{sl_{{\small 6%
}}}}\dsum\limits_{m_{s}=1-s}^{s-1}e^{+m_{s}\rho }\widetilde{\mathfrak{A}}%
_{\pm {\small (s)}}^{m_{s}}W_{m_{s}}^{(s)}$%
\end{tabular}%
\end{equation}%
which implies%
\begin{equation}
E_{\pm {\small (s)}}^{m_{s}}=e^{-m_{s}\rho }\mathfrak{A}_{\pm {\small (s)}%
}^{m_{s}}-e^{+m_{s}\rho }\widetilde{\mathfrak{A}}_{\pm {\small (s)}}^{m_{s}}
\end{equation}%
Furthermore, calculating $E_{\mu {\small (s)}}^{m_{s}}E_{\nu {\small (\sigma
)}}^{n_{\sigma }},$ we get%
\begin{equation}
\begin{tabular}{lll}
$E_{\mu {\small (s)}}^{m_{s}}E_{\nu {\small (\sigma )}}^{n_{\sigma }}$ & $=$
& $e^{-\left( m_{s}+n_{\sigma }\right) \rho }\mathfrak{A}_{\mu {\small (s)}%
}^{m_{s}}\mathfrak{A}_{\nu {\small (\sigma )}}^{n_{\sigma }}+e^{+\left(
m_{s}+n_{\sigma }\right) \rho }\widetilde{\mathfrak{A}}_{\nu {\small (s)}%
}^{m_{s}}\widetilde{\mathfrak{A}}_{\nu {\small (\sigma )}}^{n_{\sigma }}-$
\\
&  & $e^{-\left( m_{s}-n_{\sigma }\right) \rho }\mathfrak{A}_{\mu {\small (s)%
}}^{m_{s}}\widetilde{\mathfrak{A}}_{\nu {\small (\sigma )}}^{n_{\sigma
}}-e^{+\left( m_{s}-n_{\sigma }\right) \rho }\widetilde{\mathfrak{A}}_{\nu
{\small (s)}}^{m_{s}}\mathfrak{A}_{\nu {\small (\sigma )}}^{n_{\sigma }}$%
\end{tabular}
\label{EE}
\end{equation}%
Because the boundary gauge potential $\mathfrak{A}_{\mu {\small (s)}%
}^{m_{s}} $ has no radial components ($\mathfrak{A}_{\rho {\small (s)}%
}^{m_{s}}=0),$ the GK metric has the form%
\begin{equation}
G_{\mu \nu }^{{\small GK}}=\left(
\begin{array}{cc}
1 & 0 \\
0 & g_{\mathrm{\alpha \beta }}^{{\small GK}}%
\end{array}%
\right)
\end{equation}%
with%
\begin{equation}
G_{\mathrm{\alpha \beta }}^{{\small GK}}=\frac{1}{2}\dsum\limits_{s,\sigma
\in J_{sl_{{\small 6}}}\times J_{sl_{{\small 6}}}}\dsum\limits_{m_{s}=%
{\small 1-s}}^{s-1}\dsum\limits_{n_{\sigma }=1-\sigma }^{\sigma -1}E_{%
\mathrm{\alpha }{\small (s)}}^{m_{s}}E_{\mathrm{\beta }{\small (\sigma )}%
}^{n_{\sigma }}\mathcal{K}_{m_{s}n_{\sigma }}^{s,\sigma }  \label{sl1}
\end{equation}%
and $E_{\mathrm{\alpha }{\small (s)}}^{m_{s}}$ as in eq(\ref{EE}).

\paragraph{B. Fefferman-Graham higher spin metric:\newline
}

The Fefferman-Graham higher spin BTZ metric reads in terms of the
Fefferman-Graham dreibein $\boldsymbol{E}_{\mu }^{{\small FG}}$ as follows
\begin{equation}
G_{\mu \nu }^{{\small FG}}=\frac{1}{2}tr\left[ \boldsymbol{E}_{\mu }^{%
{\small FG}}\boldsymbol{E}_{\nu }^{{\small FG}}\right]   \label{EEM}
\end{equation}%
The components of $\boldsymbol{E}_{\mu }^{{\small FG}}$ are expressed in
terms of the left and the right handed bulk gauge potentials \textrm{as
follows}
\begin{equation}
\boldsymbol{E}_{\mu }^{{\small FG}}=A_{\mu }^{{\small FG}}-\tilde{A}_{\mu }^{%
{\small FG}}
\end{equation}%
with
\begin{equation}
A_{\mu }^{{\small FG}}=\mathrm{g}_{{\small FG}}^{-1}\left( \partial _{\mu }+%
\mathfrak{A}_{\mu }\right) \mathrm{g}_{{\small FG}},\qquad \tilde{A}_{\pm }^{%
{\small FG}}=\mathrm{g}_{{\small FG}}\left( \partial _{\mu }+\widetilde{%
\mathfrak{A}}_{\mu }\right) \mathrm{g}_{{\small FG}}^{-1}
\end{equation}%
and $\mathrm{g}_{{\small FG}}=e^{\rho L_{0}}e^{L_{+}}.$ By using the
relationship between the FG and the GK gauge transformations namely $\mathrm{%
g}_{{\small FG}}=\mathrm{g}_{{\small GK}}e^{L_{+}};$\ we can express the
potentials $A_{\mu }^{{\small FG}}$ and $\tilde{A}_{\mu }^{{\small FG}}$ in
terms of their GK counterparts as
\begin{equation}
A_{\mu }^{{\small FG}}=e^{-L_{+}}A_{\mu }^{{\small GK}}e^{L_{+}},\qquad
\tilde{A}_{\mu }^{{\small FG}}=e^{L_{+}}A_{\mu }^{{\small GK}}e^{-L_{+}}
\end{equation}%
Using these relations, the dreibein $\boldsymbol{E}_{\mu }^{{\small FG}}$
reads as follows%
\begin{equation}
\boldsymbol{E}_{\mu }^{{\small FG}}=e^{-ad_{L_{+}}}A_{\mu }^{{\small GK}%
}-e^{+ad_{L_{+}}}A_{\mu }^{{\small GK}}
\end{equation}%
with%
\begin{equation}
\begin{tabular}{lll}
$A_{\mu }^{{\small FG}}$ & $=$ & $\dsum\limits_{s\in I_{sl_{{\small 6}%
}}}\dsum\limits_{m_{s}=1-s}^{s-1}e^{-m_{s}\rho }\mathfrak{A}_{\mu {\small (s)%
}}^{m_{s}}e^{-ad_{L_{+}}}(W_{m_{s}}^{(s)})$ \\
$\tilde{A}_{\mu }^{{\small FG}}$ & $=$ & $\dsum\limits_{s\in I_{sl_{{\small 6%
}}}}\dsum\limits_{m_{s}=1-s}^{s-1}e^{+m_{s}\rho }\widetilde{\mathfrak{A}}%
_{\mu {\small (s)}}^{m_{s}}e^{+ad_{L_{+}}}(W_{m_{s}}^{(s)})$%
\end{tabular}
\label{AFG}
\end{equation}%
These relations allow the following computation%
\begin{equation}
\boldsymbol{E}_{\mu }^{{\small FG}}\boldsymbol{E}_{\nu }^{{\small FG}%
}=A_{\mu }^{{\small FG}}A_{\nu }^{{\small FG}}-A_{\mu }^{{\small FG}}\tilde{A%
}_{\nu }^{{\small FG}}-\tilde{A}_{\mu }^{{\small FG}}A_{\nu }^{{\small FG}}+%
\tilde{A}_{\mu }^{{\small FG}}\tilde{A}_{\nu }^{{\small FG}}
\end{equation}%
Putting back into eq(\ref{EEM}), we obtain%
\begin{equation}
\begin{tabular}{lll}
$G_{\mu \nu }^{{\small FG}}$ & $=$ & $+\frac{1}{2}tr\left( A_{\mu }^{{\small %
FG}}A_{\nu }^{{\small FG}}\right) +\frac{1}{2}tr\left( \tilde{A}_{\mu }^{%
{\small FG}}\tilde{A}_{\nu }^{{\small FG}}\right) $ \\
&  & $-\frac{1}{2}tr\left( A_{\mu }^{{\small FG}}\tilde{A}_{\nu }^{{\small FG%
}}\right) -\frac{1}{2}tr\left( \tilde{A}_{\mu }^{{\small FG}}A_{\nu }^{%
{\small FG}}\right) $%
\end{tabular}
\label{sl2}
\end{equation}%
with $A_{\mu }^{{\small FG}}$ as in (\ref{AFG}).

\subsection{BTZ black hole solution II: pattern $E_{6}\rightarrow SO_{5,5}$}

In this part, we explore the otho-exceptional based BTZ black hole solution
arising from the branching pattern $E_{6}\rightarrow SO_{5,5}$. Similarly to
the previous $E_{6}\rightarrow SL_{6}$ description, using the higher spin
content of the $E_{6}/SO_{5,5}$ gravity, we construct the ortho-exceptional
BTZ metrics as described below:

\paragraph{The ortho-exceptional BTZ metric}

To determine the ortho-exceptional higher spin BTZ metric, one can proceed
as for the linear-exceptional construction done above. We use the relation $%
G_{\mu \nu }=\frac{1}{2}tr\left[ \boldsymbol{E}_{\mu }\boldsymbol{E}_{\nu }%
\right] $ with dreibein $\boldsymbol{E}_{\mu }=A_{\mu }-\tilde{A}_{\mu }$
and evaluate the trace by using the Killing form of the ortho-exceptional
Lie algebra. However in this part, we will pursue a different method to
reach the ortho-exceptional higher spin BTZ metric.

We start by substituting the dreibeins $\boldsymbol{E}_{\mu }$ and $%
\boldsymbol{E}_{\nu }$ by their expressions, we get%
\begin{equation}
G_{\mu \nu }=\frac{1}{2}tr\left( A_{\mu }A_{\nu }\right) +\frac{1}{2}%
tr\left( \tilde{A}_{\mu }\tilde{A}_{\nu }\right) -\frac{1}{2}tr\left( A_{\mu
}\tilde{A}_{\nu }\right) -\frac{1}{2}tr\left( \tilde{A}_{\mu }A_{\nu
}\right)
\end{equation}%
As before, the left $A_{\mu }$ and the right $\tilde{A}_{\mu }$ bulk
potentials are not uniquely defined as we had encountered both ($A_{\mu }^{%
{\small GK}},\tilde{A}_{\mu }^{{\small GK}}$) and ($A_{\mu }^{{\small FG}},%
\tilde{A}_{\mu }^{{\small FG}}$). Generally speaking, the bulk potentials
are labelled by a gauge transformation $\mathrm{g}$ into $E_{6};$ and are
related to the boundary potentials $\mathfrak{A}_{\mu }$ and $\widetilde{%
\mathfrak{A}}_{\mu }$ (with $\mathfrak{A}_{\rho }=\widetilde{\mathfrak{A}}%
_{\rho }=0$) as follows
\begin{equation}
\begin{tabular}{lllllll}
$A_{\mu }^{\mathrm{g}}$ & $=$ & $\mathrm{g}^{-1}\left( \partial _{\mu }+%
\mathfrak{A}_{\mu }\right) \mathrm{g}$ & $,\qquad $ & $\tilde{A}_{\mu }^{%
\mathrm{g}}$ & $=$ & $\mathrm{g}\left( \partial _{\mu }+\widetilde{\mathfrak{%
A}}_{\mu }\right) \mathrm{g}^{-1}$ \\
& $=$ & $\mathrm{g}^{-1}\mathfrak{D}_{\mu }\mathrm{g}$ & $,\qquad $ &  & $=$
& $\mathrm{g}\mathfrak{\tilde{D}}_{\mu }\mathrm{g}^{-1}$%
\end{tabular}
\label{trs}
\end{equation}%
with $\mathfrak{A}_{\mu }$ and $\widetilde{\mathfrak{A}}_{\mu }$ restrained
by the DS\textrm{\ }gauge and the field equations of motion. As such, the
ortho- exceptional higher spin BTZ metric is labelled by the gauge
transformation $\mathrm{g}$ like
\begin{equation}
\begin{tabular}{lll}
$G_{\mu \nu }^{\mathrm{g}}$ & $=$ & $+\frac{1}{2}tr\left( A_{\mu }^{\mathrm{g%
}}A_{\nu }^{\mathrm{g}}\right) +\frac{1}{2}tr\left( \tilde{A}_{\mu }^{%
\mathrm{g}}\tilde{A}_{\nu }^{\mathrm{g}}\right) $ \\
&  & $-\frac{1}{2}tr\left( A_{\mu }^{\mathrm{g}}\tilde{A}_{\nu }^{\mathrm{g}%
}\right) -\frac{1}{2}tr\left( \tilde{A}_{\mu }^{\mathrm{g}}A_{\nu }^{\mathrm{%
g}}\right) $%
\end{tabular}%
\end{equation}%
which for $\mathrm{g=g}_{{\small GK}}$ we have $G_{\mu \nu }^{{\small GK}},$
and for $\mathrm{g=g}_{{\small FG}}$ we get $G_{\mu \nu }^{{\small FG}}.$
Being elements of\ the exceptional Lie algebra ${\Large e}_{6}\rightarrow
SO_{{\small 5,5}}$ of the gauge symmetry, we can then develop the $\mathfrak{%
A}_{\mu }$ and the $\widetilde{\mathfrak{A}}_{\mu }$ as follows%
\begin{equation}
\begin{tabular}{lll}
$\mathfrak{A}_{\mu }$ & $=$ & $\dsum\limits_{s\in J_{{\small so}_{{\small 10}%
}}}\dsum\limits_{m_{s}=1-s}^{s-1}\mathfrak{A}_{\mu {\small (s)}%
}^{m_{s}}W_{m_{s}}^{(s)}$ \\
$\widetilde{\mathfrak{A}}_{\mu }$ & $=$ & $\dsum\limits_{s\in J_{{\small so}%
_{{\small 10}}}}\dsum\limits_{m_{s}=1-s}^{s-1}\widetilde{\mathfrak{A}}_{\mu
{\small (s)}}^{m_{s}}W_{m_{s}}^{(s)}$%
\end{tabular}
\label{LR}
\end{equation}%
with generators $W_{m_{s}}^{(s)}$ obeying the typical commutation relations%
\begin{equation}
\left[ W_{m_{s}}^{(\tau )},W_{n_{\sigma }}^{(\sigma )}\right]
=\dsum\limits_{s\in J_{{\small so}_{{\small 10}}}}c_{n_{\sigma },m_{\tau
}|s}^{\tau ,\sigma }W_{m_{\tau }+n_{\sigma }}^{(s)}
\end{equation}%
where here the $c_{n_{\sigma },m_{\tau }|s}^{\tau ,\sigma }$ are the
structure constants of the ortho- exceptional ${\Large e}_{6}\rightarrow SO_{%
{\small 5,5}}$ algebra, \textrm{see appendix B.} Analogous expansions hold
also for the bulk potentials namely%
\begin{equation}
\begin{tabular}{lll}
$A_{\mu }^{\mathrm{g}}$ & $=$ & $\dsum\limits_{s\in J_{{\small so}_{{\small %
10}}}}\dsum\limits_{m_{s}=1-s}^{s-1}A_{\mu {\small (s)}}^{(\mathrm{g)}%
m_{s}}W_{m_{s}}^{(s)}$ \\
$\tilde{A}_{\mu }^{\mathrm{g}}$ & $=$ & $\dsum\limits_{s\in J_{{\small so}_{%
{\small 10}}}}\dsum\limits_{m_{s}=1-s}^{s-1}\tilde{A}_{\mu {\small (s)}}^{(%
\mathrm{g)}m_{s}}W_{m_{s}}^{(s)}$%
\end{tabular}
\label{de}
\end{equation}%
Notice that by using the form $tr(V_{-n_{\sigma }}^{(\sigma
)}W_{m_{s}}^{(s)})=\delta ^{\sigma s}\delta _{n_{\sigma }m_{s}}$ with dual
generator $V_{-n_{\sigma }}^{(\sigma )}$ proportional to $W_{-n_{\sigma
}}^{(\sigma )}$ (say $V_{-n_{\sigma }}^{(\sigma )}=\mathrm{\chi }_{n_{\sigma
}}W_{-n_{\sigma }}^{(\sigma )}$), the components $A_{\mu {\small (\sigma )}%
}^{(\mathrm{g)}n_{\sigma }}$ and $\tilde{A}_{\mu {\small (\sigma )}}^{(%
\mathrm{g)}n_{\sigma }}$ can be also defined like%
\begin{equation}
A_{\mu {\small (\sigma )}}^{(\mathrm{g)}n_{\sigma }}=tr\left( V_{-n_{\sigma
}}^{(\sigma )}A_{\mu }^{\mathrm{g}}\right) ,\qquad \tilde{A}_{\mu {\small %
(\sigma )}}^{(\mathrm{g)}n_{\sigma }}=tr\left( V_{-n_{\sigma }}^{(\sigma )}%
\tilde{A}_{\mu }^{\mathrm{g}}\right)   \label{ev}
\end{equation}%
Moreover, setting $\mathrm{g=\exp }\varpi $ with some radial function $%
\varpi \left( \rho \right) $ valued in the exceptional ${\Large e}%
_{6}\rightarrow SO_{{\small 5,5}}$; we also have the expansion $\varpi =\sum
\varpi _{{\small (s)}}^{m_{s}}W_{m_{s}}^{(s)}$. Then, by using (\ref{LR}),
we can calculate the transformation $\mathrm{g}^{-1}\mathfrak{A}_{\mu }%
\mathrm{g}$ and $\mathrm{g}\widetilde{\mathfrak{A}}_{\mu }\mathrm{g}^{-1};$\
we obtain%
\begin{equation}
\begin{tabular}{lll}
$\mathrm{g}^{-1}\mathfrak{A}_{\mu }\mathrm{g}$ & $=$ & $\dsum\limits_{s\in
J_{{\small so}_{{\small 10}}}}\dsum\limits_{m_{s}=1-s}^{s-1}\mathfrak{A}%
_{\mu {\small (s)}}^{m_{s}}\left( e^{-ad_{\varpi }}W_{m_{s}}^{(s)}\right) $
\\
$\mathrm{g}\widetilde{\mathfrak{A}}_{\mu }\mathrm{g}^{-1}$ & $=$ & $%
\dsum\limits_{s\in J_{{\small so}_{{\small 10}}}}\dsum%
\limits_{m_{s}=1-s}^{s-1}\widetilde{\mathfrak{A}}_{\mu {\small (s)}%
}^{m_{s}}\left( e^{+ad_{\varpi }}W_{m_{s}}^{(s)}\right) $%
\end{tabular}%
\end{equation}%
with%
\begin{equation}
\mathrm{g}^{-1}W_{m_{s}}^{(s)}\mathrm{g}=e^{-ad_{\varpi
}}W_{m_{s}}^{(s)},\qquad \mathrm{g}W_{m_{s}}^{(s)}\mathrm{g}%
^{-1}=e^{+ad_{\varpi }}W_{m_{s}}^{(s)}
\end{equation}%
Using these relations and (\ref{trs}) namely $A_{\mu }^{\mathrm{g}}=\mathrm{g%
}^{-1}\left( \partial _{\mu }+\mathfrak{A}_{\mu }\right) \mathrm{g}$ and $%
\tilde{A}_{\mu }^{\mathrm{g}}=\mathrm{g}\left( \partial _{\mu }+\widetilde{%
\mathfrak{A}}_{\mu }\right) \mathrm{g}^{-1}$, we can also expand the $A_{\mu
}^{\mathrm{g}}$ and the $\tilde{A}_{\mu }^{\mathrm{g}}$ like,%
\begin{equation}
\begin{tabular}{lll}
$A_{\mu }^{\mathrm{g}}$ & $=$ & $+\partial _{\mu }\varpi +\dsum\limits_{s\in
J_{{\small so}_{{\small 10}}}}\dsum\limits_{m_{s}={\small 1-s}}^{{\small s-1}%
}\mathfrak{A}_{\mu {\small (s)}}^{m_{s}}\left( e^{-ad_{\varpi
}}W_{m_{s}}^{(s)}\right) $ \\
$\tilde{A}_{\mu }^{\mathrm{g}}$ & $=$ & $-\partial _{\mu }\varpi
+\dsum\limits_{s\in J_{{\small so}_{{\small 10}}}}\dsum\limits_{m_{s}=%
{\small 1-s}}^{{\small s-1}}\widetilde{\mathfrak{A}}_{\mu {\small (s)}%
}^{m_{s}}\left( e^{+ad_{\varpi }}W_{m_{s}}^{(s)}\right) $%
\end{tabular}
\label{exp}
\end{equation}%
Equating this development with eq(\ref{de}-\ref{ev}), we get%
\begin{equation}
\begin{tabular}{lll}
$A_{\mu {\small (\sigma )}}^{(\mathrm{g)}n_{\sigma }}$ & $=$ & $+\partial
_{\mu }\varpi _{{\small (\sigma )}}^{n_{\sigma }}+\dsum\limits_{s\in J_{%
{\small so}_{{\small 10}}}}\dsum\limits_{m_{s}={\small 1-s}}^{{\small s-1}}%
\mathfrak{A}_{\mu {\small (s)}}^{m_{s}}tr\left( V_{-n_{\sigma }}^{(\sigma
)}e^{-ad_{\varpi }}W_{m_{s}}^{(s)}\right) $ \\
$\tilde{A}_{\mu {\small (\sigma )}}^{(\mathrm{g)}n_{\sigma }}$ & $=$ & $%
-\partial _{\mu }\varpi _{{\small (\sigma )}}^{n_{\sigma
}}+\dsum\limits_{s\in J_{{\small so}_{{\small 10}}}}\dsum\limits_{m_{s}=%
{\small 1-s}}^{{\small s-1}}\widetilde{\mathfrak{A}}_{\mu {\small (s)}%
}^{m_{s}}tr\left( V_{-n_{\sigma }}^{(\sigma )}e^{+ad_{\varpi
}}W_{m_{s}}^{(s)}\right) $%
\end{tabular}%
\end{equation}%
Notice that because the boundary potentials have no radial components ($%
\mathfrak{A}_{\rho {\small (s)}}^{m_{s}}=\mathfrak{A}_{\rho {\small (s)}%
}^{m_{s}}=0$); it results that we must have
\begin{equation}
\begin{tabular}{lllll}
$A_{\rho {\small (\sigma )}}^{(\mathrm{g)}n_{\sigma }}$ & $=$ & $+\frac{%
\partial \varpi _{{\small (\sigma )}}^{n_{\sigma }}}{\partial \rho }$ & $%
\equiv $ & $+L_{0}\delta _{2,\sigma }\delta ^{0,n_{\sigma }}$ \\
$\tilde{A}_{\rho {\small (\sigma )}}^{(\mathrm{g)}n_{\sigma }}$ & $=$ & $-%
\frac{\partial \varpi _{{\small (\sigma )}}^{n_{\sigma }}}{\partial \rho }$
& $\equiv $ & $-L_{0}\delta _{2,\sigma }\delta ^{0,n_{\sigma }}$%
\end{tabular}%
\end{equation}%
Notice also that if one expresses the adjoint group actions $e^{\pm
ad_{\varpi }}W_{m_{s}}^{(s)}$ in terms of the following expansions%
\begin{equation}
\begin{tabular}{lll}
$e^{-ad_{\varpi }}W_{m_{s}}^{(s)}$ & $=$ & $\dsum\limits_{\tau \in J_{%
{\small so}_{{\small 10}}}}\dsum\limits_{l_{\tau }={\small 1-\tau }}^{%
{\small \tau -1}}\mathcal{N}_{m_{s},\tau }^{(s),l_{\tau }}W_{l_{\tau
}}^{(\tau )}$ \\
$e^{+ad_{\varpi }}W_{m_{s}}^{(s)}$ & $=$ & $\dsum\limits_{\tau \in J_{%
{\small so}_{{\small 10}}}}\dsum\limits_{l_{\tau }={\small 1-\tau }}^{%
{\small \tau -1}}\widetilde{\mathcal{N}}_{m_{s},\tau }^{(s),l_{\tau
}}W_{l_{\tau }}^{(\tau )}$%
\end{tabular}%
\end{equation}%
it results that the $\mathcal{N}_{m_{s},\sigma }^{(s),n_{\sigma }}$ and $%
\widetilde{\mathcal{N}}_{m_{s},\sigma }^{(s),n_{\sigma }}$ are nothing but
\begin{equation}
\begin{tabular}{lll}
$tr\left( V_{-n_{\sigma }}^{(\sigma )}e^{-ad_{\varpi
}}W_{m_{s}}^{(s)}\right) $ & $=$ & $\mathcal{N}_{m_{s},\sigma
}^{(s),n_{\sigma }}$ \\
$tr\left( V_{-n_{\sigma }}^{(\sigma )}e^{+ad_{\varpi
}}W_{m_{s}}^{(s)}\right) $ & $=$ & $\widetilde{\mathcal{N}}_{m_{s},\sigma
}^{(s),n_{\sigma }}$%
\end{tabular}%
\end{equation}%
Consequently, we end up with the following interesting relations%
\begin{equation}
\begin{tabular}{lll}
$A_{\mu {\small (\sigma )}}^{(\mathrm{g)}n_{\sigma }}$ & $=$ & $+L_{0}\delta
_{2,\sigma }\delta ^{0,n_{\sigma }}\delta _{\mu ,\rho }+\dsum\limits_{s\in
J_{{\small so}_{{\small 10}}}}\dsum\limits_{m_{s}={\small 1-s}}^{{\small s-1}%
}\mathfrak{A}_{\mu {\small (s)}}^{m_{s}}\mathcal{N}_{m_{s},\sigma
}^{(s),n_{\sigma }}$ \\
$\tilde{A}_{\mu {\small (\sigma )}}^{(\mathrm{g)}n_{\sigma }}$ & $=$ & $%
-L_{0}\delta _{2,\sigma }\delta ^{0,n_{\sigma }}\delta _{\mu ,\rho
}+\dsum\limits_{s\in J_{{\small so}_{{\small 10}}}}\dsum\limits_{m_{s}=%
{\small 1-s}}^{{\small s-1}}\widetilde{\mathfrak{A}}_{\mu {\small (s)}%
}^{m_{s}}\widetilde{\mathcal{N}}_{m_{s},\sigma }^{(s),n_{\sigma }}$%
\end{tabular}%
\end{equation}%
and the dreibein%
\begin{equation}
\begin{tabular}{lll}
$\boldsymbol{E}_{\mu }^{\mathrm{g}}$ & $=$ & $E_{\mu {\small (2)}}^{(\mathrm{%
g)}m_{2}}W_{m_{2}}^{(2)}+E_{\mu {\small (3)}}^{(\mathrm{g)}%
m_{3}}W_{m_{3}}^{(3)}+E_{\mu {\small (4)}}^{(\mathrm{g)}%
m_{4}}W_{m_{4}}^{(4)}+$ \\
&  & $E_{\mu {\small (5)}}^{(\mathrm{g)}m_{5}}W_{m_{5}}^{(5)}+E_{\mu {\small %
(11)}}^{(\mathrm{g)}m_{11}}W_{m_{11}}^{(11)}+E_{\mu {\small (17)}}^{(\mathrm{%
g)}m_{17}}W_{m_{17}}^{(17)}$%
\end{tabular}%
\end{equation}%
with dreibeins' components as%
\begin{equation}
E_{\mu {\small (s)}}^{(\mathrm{g)}m_{s}}=A_{\mu {\small (s)}}^{(\mathrm{g)}%
m_{s}}-\tilde{A}_{\mu {\small (s)}}^{(\mathrm{g)}m_{s}}
\end{equation}%
the ortho- exceptional higher spin BTZ metric is therefore%
\begin{equation}
G_{\mu \nu }^{\mathrm{g}}=\frac{1}{2}\dsum\limits_{s,\sigma \in J_{sl_{%
{\small 6}}}\times J_{sl_{{\small 6}}}}\dsum\limits_{m_{s}={\small 1-s}%
}^{s-1}\dsum\limits_{n_{\sigma }=1-\sigma }^{\sigma -1}E_{\mu {\small (s)}%
}^{(\mathrm{g)}m_{s}}\mathcal{K}_{m_{s}n_{\sigma }}^{s,\sigma }E_{\nu
{\small (\sigma )}}^{(\mathrm{g)}n_{\sigma }}  \label{so}
\end{equation}%
In the end, notice that by taking $\mathrm{g}=\mathrm{g}_{{\small GK}}=%
\mathrm{e}^{\rho L_{0}}$, we obtain the Gutperle-Kraus higher spin metric $%
G_{\mu \nu }^{{\small GK}}$ for the ortho-exceptional model; while for $%
\mathrm{g}=\mathrm{g}_{{\small FG}}$, we get the generalisation of the
Fefferman-Graham higher spin metric $G_{\mu \nu }^{{\small FG}}$.

\subsection{BTZ black hole solution III: pattern $E_{6}\rightarrow
F_{4}\rightarrow G_{2}$}

Here, we investigate a third solution besides Levi decompositions using the
embedding pattern
\begin{equation*}
E_{6}\supset F_{4}\supset G_{2}
\end{equation*}%
This pattern relies on constructing a bridge starting from the G$_{2}$ model
towards E$_{6}$ passing through F$_{4}$. Recall that an example of
exceptional higher spin BTZ black hole solution is given by the G$_{2}$
model \cite{trunc} having two spins: $s_{1}=2$ and unknown $s_{2}=\sigma $
to determine. This feature follows from the fact that the number of spins is
just the $rank(G_{2})=2$; recall that the $\dim G_{2}=14$. Because here
there is only one unknown spin $s_{2}=(j_{2}+1)$, one can calculate it by
using the decomposition%
\begin{equation}
14=3+(2s_{2}-1)=3+11
\end{equation}%
this gives $s_{2}=6.$ However, the exceptional G$_{2}$ higher spin model can
be also obtained by truncating $hs\left[ \lambda \right] $ theory to Type II
Spin 6 gravity with $SO(3,4)$ symmetry \cite{trunc}. This approach, termed
below as \emph{Top-Down} method, takes advantage from:

\begin{description}
\item[$\left( \mathbf{i}\right) $] the triality property of the SO(4,4)
symmetry, nicely exhibited by the outer automorphism of the SO(4,4)
Tits-Satake diagram as depicted by the \textbf{Figure} \ref{fol};
\begin{figure}[ph]
\begin{center}
\includegraphics[width=14cm]{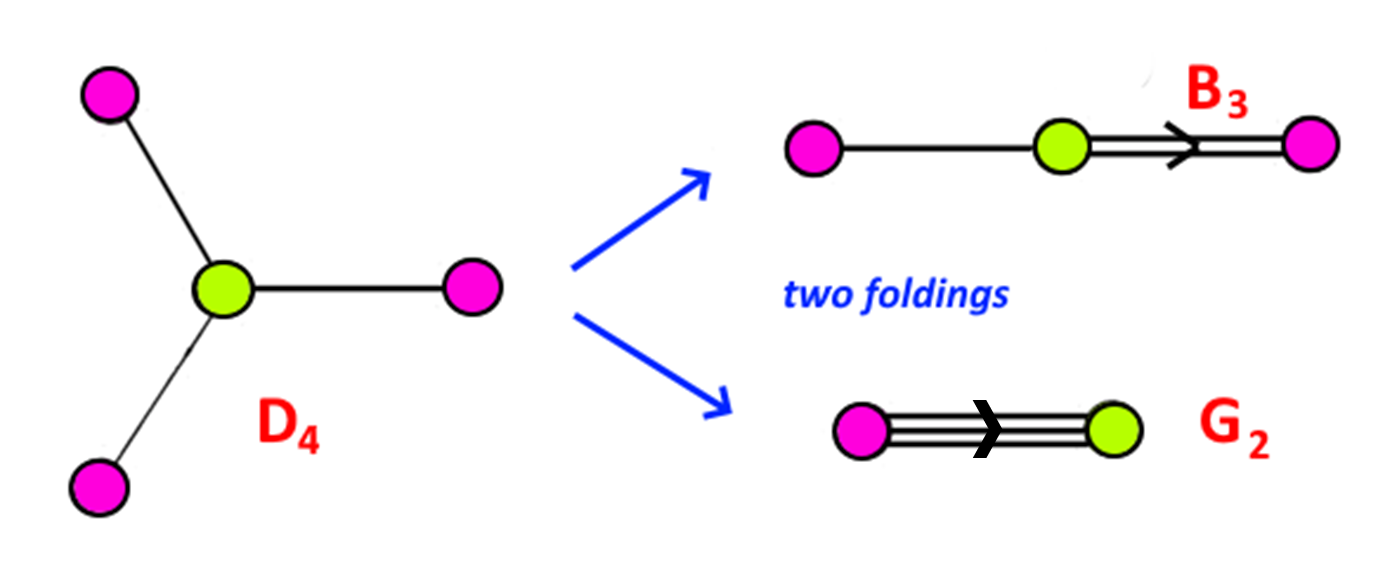}
\end{center}
\par
\vspace{-0.5cm}
\caption{the two foldings of the Dynkin diagram D4 (Tits-Satake graph) with
outer-automorphism symmetry $\mathbb{S}_{3}$.}
\label{fol}
\end{figure}

\item[$\left( \mathbf{ii}\right) $] the symmetry embeddings%
\begin{equation}
G_{2}\subset SO(7)\subset SO(8)  \label{emb}
\end{equation}%
suggesting the following dimension splittings%
\begin{equation}
\begin{tabular}{lllll}
$28$ & $=$ & $21+8$ & $=$ & $3+\dsum\limits_{n=1}^{3}\left( 2j_{n}+1\right) $
\\
$21$ & $=$ & $14+7$ & $=$ & $3+\dsum\limits_{n=1}^{2}\left( 2j_{n}+1\right) $%
\end{tabular}%
\end{equation}%
with $s_{n}=j_{n}+1$.
\end{description}

\begin{figure}[tbp]
\begin{center}
\includegraphics[width=14cm]{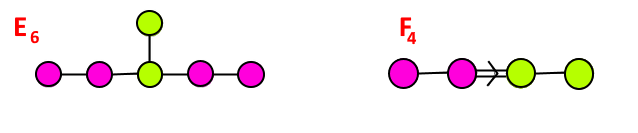}
\end{center}
\par
\vspace{-0.5cm}
\caption{The folding of the Dynkin diagram E$_{6}$ down to F$_{4}$ by using
outer-automorphism symmetry $\mathbb{Z}_{2}$.}
\label{F4}
\end{figure}

In this Top-Down approach, the 21 dimensions of the $SO(7)$ gauge symmetry
are decomposed in \textrm{SL}$_{2}$ isomultiplets like $21=3+7+11;$ thus
leading to the higher spins $s=2,4,6.$ By omitting the spin 4 generators or
equivalently the 7 dimensional isomultiplet, one obtains the two desired
spins $s=\left\{ 2,6\right\} $ of the G$_{2}$ theory in agreement with the
dimension decomposition $14=3+11.$We can take advantage of the analysis that
we have done for the exceptional G$_{2}$ model by using a Bottom-Up approach
with the following exceptional embeddings%
\begin{equation}
G_{2}\subset F_{4}\subset E_{6}  \label{exb}
\end{equation}%
as depicted by the \textbf{Figure} \textbf{\ref{F4}}.
This solution is clearly different than the $E_{6}\rightarrow SL_{6}$ and
the $E_{6}\rightarrow SO_{5,5}$ models we have considered before; it
indicates that the constructions we have built and which have been motivated
by the two Levi-decompositions of E$_{6}$ algebra are not the sole possible
solutions. Below, we comment on the exceptional embeddings (\ref{exb}) with
the following dimension decompositions%
\begin{equation*}
\begin{tabular}{lllllll}
$F_{4}$ & : & $52$ & $=$ & $3+7+11+31$ & $=$ & $3+\dsum\limits_{n=2}^{4}%
\left( 2j_{n}+1\right) $ \\
$E_{6}$ & : & $78$ & $=$ & $52+26$ & $=$ & $3+\dsum\limits_{n=2}^{6}\left(
2j_{n}+1\right) $%
\end{tabular}%
\end{equation*}%
In the first row of these relations, we have used the embedding $%
G_{2}\subset SO(7)\subset F_{4}$ predicting therefore the desired isospin $%
j_{4}$ of the fourth multiplet as it solves the condition $\left(
2j_{4}+1\right) =31$ giving $j_{4}=15.$Thus the four higher conformal spins
content of the F$_{4}$ theory are as follows
\begin{equation}
\left( s_{1},s_{2},s_{3},s_{4}\right) =\left( 2,4,6,16\right) ,\qquad \left(
j_{1},j_{2},j_{3},j_{4}\right) =\left( 1,3,5,15\right)
\end{equation}%
In the second row, we have given constrained data on the two missing SL$_{2}$
isomultiplets with isospin $j_{5}$ and $j_{6}$ required by the rank of E$%
_{6} $. The values of these isospins are constrained by the decomposition of
the E$_{6}$ dimension as $78=52+26;$ this leads to the constraint $\left(
2j_{5}+1\right) +\left( 2j_{6}+1\right) =26$ having several solutions. A
simple candidate solution is that given by taking $j_{5}=j_{6}=6$ and then
the conformal spins $s_{5}=s_{6}=7$ indicating the existence of two currents
with same conformal spin. Other solutions are given by%
\begin{equation}
\begin{tabular}{llllllll}
$2j_{5}+1$ & : & 13 & 11 & 9 & 7 & 5 & 3 \\
$2j_{6}+1$ & : & 13 & 15 & 17 & 19 & 21 & 23%
\end{tabular}%
\end{equation}%
and then%
\begin{equation}
\begin{tabular}{llllllll}
$s_{5}$ & : & 7 & 6 & 5 & 4 & 3 & 2 \\
$s_{6}$ & : & 7 & 8 & 9 & 10 & 11 & 12%
\end{tabular}%
\end{equation}

\section{Thermodynamics of BTZ black hole in E$_{6}$ gravity}

\quad For completeness, we study in this section the thermodynamics of the
HS BTZ black holes for the two solutions $E_{6}\rightarrow SL_{6}$ and $%
E_{6}\rightarrow SO_{5,5}$. We give the partition functions associated to
each conformal spin spectrum for the two spin models using a one to one
correspondence between the root system and the factors of the vacuum
character of the asymptotic W-algebra.

The partition function of an Euclidean BTZ black hole with 2-torus topology
at its asymptote is given by \cite{t1,t2,t3,spin3}:%
\begin{equation}
Z_{\text{{\small BTZ}}}\left( \tau ,\tilde{\tau}\right) =tr(q^{L_{0}}\tilde{q%
}^{\tilde{L}_{0}})
\end{equation}%
For the $SL_{{\small 3}}$ based theory, the partition function $Z_{\text{%
{\small HS-BTZ}}}^{{\small SL}_{{\small 3}}}$ of the spin 3 BTZ black hole
includes a contribution from the chemical potentials conjugate to the spin-3
currents as follows \textrm{\cite{spin3}},%
\begin{equation}
Z_{\text{{\small HS-BTZ}}}^{{\small SL}_{{\small 3}}}\left( \tau ,\alpha ,%
\tilde{\tau},\tilde{\alpha}\right) =tr\left[ q^{L_{0}-\frac{c}{24}}\tilde{q}%
^{-\tilde{L}_{0}-\frac{c}{24}}y^{W_{0}}\tilde{y}^{-\tilde{W}_{0}}\right]
\end{equation}%
\textrm{where} $q=e^{2i\pi \mathrm{\tau }}$ and $y=e^{2i\pi \mathrm{\alpha }}
$ with $\left( \mathrm{\tau },\mathrm{\tilde{\tau}}\right) $ being\ the
complex parameters of the boundary 2-torus ($\mathrm{\tilde{\tau}}=-\mathrm{%
\tau }$) ; and $\left( \mathrm{\alpha },\mathrm{\tilde{\alpha}}\right) $ the
spin-3 chemical potentials such that $\mathrm{\alpha }=-\mathrm{\tau }%
\mathcal{W}_{+}^{--}$ and $\mathrm{\tilde{\alpha}}=\mathrm{\tau }\mathcal{%
\tilde{W}}_{-}^{--}.$ For a higher spin theory with gauge symmetry $\mathcal{%
G}$ with spins $s\geq 3$ in some set $J_{hs}^{\mathcal{G}},$ the
construction given above can be generalised to $Z_{\text{{\small HS-BTZ}}}^{%
\mathcal{G}}$ given by
\begin{equation}
Z_{\text{{\small HS-BTZ}}}^{\mathcal{G}}\left( \tau ,\alpha _{J},\tilde{\tau}%
,\tilde{\alpha}_{J}\right) =tr\left[ q^{L_{0}-\frac{c}{24}}\tilde{q}^{-%
\tilde{L}_{0}-\frac{c}{24}}\prod_{s\in J_{hs}}y_{s}^{W_{{\small 0(s)}}}%
\tilde{y}_{s}^{-\tilde{W}_{{\small 0(s)}}}\right]
\end{equation}%
Building on the preceding partition function, we construct the partition
functions of the exceptional E$_{6}$ higher spin BTZ black hole for the E$%
_{6}\rightarrow SL_{6}$ and the E$_{6}\rightarrow SO_{5,5}$ models
respectively denoted like $\mathcal{Z}_{{\small E}_{{\small 6}}}^{{\small SL}%
_{{\small 6}}}$ and $\mathcal{Z}_{{\small E}_{{\small 6}}}^{{\small SO}_{%
{\small 5,5}}}$. As we have seen before, the E$_{6}$ higher spin theory
involves a rich set of chemical potentials \{$\mathfrak{A}_{+{\small (s)}%
}^{m_{s}}$\} compared to the well studied SL$_{3}$ theory having \{$%
\mathfrak{A}_{+{\small (2)}}^{m_{2}}$, $\mathfrak{A}_{+{\small (3)}%
}^{m_{3}}\}$. This richness of the chemical potentials leads to a growth of
the partition function \cite{spin3} which can be defined like%
\begin{equation}
\mathcal{Z}_{E_{6}}=tr\left[ \dprod\nolimits_{s}q_{{\small (s)}}^{W_{\mathrm{%
0}}^{{\small (s)}}}\tilde{q}_{{\small (s)}}^{W_{\mathrm{0}}^{{\small (s)}}}%
\right]   \label{6e}
\end{equation}%
with
\begin{equation}
\begin{tabular}{lllllll}
$q_{{\small (s)}}$ & $=$ & $e^{2i\pi \tau \mathcal{A}_{+}^{{\small (s-1)}}}$
& $,\qquad $ & $\mathcal{A}_{+}^{{\small (s-1)}}$ & $=$ & $\delta
_{s-1,m_{s}}\mathfrak{A}_{+{\small (s)}}^{m_{s}}$ \\
$\tilde{q}_{{\small (s)}}$ & $=$ & $e^{2i\pi \tau \widetilde{\mathcal{A}}%
_{+}^{{\small (s-1)}}}$ & $,\qquad $ & $\widetilde{\mathcal{A}}_{+}^{{\small %
(s-1)}}$ & $=$ & $\delta _{s-1,m_{s}}\widetilde{\mathfrak{A}}_{+{\small (s)}%
}^{m_{s}}$%
\end{tabular}%
\end{equation}%
where $\mathfrak{A}_{+{\small (s)}}^{m_{s}}$ and $\widetilde{\mathfrak{A}}_{+%
{\small (s)}}^{m_{s}}$ are the left and right chemical potentials associated
with the conformal spins. In general, the spin-3 charge contribution $%
y^{W_{0}^{{\small (3)}}}$ to the sl(3) HS-BTZ black hole's partition
function $\mathcal{Z}_{{\small SL}_{{\small 3}}}$ was made via its chemical
potential $\mathfrak{A}_{+{\small (3)}}^{++}=\mathcal{W}_{+}^{++}$ with $q_{%
{\small (s)}}$ given by $y=e^{2i\pi \mathrm{\alpha }}$ and $\mathrm{\alpha }%
=-\tau \mathcal{W}_{+}^{++}$ \cite{spin3}.

\subsection{Linear-exceptional model}

Expanding upon this, we construct the exceptional E$_{6}$ HS-partition
function by considering the sum of contributions stemming from the set of
conformal spin charges.\newline
For the E$_{6}\rightarrow SL_{6}$ model, the\emph{\ holomorphic} part of
partition function is given by
\begin{equation}
Z_{{\small E}_{{\small 6}}}^{{\small SL}_{{\small 6}}}=tr\left[
\dprod\nolimits_{s\in J_{{\small Sl}_{{\small 6}}}}q_{{\small (s)}}^{W_{%
\mathrm{0}}^{{\small (s)}}}\right] ,\qquad W_{0}^{{\small (2)}}=L_{0}-\frac{c%
}{24}
\end{equation}%
with higher spin content given by the set $J_{{\small SL}_{{\small 6}%
}}=\{2,3,4,5,6,22\}.$ The six conformal spins contribute in $Z_{{\small SL}_{%
{\small 6}}}^{E_{6}}$ with the factors $q_{{\small (s)}}^{W_{\mathrm{0}}^{%
{\small (s)}}}$ expanding like%
\begin{equation}
Z_{{\small E}_{{\small 6}}}^{{\small SL}_{{\small 6}}}[\tau ,\theta _{%
{\small (s)}}]=tr\left[ q^{W_{0}^{{\small (2)}}}y^{W_{0}^{(3)}}p^{W_{0}^{%
\left( 4\right) }}x^{W_{0}^{\left( 5\right) }}u^{W_{0}^{\left( 6\right)
}}v^{W_{0}^{\left( 22\right) }}\right]
\end{equation}%
with%
\begin{equation}
\begin{tabular}{llll}
$s=2$ & $\rightarrow $ & $q^{W_{0}^{{\small (2)}}}$ & $=e^{2i\pi \tau \left(
L_{0}-\frac{c}{24}\right) }$ \\
$s=3$ & $\rightarrow $ & $y^{W_{0}^{(3)}}$ & $=e^{2i\pi \theta _{{\small (3)}%
}W_{0}^{{\small (3)}}}$ \\
$s=4$ & $\rightarrow $ & $p^{W_{0}^{\left( 4\right) }}$ & $=e^{2i\pi \theta
_{{\small (4)}}W_{0}^{(4)}}$ \\
$s=5$ & $\rightarrow $ & $x^{W_{0}^{\left( 5\right) }}$ & $=e^{2i\pi \theta
_{{\small (5)}}W_{0}^{\left( 5\right) }}$ \\
$s=6$ & $\rightarrow $ & $u^{W_{0}^{\left( 11\right) }}$ & $=e^{2i\pi \theta
_{{\small (6)}}W_{0}^{\left( 6\right) }}$ \\
$s=22$ & $\rightarrow $ & $v^{W_{0}^{\left( 22\right) }}$ & $=e^{2i\pi
\theta _{{\small (22)}}W_{0}^{\left( 22\right) }}$%
\end{tabular}%
\end{equation}%
where we have set\textrm{\ }$\theta _{{\small (s)}}=-\tau \mathfrak{A}_{+%
{\small (s)}}^{{\small s-1}}.$

Notice that in eq(\ref{es}) we only express the holomorphic partition
function $Z_{{\small E}_{{\small 6}}}^{{\small SL}_{{\small 6}}}$ to prevent
clutter. But there is also a similar contribution with a tilted function $%
\bar{Z}_{{\small E}_{{\small 6}}}^{{\small SL}_{{\small 6}}}$ describing the
anti-holomorphic sector such that the full partition function $\mathcal{Z}_{%
{\small E}_{{\small 6}}}^{{\small SL}_{{\small 6}}}$ of the E$%
_{6}\rightarrow SL_{6}$ model includes the block%
\begin{equation}
Z_{{\small E}_{{\small 6}}}^{{\small SL}_{{\small 6}}}\times \bar{Z}_{%
{\small E}_{{\small 6}}}^{{\small SL}_{{\small 6}}}
\end{equation}%
with $Z_{{\small E}_{{\small 6}}}^{{\small SL}_{{\small 6}}}$ as in eq(\ref%
{es}) and $\bar{Z}_{{\small E}_{{\small 6}}}^{{\small SL}_{{\small 6}}}$
given by
\begin{equation}
\bar{Z}_{{\small E}_{{\small 6}}}^{{\small SL}_{{\small 6}}}=tr\left[
\dprod\nolimits_{s\in J_{{\small SL}_{{\small 6}}}}\tilde{q}_{{\small (s)}%
}^{W_{\mathrm{0}}^{{\small (s)}}}\right]
\end{equation}%
with $\tilde{q}_{{\small (s)}}=e^{-2i\pi \tilde{\theta}_{{\small (s)}}}$ and
$\theta _{{\small (s)}}=-\tau \widetilde{\mathfrak{A}}_{+{\small (s)}}^{%
{\small s-1}}$ where $\tau \ $is the parameter of the boundary 2-torus. A
remarkable beneficial one to one correspondence between the root system of
the Lie algebra and the vacuum character of the associated W-algebra was
developed in \cite{son} and will be used subsequently to derive the
expression of $Z_{{\small E}_{{\small 6}}}^{{\small SL}_{{\small 6}}}\ $in a
particular setting by generalising the\textrm{\ }computation of $Z_{{\small %
sl}_{{\small 3}}}^{{\small sl}_{{\small 2}}}$ which is known to factorise as
follows
\begin{equation}
Z_{{\small sl}_{{\small 3}}}^{{\small sl}_{{\small 2}}}=\left\vert \mathbf{%
\chi }_{1}^{sl_{3}}\right\vert ^{2}  \label{Z}
\end{equation}%
where $\mathbf{\chi }_{1}^{sl_{3}}$ is the vacuum character of the $WSL_{3}$%
- algebra.

For the rest of the analysis, we restrict to the case of vanishing higher
spin chemical potentials $W_{0}^{(s\geq 3)}=0$ as vacuum characters of the
associated W-algebra are only explicitly known for this particular instance.
Indeed, following \cite{35}-\cite{38} and using thermal AdS$_{3}$
formulation, the one-loop contribution of the boundary conformal spins $2$
and $3$ to the partition function $Z_{{\small sl}_{{\small 3}}}^{{\small sl}%
_{{\small 2}}}$ can be expressed as in (\ref{Z}) with $\mathbf{\chi }%
_{1}^{sl_{3}}$ given by%
\begin{equation}
\mathbf{\chi }_{1}^{sl_{3}}=q^{-\frac{c}{24}}\dprod\limits_{s=2}^{3}\left(
\dprod\limits_{n=s}^{\infty }\frac{1}{1-q^{n}}\right)
\end{equation}%
also reading as%
\begin{equation}
\mathbf{\chi }_{1}^{sl_{3}}=\frac{1}{\mathbf{\eta }\left( q\right) ^{2}}q^{-%
\frac{c-2}{24}}\left( 1-q\right) ^{2}\left( 1-q^{2}\right)  \label{ZZ}
\end{equation}%
where $\mathbf{\eta }\left( q\right) =q^{\frac{1}{24}}\dprod%
\nolimits_{n=1}^{\infty }\left( 1-q^{n}\right) $ is the Dedekind eta
function \textrm{\cite{33A}-\cite{33C}.} The above formula\textrm{\ }%
involves three factors namely: $\left( i\right) $ a squared $\left(
1-q\right) $ each in one to one correspondence with the two simple roots $%
\alpha _{1},$ $\alpha _{2}$ of the positive root system $\Phi _{sl_{3}}^{+}$%
; and $\left( ii\right) $ one factor $\left( 1-q^{2}\right) $ associated to
the positive root $\alpha _{1}+\alpha _{2}.$ Recall that $\Phi _{sl_{3}}^{+}$
has three roots \{$\alpha _{1},$ $\alpha _{2},\alpha _{1}+\alpha _{2}$\}. In
terms of the splitting $(sl_{3}\backslash sl_{2})\oplus sl_{2},$ the vacuum
character $\mathbf{\chi }_{1}^{sl_{3}}$ can be factorised like%
\begin{equation}
\mathbf{\chi }_{1}^{sl_{3}}=\mathbf{\chi }_{1}^{(sl_{3}\backslash sl_{2})}.%
\mathbf{\chi }_{1}^{sl_{2}}\qquad ,\qquad \mathbf{\chi }_{1}^{(sl_{3}%
\backslash sl_{2})}=\frac{\mathbf{\chi }_{1}^{sl_{3}}}{\mathbf{\chi }%
_{1}^{sl_{2}}}
\end{equation}%
Because $\mathbf{\chi }_{1}^{sl_{2}}=q^{-\frac{c}{24}}q^{+\frac{1}{24}%
}\left( 1-q\right) /\mathbf{\eta }\left( q\right) $, we can then calculate
the contribution of the boundary conformal spin $3$ to the character $%
\mathbf{\chi }_{1}^{sl_{3}},$ we find%
\begin{equation}
\mathbf{\chi }_{1}^{(sl_{3}\backslash sl_{2})}\left( \tau \right) =\frac{1}{%
\mathbf{\eta }\left( q\right) }q^{\frac{1}{24}}\left( 1-q\right) \left(
1-q^{2}\right)  \label{A21}
\end{equation}%
Before proceeding, notice that the contribution $q^{+\frac{1}{24}}\left(
1-q\right) /\mathbf{\eta }\left( q\right) $ of $\mathbf{\chi }_{1}^{sl_{2}}$
can be interpreted in terms of the sl$_{2}$ generators as follows: the $q^{%
\frac{1}{24}}/\mathbf{\eta }\left( q\right) $ is associated with the Cartan
generator H$_{\alpha _{1}}$ while the factor $\left( 1-q\right) $ correspond
to the step generator E$_{+\alpha _{1}}$. Therefore\textrm{\ }by comparing
the character $\mathbf{\chi }_{1}^{(sl_{3}\backslash sl_{2})}$ to $\mathbf{%
\chi }_{1}^{sl_{2}}$, we learn the following: $\left( i\right) $ the factor $%
q^{\frac{1}{24}}/\mathbf{\eta }\left( q\right) $ is associated with the
second Cartan generator H$_{\alpha _{2}}$ of sl$_{3}$. $\left( ii\right) $
And the two factors $\left( 1-q\right) $ and $\left( 1-q^{2}\right) $ relate
respectively to the two positive roots $\alpha _{2}$ and $\alpha _{1}+\alpha
_{2}$ generating $\Phi _{sl_{3}}^{+}/\Phi _{sl_{2}}^{+}.$

Given this description regarding the partition function $Z_{{\small sl}_{%
{\small 3}}}^{{\small sl}_{{\small 2}}}$ for the higher spin model $%
SL_{3}\rightarrow SL_{2},$ we turn now to the computation of $Z_{{\small E}_{%
{\small 6}}}^{{\small SL}_{{\small 6}}}$ of the higher spin model $E_{%
{\small 6}}\rightarrow SL_{6}$.\newline
First, notice that the generalisation of eqs(\ref{Z}-\ref{ZZ}) to the
special linear $SL\left( 6,\mathbb{R}\right) $ gives the partition function $%
Z_{{\small sl}_{{\small 6}}}^{{\small sl}_{{\small 2}}}=|\mathbf{\chi }_{1}^{%
{\small sl}_{{\small 6}}}\mathbf{|}^{2}$ with vacuum character $\mathbf{\chi
}_{1}^{{\small sl}_{{\small 6}}}$ as%
\begin{equation}
\mathbf{\chi }_{1}^{{\small sl}_{{\small 6}}}=q^{-\frac{c}{24}%
}\dprod\limits_{s=1}^{5}\left( \dprod\limits_{n=s}^{\infty }\frac{1}{1-q^{n}}%
\right)  \label{Ann}
\end{equation}%
reading in terms of the Dedekind eta function like%
\begin{equation}
\mathbf{\chi }_{1}^{{\small sl}_{{\small 6}}}=q^{-\frac{c}{24}}\frac{q^{%
\frac{5}{24}}}{\left[ \mathbf{\eta }\left( q\right) \right] ^{5}}%
\dprod\limits_{j=1}^{5}\left( 1-q^{j}\right) ^{5-j}  \label{An}
\end{equation}%
Notice also that the factor $q^{\frac{5}{24}}=\tprod\nolimits_{j=1}^{5}q^{%
\frac{1}{24}},$ divided by $[\mathbf{\eta }\left( q\right) ]^{5},$ is
associated with the five Cartan generators H$_{\alpha _{i}}$ of the Lie
algebra $sl_{6};$ while the factors in the product $\tprod%
\nolimits_{j=1}^{5}\left( 1-q^{j}\right) ^{5-j}$ can be put in
correspondence with the \emph{15} positive roots of in $\Phi _{{\small sl}_{%
{\small 6}}}^{+}$ like%
\begin{table}[h]
  \centering
  \begin{tabular}{c||c|c|c|c|c}
{\small factor} & $\left( 1-q\right) ^{5}$ & $\left( 1-q^{2}\right) ^{4}$ & $%
\left( 1-q^{3}\right) ^{3}$ & $\left( 1-q^{4}\right) ^{2}$ & $\left(
1-q^{5}\right) $ \\ \hline
{\small root} & $\alpha _{i}$ & $\alpha _{i}+\alpha _{i+1}$ & $\alpha
_{i}+\alpha _{i+1}+\alpha _{i+2}$ & $\alpha _{i}+...+\alpha _{i+3}$ & $%
\alpha _{1}+...+\alpha _{5}$ \\ \hline
{\small number} & ${\small 5}$ & ${\small 4}$ & ${\small 3}$ & ${\small 2}$
& ${\small 1}$ \\ \hline
\end{tabular}
\caption{Partition of the 15 positive roots and the associated factors of the vacuum character.}
\label{tab}
\end{table}
To get the expression of the partition function $Z_{{\small E}_{{\small 6}}}$
factorising like $|\mathbf{\chi }_{1}^{{\small E}_{{\small 6}}}|^{2}$, we
proceed as for $Z_{{\small sl}_{{\small 3}}}$ case; we use the relation%
\begin{equation}
\mathbf{\chi }_{1}^{{\small E}_{{\small 6}}}=\mathbf{\chi }_{1}^{{\small sl}%
_{{\small 6}}}.\mathbf{\chi }_{1}^{({\small E}_{{\small 6}}/{\small sl}_{%
{\small 6}})}
\end{equation}%
with $\mathbf{\chi }_{1}^{{\small sl}_{{\small 6}}}$ as in (\ref{Ann}-\ref%
{An}). To determine the missing $\mathbf{\chi }_{1}^{({\small E}_{{\small 6}%
}/{\small sl}_{{\small 6}})},$ we use the algorithm of the table (\ref{tab}%
). In this regard, notice that the \emph{36} positive roots split like \emph{%
15+21}; the \emph{15} roots of sl$_{6}$ are as in (\ref{tab}); they
contribute by (\ref{An}). The \emph{21} roots are given by the coset $\Phi _{%
{\small E}_{{\small 6}}}^{+}/\Phi _{{\small sl}_{{\small 6}}}^{+}$; they
contribute as follows:%
\begin{table}[h!]
\centering
\begin{tabular}{ll}
\ \ \ \ \ \ \ \ \ root & \ \ \ contribution \ \ \  \\ \hline\hline
$\ \ \ \ \ \ \ \ \ \ \left. \alpha _{6}\right. $ & \ \ \ $\ \ \ \ \left(
1-q\right) $ \\
$\ \ \ \ \ \ \ \left. \alpha _{3}+\alpha _{6}\right. $ & \ \ \ $\ \ \ \
\left( 1-q^{2}\right) $ \\
\ \ \  &  \\
$\left.
\begin{array}{c}
\alpha _{3}+\alpha _{2}+\alpha _{6} \\
\alpha _{4}+\alpha _{3}+\alpha _{6}%
\end{array}%
\right. $ & \ \ \ $\ \ \ \ \left( 1-q^{3}\right) ^{2}$ \\
&  \\
$\left.
\begin{array}{c}
\alpha _{1}+\alpha _{3}+\alpha _{2}+\alpha _{6} \\
\alpha _{5}+\alpha _{4}+\alpha _{3}+\alpha _{6}%
\end{array}%
\right. $ & \ \ \ $\ \ \ \ \left( 1-q^{4}\right) ^{2}$ \\ \hline\hline
\end{tabular}
\caption{The first set of roots and the associated vacuum character factors}
\label{l1}
\end{table}%
\begin{table}[h!]
\centering
\begin{tabular}{ll}
\ \ \ \ \ \ \ \ \ \ \ \ \ \ \ root & \ \ \ contribution \ \ \ \  \\
\hline\hline
$\left.
\begin{array}{c}
\alpha _{4}+2\alpha _{3}+\alpha _{2}+\alpha _{6} \\
\alpha _{4}+\alpha _{3}+2\alpha _{2}+\alpha _{6} \\
\alpha _{5}+\alpha _{4}+\alpha _{3}+\alpha _{2}+\alpha _{6} \\
\alpha _{1}+\alpha _{4}+\alpha _{3}+\alpha _{2}+\alpha _{6}%
\end{array}%
\right. $ & $\ \ \ \ \ \ \left( 1-q^{5}\right) ^{4}$ \\
\ \ \ \  &  \\
$\left.
\begin{array}{c}
\alpha _{5}+\alpha _{4}+2\alpha _{3}+\alpha _{2}+\alpha _{6} \\
\alpha _{1}+\alpha _{4}+2\alpha _{3}+\alpha _{2}+\alpha _{6} \\
\alpha _{1}+\alpha _{5}+\alpha _{4}+\alpha _{3}+\alpha _{2}+\alpha _{6}%
\end{array}%
\right. $ & \ \ \ $\ \ \ \ \left( 1-q^{6}\right) ^{3}$ \\ \hline\hline
\end{tabular}%
\caption{The second set of roots and the associated vacuum character factors}
\end{table}%
\begin{table}[h!]
\centering
\begin{tabular}{ll}
\ \ \ \ \ \ \ \ \ \ \ \ \ \ \ root & \ \ \ contribution \ \ \  \\
\hline\hline
$\left.
\begin{array}{c}
\alpha _{5}+2\alpha _{4}+2\alpha _{3}+\alpha _{2}+\alpha _{6} \\
\alpha _{1}+\alpha _{4}+2\alpha _{3}+2\alpha _{2}+\alpha _{6} \\
\alpha _{1}+\alpha _{5}+\alpha _{4}+2\alpha _{3}+\alpha _{2}+\alpha _{6}%
\end{array}%
\right. $ & \multicolumn{1}{c}{$\ \ \ \ \left( 1-q^{7}\right) ^{3}$} \\
\ \ \ \ \  &  \\
$\left.
\begin{array}{c}
\alpha _{1}+\alpha _{5}+\alpha _{4}+2\alpha _{3}+2\alpha _{2}+\alpha _{6} \\
\alpha _{1}+\alpha _{5}+2\alpha _{4}+2\alpha _{3}+\alpha _{2}+\alpha _{6}%
\end{array}%
\right. $ & $\ \ \ \ \ \ \ \left( 1-q^{8}\right) ^{2}$ \\
\ \ \ \ \  &  \\
$\alpha _{1}+\alpha _{5}+2\alpha _{4}+2\alpha _{3}+2\alpha _{2}+\alpha _{6}$
& \ \ \ $\ \ \ \ \left( 1-q^{9}\right) $ \\
$\alpha _{1}+\alpha _{5}+2\alpha _{4}+3\alpha _{3}+2\alpha _{2}+\alpha _{6}$
& \ \ \ $\ \ \ \ \left( 1-q^{10}\right) $ \\
$\alpha _{1}+\alpha _{5}+2\alpha _{4}+3\alpha _{3}+2\alpha _{2}+2\alpha _{6}$
& \ \ \ $\ \ \ \ \left( 1-q^{11}\right) $ \\ \hline\hline
\end{tabular}
\caption{The third set of roots and the associated vacuum character factors}
\label{l3}
\end{table}
\newpage
\subsection{Ortho-exceptional model}

The above construction of the linear-exceptional theory (E$_{6}\rightarrow
SL_{6}$) partition function applies also for the branching pattern E$%
_{6}\rightarrow SO_{5,5}$ of the ortho-exceptional gauge symmetry. The
partition function $\mathcal{Z}_{{\small E}_{{\small 6}}}^{{\small SO}_{%
{\small 5,5}}}$ of this model is given by (\ref{6e}) namely
\begin{equation}
\mathcal{Z}_{{\small E}_{{\small 6}}}^{{\small SO}_{{\small 5,5}}}=tr\left[
\prod\limits_{s\in J_{{\small SO}_{{\small 5,5}}}}q_{{\small (s)}}^{W_{%
\mathrm{0}}^{{\small (s)}}}\tilde{q}_{{\small (s)}}^{W_{\mathrm{0}}^{{\small %
(s)}}}\right]
\end{equation}%
with higher spin content given by the set $J_{{\small SO}_{{\small 5,5}%
}}=\{2,3,4,5,11,17\}.$ For this higher spin model, the\emph{\ holomorphic}
part of the partition function is given by
\begin{equation}
Z_{{\small E}_{{\small 6}}}^{{\small SO}_{{\small 5,5}}}=tr\left[
\prod\limits_{s\in J_{{\small SO}_{{\small 5,5}}}}q_{{\small (s)}}^{W_{%
\mathrm{0}}^{{\small (s)}}}\right] ,\qquad W_{0}^{{\small (2)}}=L_{0}-\frac{c%
}{24}  \label{es}
\end{equation}%
The six conformal spin blocks of the E$_{6}$ theory contribute in the $Z_{%
{\small E}_{{\small 6}}}^{{\small SO}_{{\small 5,5}}}$ with six bloc factors
$q_{{\small (s)}}^{W_{\mathrm{0}}^{{\small (s)}}}$ in the sense that it
expands like%
\begin{equation}
Z_{{\small E}_{{\small 6}}}^{{\small SO}_{{\small 5,5}}}[\tau ,\theta _{%
{\small (s)}}]=tr\left[ q^{W_{0}^{{\small (2)}}}y^{W_{0}^{(3)}}p^{W_{0}^{%
\left( 4\right) }}x^{W_{0}^{\left( 5\right) }}u^{W_{0}^{\left( 11\right)
}}v^{W_{0}^{\left( 17\right) }}\right]   \label{BH2}
\end{equation}%
with%
\begin{equation}
\begin{tabular}{llll}
$s=2$ & $\rightarrow $ & $q^{W_{0}^{{\small (2)}}}$ & $=e^{2i\pi \tau \left(
L_{0}-\frac{c}{24}\right) }$ \\
$s=3$ & $\rightarrow $ & $y^{W_{0}^{(3)}}$ & $=e^{2i\pi \theta _{{\small (3)}%
}W_{0}^{{\small (3)}}}$ \\
$s=4$ & $\rightarrow $ & $p^{W_{0}^{\left( 4\right) }}$ & $=e^{2i\pi \theta
_{{\small (4)}}W_{0}^{(4)}}$ \\
$s=5$ & $\rightarrow $ & $x^{W_{0}^{\left( 5\right) }}$ & $=e^{2i\pi \theta
_{{\small (5)}}W_{0}^{\left( 5\right) }}$ \\
$s=11$ & $\rightarrow $ & $u^{W_{0}^{\left( 11\right) }}$ & $=e^{2i\pi
\theta _{{\small (11)}}W_{0}^{\left( 11\right) }}$ \\
$s=17$ & $\rightarrow $ & $v^{W_{0}^{\left( 17\right) }}$ & $=e^{2i\pi
\theta _{{\small (17)}}W_{0}^{\left( 17\right) }}$%
\end{tabular}%
\end{equation}%
with\textrm{\ }$\theta _{{\small (s)}}=-\tau \mathfrak{A}_{+{\small (s)}}^{%
{\small s-1}}.$ To obtain the expression of the holomorphic partition
function $Z_{{\small E}_{{\small 6}}}^{{\small SO}_{{\small 5,5}}}$
factorising like $|\mathbf{\chi }_{1}^{{\small E}_{{\small 6}}}|^{2}$, we
proceed as for the partition functions $Z_{{\small SL}_{{\small 3}}}^{%
{\small SL}_{{\small 2}}}$ and $Z_{{\small E}_{{\small 6}}}^{{\small SL}_{%
{\small 6}}}$ studied above. To get $Z_{{\small E}_{{\small 6}}}^{{\small SO}%
_{{\small 5,5}}}$ we use\ the following factorisation%
\begin{equation}
\mathbf{\chi }_{1}^{{\small E}_{{\small 6}}}=\mathbf{\chi }_{1}^{{\small SO}%
_{{\small 5,5}}}.\mathbf{\chi }_{1}^{({\small E}_{{\small 6}}/{\small SO}_{%
{\small 5,5}})}
\end{equation}%
where $\mathbf{\chi }_{1}^{{\small SO}_{{\small 5,5}}}$ is given by \textrm{%
\cite{son}}
\begin{equation}
\mathbf{\chi }_{1}^{{\small SO}_{{\small 5,5}}}=q^{-\frac{c}{24}}\frac{q^{%
\frac{5}{24}}}{[\mathbf{\eta }\left( q\right) ]^{5}}\frac{\left(
1-q^{7}\right) \left( 1-q^{8}\right) }{\left( 1-q\right) \left(
1-q^{2}\right) }\dprod\limits_{n=1}^{8}\left( 1-q^{n}\right) ^{7-n}
\label{o5}
\end{equation}%
Notice that in this expression, the factor $q^{\frac{5}{24}}/[\mathbf{\eta }%
\left( q\right) ]^{5}$ only translates the contribution of the five Cartan
generators of ${\small SO}_{{\small 5,5}}$ descending from the
Levi-decomposition after cutting the simple root $\alpha _{1}$ in the
Tits-Satake diagram \textbf{EI}; namely the diagonal generators H$_{\alpha
_{2}},$ H$_{\alpha _{3}},$ H$_{\alpha _{4}},$ H$_{\alpha _{5}},$ H$_{\alpha
_{6}}$. The other factors in (\ref{o5}) are associated with the \emph{10 }%
positive\ step operators E$_{+\beta }$ of ${\small SO}_{{\small 5,5}};$
these positive roots $\beta $ are directly read from $\Phi _{e_{6}}^{+}$;
they obey the property%
\begin{equation}
\frac{\partial \beta }{\partial \alpha _{1}}=0\qquad ,\qquad \beta \in \Phi
_{e_{6}}^{+}
\end{equation}%
Notice also that the contribution of $\mathbf{\chi }_{1}^{({\small E}_{%
{\small 6}}/{\small SO}_{{\small 5,5}})}$ can be obtained as before; it is
calculated by using the splitting $78=45+1+2\times 16$ under the branching $%
E_{6}\rightarrow {\small SO}_{{\small 5,5}}$. In this splitting, the number $%
45$ refers to the dimensions of ${\small SO}_{{\small 5,5}}$ with
contribution as in eq(\ref{o5}). The number 1 is associated with the Cartan
generator H$_{\alpha _{1}}$ used in the Levi-decomposition; it contributes
with the factor $q^{\frac{1}{24}}/\mathbf{\eta }\left( q\right) .$ Finally,
the number $2\times 16$ is associated with the 32 roots $\beta $ in the
sub-root system $\Phi _{e_{6}}/\Phi _{{\small SO}_{{\small 5,5}}}$ labeling
the \emph{16+16} step operators E$_{\pm \beta }.$ These roots are given by
those $\beta $ belonging to $\Phi _{e_{6}}$ such that%
\begin{equation}
\frac{\partial \beta }{\partial \alpha _{1}}\neq 0\qquad ,\qquad \beta \in
\Phi _{e_{6}}
\end{equation}%
Their contribution follows from the same logic as done for eqs(\ref{l1}-\ref%
{l3}) giving the same exact results such that $Z_{{\small E}_{{\small 6}}}^{%
{\small SO}_{{\small 5,5}}}=Z_{{\small E}_{{\small 6}}}^{{\small SL}_{6}}$.

\section{Conclusion}

\qquad In this paper, we \textrm{investigated} exceptional higher spin BTZ
black holes and their thermodynamical properties while focusing on the E$%
_{6} $ theory. After introducing the Landscape of higher spin black holes as
depicted by the Figure \textbf{\ref{L}}, we focused on the exceptional
corner and worked out three solutions relying on the following decomposition
of E$_{6}$: $\left( \mathbf{i}\right) $ the linear Levi-decomposition $%
E_{6}\rightarrow A_{5};$ $\left( \mathbf{ii}\right) $ the spinorial
Levi-decomposition $E_{6}\rightarrow D_{5};$ and $\left( \mathbf{iii}\right)
$ the exceptional decomposition $E_{6}\rightarrow F_{4}\rightarrow G_{2}$.%
\newline
As a front matter, we reviewed the higher spin-2 BTZ black hole with SL(2,$%
\mathbb{R}$) gauge symmetry with a metric in the generalised
Fefferman-Graham form upon imposing the most general GR boundary conditions.
The extension to higher dimensional, more exotic symmetries was seamless but
not quite straightforward.

We first defined a Landscape of higher spin symmetries giving rise to higher
spin BTZ black holes which can be constructed by principally embedding the
core SL(2,$\mathbb{R}$) with the help of Tits-Satake graphs of real forms
and the higher spin algebraic framework. We were therefore able to identify
an underdeveloped corner within the 3D higher spin black hole Landscape that
ought to be built upon exceptional higher spin symmetries. In fact, aside
from the G$_{2}$ BTZ black hole, the F$_{4}$ and E$_{r=6,7,8}$ higher spin
black holes were passed over except for a few notable mentions. For this
reason, we selected the E$_{6}$ as a higher spin symmetry to develop models
for exceptional HS-BTZ black holes.

Among the four Tits-Satake graphs of real forms of the complex E$_{6}$ Lie
algebra, we \textrm{considered} the split real form E$_{6_{\left( 6\right) }}
$ as a direct generalisation of the standard SL(2,$\mathbb{R}$) model. A
graphical description of the principal embedding on an all white node
Tits-Stake graph of E$_{6_{\left( 6\right) }}$ allow\textrm{ed} us to define
two theories corresponding to the two possible Levi's decompositions. We swap%
\textrm{ed} the usual canonical basis with Chevalley basis to be able to
directly exploit the root system partitioning. And for each one of these two
models, we examin\textrm{ed} the higher spin content, deriv\textrm{ed} the
associated bulk and boundary gauge connections and computed the
corresponding GK and generalised FG metrics.

For completeness, we investigat\textrm{ed} other types of solutions beyond
the Levi's decompositions and gave an overview of an exceptional-exceptional
model using the embedding G$_{2}\subset $ F$_{4}\subset $ E$_{6}$. We also
calculat\textrm{ed} the partition function of the BTZ black hole of the
exceptional higher spin E$_{6}$ gravity as a way to show the equivalence
between the two models which is exhibited by the identical partition
functions and the ensuing number of states.

\textrm{This work served as an authentication of numerous points}; we were
able to extend the GR's gauge and boundary conditions to higher spin
settings with exceptional symmetries. Our study also showed the validity of
adopting the HS-BTZ black hole with higher spin split real forms as a
solution of HS-AdS$_{3}$ gravity. And we managed out to describe a step by
step process to explicitely compute the HS-BTZ black hole with exceptional
symmetries for various types of embeddings. As a continuation of this work,
it will be interesting to investigate an expansion of the above Landscape by
studying, for instance, the possibility of principally embedding the
fundamental sl(2,$\mathbb{R}$) in other real forms besides the split real
forms. As well, one may ponder about the ensuing entropy for these models
and how the different types of real forms might affect the entropy's formula
and its computation.

\section*{Acknowledgements}

The work of Rajae Sammani is funded by the National Center for Scientific and
Technical Research (CNRST) under the PhD-ASsociate Scholarship -- PASS.

\section{Appendix A: HS-E$_{6}$ symmetry and descendants}

Here we construct a new higher spin (HS) basis for the principal embedding
of sl(2) into the E$_{6}$ symmetry of exceptional BTZ black holes and its
Levi-decompositions towards A$_{5}$ and D$_{5}$ symmetries. We start by
describing the canonical HS basis of the sl(N,$\mathbb{R}$) Lie algebra as a
guide for our calculations; \textrm{then we }discuss the relevant cases for
our study namely A$_{5},$ E$_{6}$ and D$_{5}$ higher spin theories.

\subsection{Canonical HS basis of the sl(N,$\mathbb{R}$)}

In the canonical basis of the sl(N,$\mathbb{R}$) Lie algebra of the
underlying gauge symmetry SL(N,R) of the BTZ black holes, the $N^{2}-1$
higher spin operators $W_{m_{s}}^{{\small (s)}}$ are labeled by two integers
($s,m_{s}$): $1-s\leq m_{{\small s}}\leq s-1$ and $s=2,...,N.$ Their $%
N\times N$ matrix representation are built as follows:

\begin{description}
\item[$\left( \mathbf{1}\right) $] single out the three $L_{0,\pm
}=W_{m_{2}}^{{\small (2)}}$ generating the principal sl(2,$\mathbb{R}$)
within sl(N,$\mathbb{R}$) corresponding to spin 2 gravity. Using the
canonical vectors $\mathbf{\epsilon }_{j}=\left\vert j\right\rangle $, the $%
L_{0,\pm }$ operators expand like $\sum_{j,k}\left\vert j\right\rangle
(L_{0,\pm })_{jk}\left\langle k\right\vert $ with representative matrices $%
(L_{0,\pm })_{jk}=\left\langle j|L_{0,\pm }|k\right\rangle $ given by
\begin{equation}
\begin{tabular}{ccc}
$\left( L_{-}\right) _{jk}$ & $=$ & $-\sqrt{j\left( N-j\right) }\delta
_{j+1,k}$ \\
$\left( L_{+}\right) _{jk}$ & $=$ & $+\sqrt{k\left( N-k\right) }\delta
_{j,k+1}$ \\
$\left( L_{0}\right) _{jk}$ & $=$ & $+\frac{1}{2}\left( N+1-2j\right) \delta
_{j,k}$%
\end{tabular}
\label{sn}
\end{equation}%
with $Tr\left( L_{0}L_{0}\right) =\frac{N}{12}\left( N^{2}-1\right) \equiv
\epsilon _{N}.$%
\begin{equation*}
\end{equation*}

\item[$\left( \mathbf{2}\right) $] use this realisation to calculate the
powers
\begin{eqnarray}
\left( -L_{+}\right) ^{s-1} &=&\sum_{l=1}^{N-s-1}\left(
\dprod\limits_{n=0}^{s-2}\sqrt{\left( l+n\right) \left( N-l-n\right) }%
\right) \left\vert l\right\rangle \left\langle l+s-1\right\vert \\
\left( L_{-}\right) ^{s-1} &=&\sum_{l=1}^{N-s-1}\left(
\dprod\limits_{n=0}^{s-2}\sqrt{\left( l+n\right) \left( N-l-n\right) }%
\right) \left\vert l+s-1\right\rangle \left\langle l\right\vert
\end{eqnarray}%
from which we learn the nilpotent properties $\left( L_{+}\right)
^{N}=\left( L_{-}\right) ^{N}=0.$

\item[$\left( \mathbf{3}\right) $] think about the higher spin $W_{m_{s}}^{%
{\small (s)}}$ labeled by $s\geq 3$ in terms of sl(2,$\mathbb{R}$)
representations of isospins $j=s-1$; they are built by starting from the
highest weight state $W_{+{\small (s-1)}}^{{\small (s)}}=\left( L_{+}\right)
^{s-1}$ and successively acting on them by $\left( ad_{L_{-}}\right)
^{s-1-m_{s}}$ with normalisation as follows%
\begin{equation}
W_{m_{s}}^{{\small (s)}}=2\left( -\right) ^{s-m_{s}-1}\frac{\left(
s-m_{s}-1\right) !}{\left( 2s-2\right) !}\left( ad_{L_{-}}\right)
^{s-1-m_{s}}\left( L_{+}\right) ^{s-1}  \label{sm}
\end{equation}%
with the property%
\begin{equation}
tr(W_{0}^{{\small (s)}}W_{0}^{{\small (s)}})=\frac{48\epsilon _{N}\left[
\left( s-1\right) !\right] ^{4}}{\left( 2s-1\right) !\left( 2s-2\right) !}%
\dprod\limits_{j=2}^{s-1}\left( N^{2}-j^{2}\right) :=\epsilon _{N}\varpi _{s}
\label{ns}
\end{equation}%
In what follows, we introduce our basis for higher spins to deal with
exceptional BTZ black holes.
\end{description}

\subsection{HS basis using root systems}

In our present investigation, we will use the root systems of Lie algebras
to study the HS exceptional BTZ black holes. For that, we re-express the
above realisation (\ref{sn}-\ref{ns}) of sl(N,R) in terms of Cartan-Weyl
generators \{$\boldsymbol{H}_{\alpha _{l}},\boldsymbol{E}_{\pm \alpha _{l}}$%
\} using the root system $\Phi _{{\small sl}_{{\small N}}}=\{\mathbf{\alpha }%
=\sum n_{l}\mathbf{\alpha }_{l}\}$ generated by the $N-1$ simple roots%
\begin{equation}
\mathbf{\alpha }_{l}=\mathbf{\epsilon }_{l}-\mathbf{\epsilon }_{l+1}\quad
,\quad \mathbf{\alpha }_{i}.\mathbf{\alpha }_{j}=\left( K_{{\small sl}_{%
{\small N}}}\right) _{ij}\quad ,\quad \left( K_{{\small sl}_{{\small N}%
}}\right) _{ij}=2\delta _{i,j}-\delta _{i+1,j}-\delta _{i,j+1}
\end{equation}%
First, notice that using the projectors $\boldsymbol{P}_{l}=\left\vert
l\right\rangle \left\langle l\right\vert $ ($l=1,...,N$) and the basic step
operators $\boldsymbol{E}_{+\alpha _{l}}=\left\vert l\right\rangle
\left\langle l+1\right\vert $ and $\boldsymbol{E}_{-\alpha _{l}}=\left\vert
l+1\right\rangle \left\langle l\right\vert $ with the properties $%
\boldsymbol{E}_{+\alpha _{l}}\boldsymbol{E}_{-\alpha _{l}}=\boldsymbol{P}%
_{l} $ and $\boldsymbol{E}_{-\alpha _{l}}\boldsymbol{E}_{+\alpha _{l}}=%
\boldsymbol{P}_{l+1}$, the three $sl_{2}$ generators $L_{0,\pm }$ expand like%
\begin{equation}
\begin{tabular}{lll}
$L_{\mp }$ & $=$ & $\mp \dsum\limits_{l=1}^{N-1}\sqrt{l\left( N-l\right) }%
\boldsymbol{E}_{\pm \alpha _{l}}$ \\
$L_{0}$ & $=$ & $\frac{1}{2}\dsum\limits_{l=1}^{N}\left( N+1-2l\right)
\boldsymbol{P}_{l}$%
\end{tabular}%
\end{equation}%
For the useful sl(6,$\mathbb{R}$) theory ($N=6$), the gauge symmetry has
five simple roots ($\alpha _{1},\alpha _{2},\alpha _{3},\alpha _{4},\alpha
_{5}$) with Cartan-Weyl operators ($\boldsymbol{H}_{\alpha _{i}},\boldsymbol{%
E}_{\pm \alpha _{i}}$). Using these quantities, the above $L_{0,\pm }$'s
read as follows
\begin{equation}
\begin{tabular}{lll}
$L_{+}$ & $=$ & $+\sqrt{5}\boldsymbol{E}_{-\alpha _{1}}+\sqrt{8}\boldsymbol{E%
}_{-\alpha _{2}}+3\boldsymbol{E}_{-\alpha _{3}}+\sqrt{8}\boldsymbol{E}%
_{-\alpha _{4}}+\sqrt{5}\boldsymbol{E}_{-\alpha _{5}}$ \\
$L_{-}$ & $=$ & $-\sqrt{5}\boldsymbol{E}_{+\alpha _{1}}-\sqrt{8}\boldsymbol{E%
}_{+\alpha _{2}}-3\boldsymbol{E}_{+\alpha _{3}}-\sqrt{8}\boldsymbol{E}%
_{+\alpha _{4}}-\sqrt{5}\boldsymbol{E}_{+\alpha _{5}}$ \\
$2L_{0}$ & $=$ & $+5\boldsymbol{H}_{\alpha _{1}}+8\boldsymbol{H}_{\alpha
_{2}}+9\boldsymbol{H}_{\alpha _{3}}+8\boldsymbol{H}_{\alpha _{4}}+5%
\boldsymbol{H}_{\alpha _{5}}$%
\end{tabular}
\label{re}
\end{equation}%
The matrix representations of these operators are given by%
\begin{equation*}
L_{-}=-\left(
\begin{array}{cccccc}
{\small 0} & \sqrt{{\small 5}} & {\small 0} & {\small 0} & {\small 0} &
{\small 0} \\
{\small 0} & {\small 0} & \sqrt{{\small 8}} & {\small 0} & {\small 0} &
{\small 0} \\
{\small 0} & {\small 0} & {\small 0} & {\small 3} & {\small 0} & {\small 0}
\\
{\small 0} & {\small 0} & {\small 0} & {\small 0} & \sqrt{{\small 8}} &
{\small 0} \\
{\small 0} & {\small 0} & {\small 0} & {\small 0} & {\small 0} & \sqrt{%
{\small 5}} \\
{\small 0} & {\small 0} & {\small 0} & {\small 0} & {\small 0} & {\small 0}%
\end{array}%
\right) \quad ,\quad L_{+}=\left(
\begin{array}{cccccc}
0 & 0 & 0 & 0 & 0 & 0 \\
\sqrt{{\small 5}} & 0 & 0 & 0 & 0 & 0 \\
0 & \sqrt{{\small 8}} & 0 & 0 & 0 & 0 \\
0 & 0 & 3 & 0 & 0 & 0 \\
0 & 0 & 0 & \sqrt{{\small 8}} & 0 & 0 \\
0 & 0 & 0 & 0 & \sqrt{{\small 5}} & 0%
\end{array}%
\right)
\end{equation*}%
\begin{equation}
2L_{0}=\left(
\begin{array}{cccccc}
{\small 5} & {\small 0} & {\small 0} & {\small 0} & {\small 0} & {\small 0}
\\
{\small 0} & {\small 3} & {\small 0} & {\small 0} & {\small 0} & {\small 0}
\\
{\small 0} & {\small 0} & {\small 1} & {\small 0} & {\small 0} & {\small 0}
\\
{\small 0} & {\small 0} & {\small 0} & {\small -1} & {\small 0} & {\small 0}
\\
{\small 0} & {\small 0} & {\small 0} & {\small 0} & {\small -3} & {\small 0}
\\
{\small 0} & {\small 0} & {\small 0} & {\small 0} & {\small 0} & {\small -5}%
\end{array}%
\right) \qquad ,\qquad Tr\left( L_{0}L_{0}\right) =\frac{35}{2}
\end{equation}%
From the realisation (\ref{re}), notice the following useful features:

\begin{description}
\item[$\left( \mathbf{i}\right) $] Eqs(\ref{re}) have a manifest discrete $%
\mathbb{Z}_{2}^{aut}$ symmetry acting on the simple roots as follows%
\begin{equation}
\mathbb{Z}_{2}^{aut}:(\mathbf{\alpha }_{1},\mathbf{\alpha }_{2},\mathbf{%
\alpha }_{3},\mathbf{\alpha }_{4},\mathbf{\alpha }_{5})\quad \rightarrow
\quad (\mathbf{\alpha }_{5},\mathbf{\alpha }_{4},\mathbf{\alpha }_{3},%
\mathbf{\alpha }_{2},\mathbf{\alpha }_{1})  \label{au}
\end{equation}%
leaving $\mathbf{\alpha }_{3}$ invariant while exchanging $\mathbf{\alpha }%
_{i}\leftrightarrow \mathbf{\alpha }_{6-i}$. This discrete symmetry is just
the well known outer automorphism symmetry of the Dynkin diagram of sl(6,$%
\mathbb{R}$). The invariance (\ref{au}) will be used below for the study of
HS spin theories concerning orthogonal and exceptional BTZ black holes.

\item[$\left( \mathbf{ii}\right) $] The highest weight states $\left(
L_{+}\right) ^{s-1}$ of the sl(2,$\mathbb{R}$) representations making sl(6,$%
\mathbb{R}$) are collected in the following table%
\begin{table}[h!]
\centering
\begin{tabular}{l|l|l}
HS spin s & $\left( L_{+}\right) ^{s-1}$ & HW states in terms of step
operators \\ \hline
$s=2$ & $L_{+}$ & $\sqrt{5}\left( \boldsymbol{E}_{-\alpha _{1}}+\boldsymbol{E%
}_{-\alpha _{5}}\right) +\sqrt{8}\left( \boldsymbol{E}_{-\alpha _{2}}+%
\boldsymbol{E}_{-\alpha _{4}}\right) +3\boldsymbol{E}_{-\alpha _{3}}$ \\
$s=3$ & $\left( L_{+}\right) ^{2}$ & $\sqrt{40}\left( \boldsymbol{E}%
_{-\alpha _{1}-\alpha _{2}}+\boldsymbol{E}_{-\alpha _{4}-\alpha _{5}}\right)
+\sqrt{72}\left( \boldsymbol{E}_{-\alpha _{2}-\alpha _{3}}+\boldsymbol{E}%
_{-\alpha _{3}-\alpha _{4}}\right) $ \\
$s=4$ & $\left( L_{+}\right) ^{3}$ & $\sqrt{360}\left( \boldsymbol{E}%
_{-\alpha _{1}-\alpha _{2}-\alpha _{3}}+\boldsymbol{E}_{-\alpha _{3}-\alpha
_{4}-\alpha _{5}}\right) +24\boldsymbol{E}_{-\alpha _{2}-\alpha _{3}-\alpha
_{4}}$ \\
$s=5$ & $\left( L_{+}\right) ^{4}$ & $24\sqrt{5}\left( \boldsymbol{E}%
_{-\alpha _{1}-\alpha _{2}-\alpha _{3}-\alpha _{4}}+\boldsymbol{E}_{-\alpha
_{2}-\alpha _{3}-\alpha _{4}-\alpha _{5}}\right) $ \\
$s=6$ & $\left( L_{+}\right) ^{5}$ & $120\boldsymbol{E}_{-\alpha _{1}-\alpha
_{2}-\alpha _{3}-\alpha _{4}-\alpha _{5}}$ \\ \hline
\end{tabular}
\caption{HW states of the HS spectrum in terms of step operators}
\label{tb}
\end{table}%
\item[$\left( \mathbf{iii}\right) $] Eqs(\ref{re}) and (\ref{tb}) are
invariant under the $\mathbb{Z}_{2}$ outer automorphism symmetry (\ref{au})
of the Dynkin diagram of sl(6,$\mathbb{R}$). This feature is remarkable and
very interesting; it suggests that the generators $L_{0},L_{\pm }$ of the
sl(2,$\mathbb{R}$) principal subalgebra of the exceptional E$_{6}$ has the
following structure%
\begin{equation}
\begin{tabular}{lll}
$L_{-}$ & $=$ & $-x_{_{15}}\left( \boldsymbol{E}_{+\alpha _{1}}+\boldsymbol{E%
}_{+\alpha _{5}}\right) -x_{_{24}}\left( \boldsymbol{E}_{+\alpha _{2}}+%
\boldsymbol{E}_{+\alpha _{4}}\right) $ \\
&  & $-x_{_{36}}\left( \kappa _{_{3}}\boldsymbol{E}_{+\alpha _{3}}+\kappa
_{_{6}}\boldsymbol{E}_{+\alpha _{6}}\right) $ \\
$L_{+}$ & $=$ & $+x_{_{15}}\left( \boldsymbol{E}_{-\alpha _{1}}+\boldsymbol{E%
}_{-\alpha _{5}}\right) -x_{_{24}}\left( \boldsymbol{E}_{-\alpha _{2}}+%
\boldsymbol{E}_{-\alpha _{4}}\right) $ \\
&  & $-x_{_{36}}\left( \kappa _{_{3}}\boldsymbol{E}_{-\alpha _{3}}+\kappa
_{_{6}}\boldsymbol{E}_{-\alpha _{6}}\right) $ \\
$2L_{0}$ & $=$ & $+y_{_{15}}\left( \boldsymbol{H}_{\alpha _{1}}+\boldsymbol{H%
}_{\alpha _{5}}\right) +y_{_{24}}\left( \boldsymbol{H}_{\alpha _{2}}+%
\boldsymbol{H}_{\alpha _{4}}\right) $ \\
&  & $+y_{_{36}}\left( \upsilon _{_{3}}\boldsymbol{H}_{\alpha _{3}}+\upsilon
_{_{6}}\boldsymbol{H}_{\alpha _{6}}\right) $%
\end{tabular}
\label{E6}
\end{equation}%
where $x_{_{ij}}$'s and the $y_{_{ij}}$'s are real numbers and where the
real $\kappa _{3}$ and $\kappa _{6}$ are motivated by the Levi decomposition
of E$_{6}$ down to sl(6,$\mathbb{R}$). The numbers ($x_{ij},\kappa _{i}$)
and ($y_{ij},\upsilon _{i}$) are related to each others by the commutation
relations $\left[ L_{+},L_{-}\right] =2L_{0}$; we have%
\begin{equation}
\upsilon _{i}y_{ij}=\kappa _{i}^{2}x_{ij}^{2}
\end{equation}%
By setting $\kappa _{6}=0$ and $\kappa _{3}=1$ and consequently $\upsilon
_{6}=0$ and $\upsilon _{3}=1,$ the terms $\kappa _{_{6}}\boldsymbol{E}_{\pm
\alpha _{6}}$ and $\upsilon _{_{6}}\boldsymbol{H}_{\alpha _{6}}$ in (\ref{E6}%
) are killed; and one recovers the realisation (\ref{re}).

\item[$\left( \mathbf{iv}\right) $] Eqs(\ref{re}) and (\ref{tb}) suggests
also that the generators $L_{0},L_{\pm }$ of the sl(2,$\mathbb{R}$)
principal subalgebra of the D$_{5}$ Lie algebra contained inside E$_{6}$ has
the structure%
\begin{equation}
\begin{tabular}{lll}
$L_{-}$ & $=$ & $-x_{1}^{\prime }\boldsymbol{E}_{+\alpha _{1}}^{\prime
}-x_{2}^{\prime }\boldsymbol{E}_{+\alpha _{2}}^{\prime }-x_{3}^{\prime }%
\boldsymbol{E}_{+\alpha _{3}}-x_{45}\left( \boldsymbol{E}_{+\alpha
_{4}}^{\prime }+\boldsymbol{E}_{+\alpha _{5}}^{\prime }\right) $ \\
$L_{+}$ & $=$ & $+x_{1}^{\prime }\boldsymbol{E}_{-\alpha _{1}}^{\prime
}+x_{2}^{\prime }\boldsymbol{E}_{-\alpha _{2}}^{\prime }+x_{3}^{\prime }%
\boldsymbol{E}_{-\alpha _{3}}+x_{45}\left( \boldsymbol{E}_{-\alpha
_{4}}^{\prime }+\boldsymbol{E}_{-\alpha _{5}}^{\prime }\right) $ \\
$2L_{0}$ & $=$ & $+y_{1}^{\prime }\left( \boldsymbol{H}_{\alpha _{1}}\right)
+y_{2}^{\prime }\left( \boldsymbol{H}_{\alpha _{2}}\right) +y_{3}^{\prime }%
\boldsymbol{H}_{\alpha _{3}}+y_{45}\left( \boldsymbol{H}_{\alpha
_{4}}^{\prime }+\boldsymbol{H}_{\alpha _{5}}^{\prime }\right) $%
\end{tabular}
\label{D5}
\end{equation}%
where $x_{i}^{\prime }$ and $y_{i}^{\prime }$ are some numbers with $%
y_{i}^{\prime }=(x_{i}^{\prime })^{2}$. This realisation is invariant under
the $\mathbb{Z}_{2}$ automorphism symmetry acting on the simple roots as
follows%
\begin{equation}
(\mathbf{\alpha }_{1},\mathbf{\alpha }_{2},\mathbf{\alpha }_{3};\mathbf{%
\alpha }_{4},\mathbf{\alpha }_{5})\quad \rightarrow \quad (\mathbf{\alpha }%
_{1},\mathbf{\alpha }_{2},\mathbf{\alpha }_{3};\mathbf{\alpha }_{5},\mathbf{%
\alpha }_{4})  \label{5D}
\end{equation}%
leaving $\mathbf{\alpha }_{1},\mathbf{\alpha }_{2},\mathbf{\alpha }_{3}$
invariant while permuting $\mathbf{\alpha }_{4}$ and $\mathbf{\alpha }_{5}.$
\newline
To get more insight into the realisation (\ref{D5}), we rewrite the diagonal
$2L_{0}$ charge matrix in (\ref{E6}) like $2L_{0}=\sum z_{i}\boldsymbol{H}%
_{\alpha _{i}}$ expanding as%
\begin{equation}
2L_{0}=z_{1}\boldsymbol{H}_{\alpha _{1}}+z_{2}\boldsymbol{H}_{\alpha
_{2}}+z_{3}\boldsymbol{H}_{\alpha _{3}}+z_{4}\boldsymbol{H}_{\alpha
_{4}}+z_{5}\boldsymbol{H}_{\alpha _{5}}+z_{_{6}}\boldsymbol{H}_{\alpha _{6}}
\end{equation}%
Then using the relation $Tr(\boldsymbol{H}_{\alpha _{i}}\boldsymbol{H}%
_{\alpha _{j}})=(K_{E_{6}})_{ij}$, we first have $4Tr\left(
L_{0}L_{0}\right) =z_{i}(K_{E_{6}})_{ij}z_{j}.$ By substituting the Cartan
matrix $(K_{E_{6}})_{ij}$ in terms of the simple roots namely $\mathbf{%
\alpha }_{i}.\mathbf{\alpha }_{j}$, we put the previous relation into the
form $4Tr\left( L_{0}L_{0}\right) =\mathbf{z}_{{\small E}_{{\small 6}}}^{2}$
where we have set%
\begin{equation}
\mathbf{z}_{{\small E}_{{\small 6}}}=z_{1}\mathbf{\alpha }_{1}+z_{2}\mathbf{%
\alpha }_{2}+z_{3}\mathbf{\alpha }_{3}+z_{4}\mathbf{\alpha }_{4}+z_{5}%
\mathbf{\alpha }_{5}+z_{_{6}}\mathbf{\alpha }_{6}  \label{z}
\end{equation}%
To obtain the rank 5 symmetry D5 having five simple roots ($\mathbf{\alpha }%
_{1}^{\prime },\mathbf{\alpha }_{2}^{\prime },\mathbf{\alpha }_{3}^{\prime },%
\mathbf{\alpha }_{4}^{\prime },\mathbf{\alpha }_{5}^{\prime }$) from the
parent rank 6 symmetry E6 with the six simple roots ($\mathbf{\alpha }_{1},%
\mathbf{\alpha }_{2},\mathbf{\alpha }_{3},\mathbf{\alpha }_{4},\mathbf{%
\alpha }_{5},\mathbf{\alpha }_{6}$), we can do it in two ways:

\begin{itemize}
\item Either we cut the root $\mathbf{\alpha }_{1}$ leaving the five roots ($%
\mathbf{\alpha }_{2},\mathbf{\alpha }_{3},\mathbf{\alpha }_{4},\mathbf{%
\alpha }_{5},\mathbf{\alpha }_{6}$) which we have to identify with the set ($%
\mathbf{\alpha }_{1}^{\prime },\mathbf{\alpha }_{2}^{\prime },\mathbf{\alpha
}_{3}^{\prime },\mathbf{\alpha }_{4}^{\prime },\mathbf{\alpha }_{5}^{\prime
} $);

\item Or we cut the root $\mathbf{\alpha }_{5}$ leaving the five roots ($%
\mathbf{\alpha }_{1},\mathbf{\alpha }_{2},\mathbf{\alpha }_{3},\mathbf{%
\alpha }_{4},\mathbf{\alpha }_{6}$) that must be identified with those of
the D$_{5}$ namely ($\mathbf{\alpha }_{1}^{\prime },\mathbf{\alpha }%
_{2}^{\prime },\mathbf{\alpha }_{3}^{\prime },\mathbf{\alpha }_{4}^{\prime },%
\mathbf{\alpha }_{5}^{\prime }$).
\end{itemize}

Focussing on the second picture and taking advantage of the fact that all
roots $\mathbf{\alpha }=\sum n_{i}\mathbf{\alpha }_{i}$ in the simply laced E%
$_{6}$ have length 2, we can realise the cutting by using folding ideas. In
this case, the folding is given by $\left( i\right) $ requiring $z_{4}=z_{5}$
in eq(\ref{z}); and $\left( ii\right) $ substitute the composite root $%
\mathbf{\alpha }_{4}+\mathbf{\alpha }_{5}$ by $\mathbf{\alpha }_{4}^{\prime
};$ that%
\begin{equation}
\left\{
\begin{tabular}{lll}
$\mathbf{\alpha }_{4}$ & $=$ & $\mathbf{\epsilon }_{4}-\mathbf{\epsilon }_{5}
$ \\
$\mathbf{\alpha }_{5}$ & $=$ & $\mathbf{\epsilon }_{5}-\mathbf{\epsilon }_{6}
$%
\end{tabular}%
\right. \quad \Rightarrow \quad \left\{
\begin{tabular}{lll}
$\mathbf{\alpha }_{4}^{\prime }$ & $=$ & $\mathbf{\alpha }_{4}+\mathbf{%
\alpha }_{5}$ \\
& $=$ & $\mathbf{\epsilon }_{4}-\mathbf{\epsilon }_{6}$%
\end{tabular}%
\right.
\end{equation}%
still obeying ($\mathbf{\alpha }_{4}^{\prime })^{2}=(\mathbf{\alpha }_{4}+%
\mathbf{\alpha }_{5})^{2}=2.$ So, we are left with the five simple roots ($%
\mathbf{\alpha }_{1},\mathbf{\alpha }_{2},\mathbf{\alpha }_{3},\mathbf{%
\alpha }_{4}^{\prime },\mathbf{\alpha }_{6}$) to describe D$_{5}$ symmetry
and then $\mathbf{z}_{{\small D}_{{\small 5}}}=z_{1}\mathbf{\alpha }%
_{1}+z_{2}\mathbf{\alpha }_{2}+z_{3}\mathbf{\alpha }_{3}+z_{1}\mathbf{\alpha
}_{4}^{\prime }+z_{_{6}}\mathbf{\alpha }_{6}$ where we have used $z_{4}=z_{5}
$ and $z_{1}=z_{5}.$ Moreover, using the outer automorphism symmetry of D$%
_{5}$\ requiring the invariance under the permutation $(\mathbf{\alpha }%
_{4}^{\prime },\mathbf{\alpha }_{6})\leftrightarrow (\mathbf{\alpha }_{6},%
\mathbf{\alpha }_{4}^{\prime })$ while preserving $(\mathbf{\alpha }_{1},%
\mathbf{\alpha }_{2},\mathbf{\alpha }_{3}),$ we must have $z_{_{6}}=z_{1}.$
\end{description}

\section{Appendix B: HS $E_{6}/A_{5}$ and $E_{6}/SO_{5,5}$ theories}

In this appendix, we overview the algebraic framework of the E$_{6}$ Lie
algebra as a higher spin symmetry and derive in detail the higher spin
content for each type of Levi- decompositions namely the patterns $%
A_{5}\rightarrow E_{6}$ and $E_{6}\rightarrow SO_{5,5}$ with the help of
Hesse diagrams$.$ We give the associated higher spin algebras and the
explicit expansions of the boundary connections.

\subsection{Higher spins in the exceptional $E_{6}/A_{5}$ BTZ theory}

\begin{figure}[ph]
\begin{center}
\includegraphics[width=17cm]{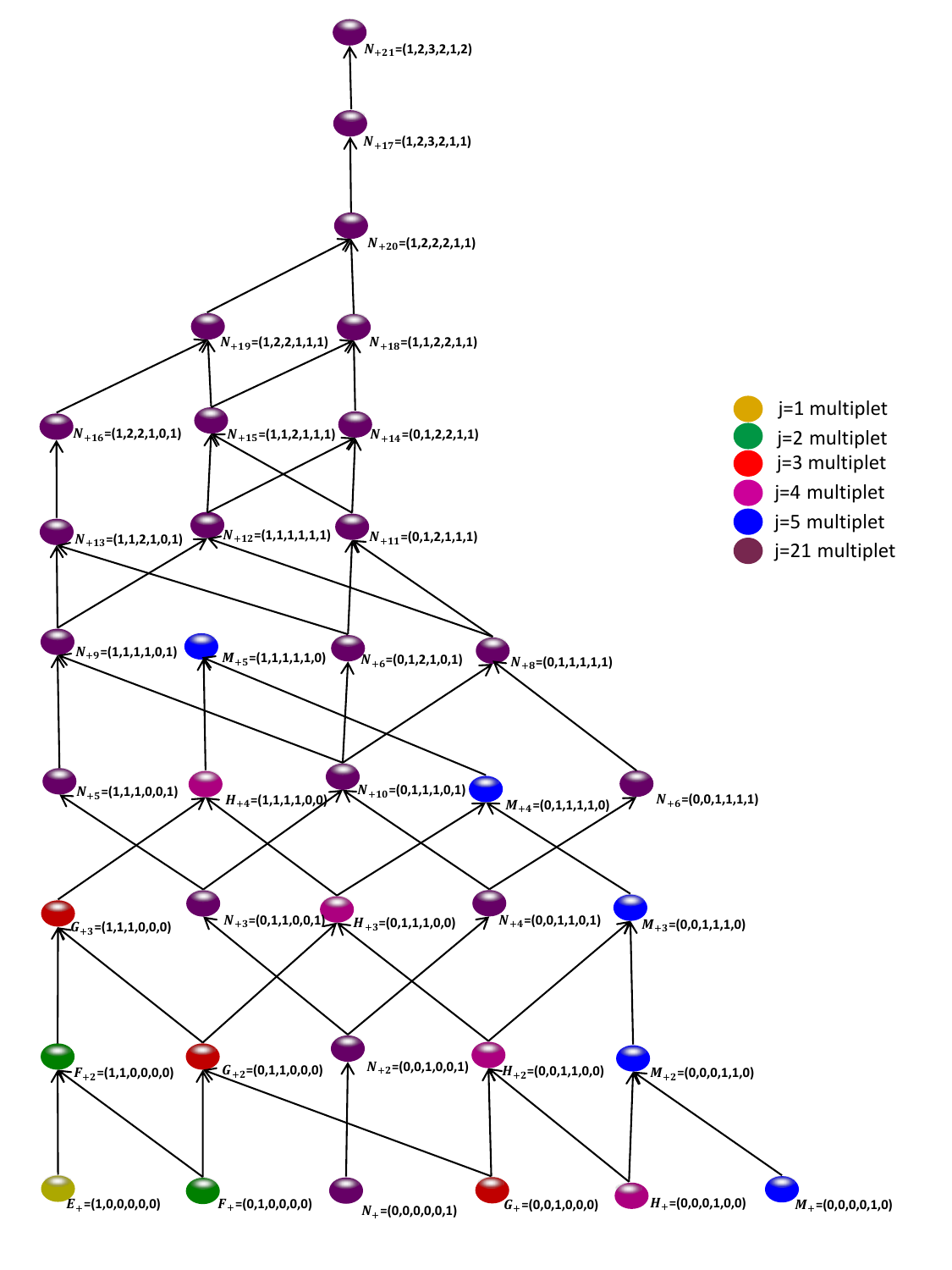}
\end{center}
\par
\vspace{-0.5cm}
\caption{The six isomultiplets within E$_{6}$ and the corresponding
generators for the linear based splitting. Nodes with same color sit in the
same isomultiplet.}
\label{multii}
\end{figure}

The branching pattern $E_{6}\rightarrow A_{5}$ corresponds to cutting the
green node $\mathfrak{N}_{6}$ as shown on the \textbf{Figure \ref{F2}}; the
resulting theory has a $sl\left( 6,\mathbb{R}\right) $ subsymmetry as shown
by the associated Levi- decomposition%
\begin{equation}
{\Large e}_{6}\rightarrow gl_{1}\oplus sl_{6}\oplus \boldsymbol{21}%
_{+}\oplus \boldsymbol{21}_{-}  \label{linlevi}
\end{equation}%
with $gl_{1}\simeq \mathbb{R}^{\ast }$ and $\boldsymbol{21}_{\pm }$ are
multiplets of $sl_{6}$ (symmetric representation $6\vee 6$) generating
nilpotent superalgebras. Under this decomposition, the \emph{36+36} step
operators $E_{\pm \beta }$ of the exceptional ${\Large e}_{6}$ splits into
two subsets: $\left( \mathbf{i}\right) $ a subset of \emph{15+15} step
operators of $sl_{6}$ (antisymmetric representation $6\wedge 6$) with
labelled roots $\beta _{sl_{6}}$ generated by the five simple roots $\alpha
_{1},\alpha _{2},\alpha _{3},\alpha _{4},\alpha _{5}$ of the system $\Phi
_{sl_{6}}$.
\begin{figure}[h!]
\begin{center}
\includegraphics[width=14cm]{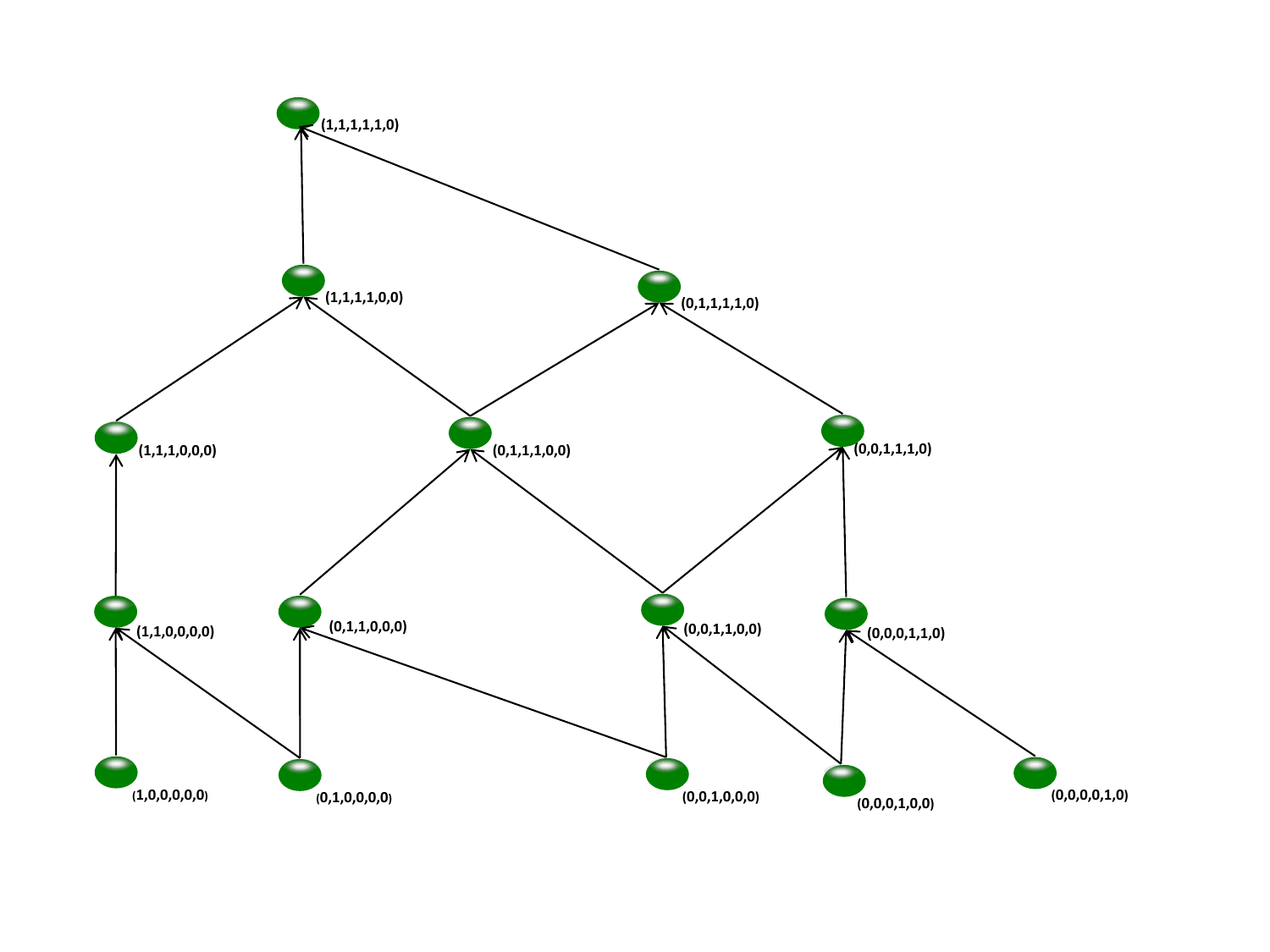}
\end{center}
\par
\vspace{-0.5cm}
\caption{Hasse diagram of the linear $sl(6)$ Levi-subalgebra of the $%
{\protect\Large e}_{6}$ exceptional model. Nodes in this diagram give the 15
positive roots of sl$_{6}$; these nodes are generated by the simple roots ($%
\protect\alpha _{1},\protect\alpha _{2},\protect\alpha _{3},\protect\alpha %
_{4},\protect\alpha _{5}$) aligned in the bottom row.}
\label{Sl6}
\end{figure}
$\left( \mathbf{ii}\right) $ a subset of \emph{21+21} step operators of the
coset space ${\Large e}_{6}/sl_{6}$ with labelled roots $\beta _{{\Large e}%
_{6}/sl_{6}}$ generated by all of the six simple roots of the system $\Phi _{%
{\Large e}_{6}}$ with the following characteristic property; see \textbf{%
Figure} \textbf{\ref{Sl6}} for $sl_{6}$ and the \textbf{Figure} \textbf{\ref%
{21} }for ${\Large e}_{6}/sl_{6}$.
\begin{equation}
\frac{\partial \beta _{{\Large e}_{6}/sl_{6}}}{\partial \alpha _{6}}\neq
0\qquad ,\qquad \frac{\partial \beta _{sl_{6}}}{\partial \alpha _{6}}=0
\end{equation}%
\begin{figure}[h!]
\begin{center}
\includegraphics[width=14cm]{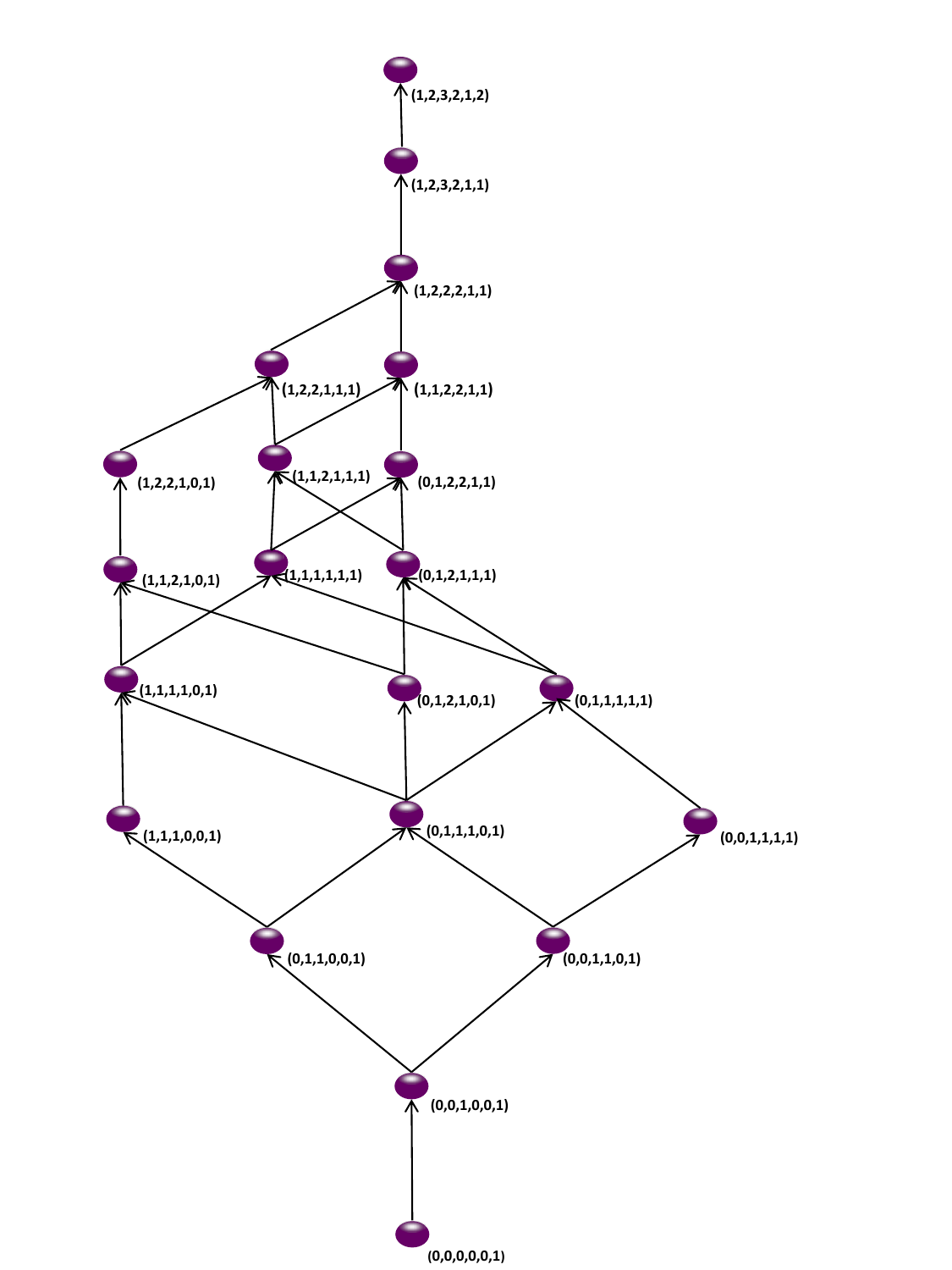}
\end{center}
\par
\vspace{-0.5cm}
\caption{Hasse diagram of the positive roots of the 21- dimensional
nilpotent subalgebra of the ${\protect\Large e}_{6}$ exceptional model. The
root in the bottom is given by the cutted node $\protect\alpha _{6}$.}
\label{21}
\end{figure}
As exhibited by the \textbf{Figure} \textbf{\ref{rs}}, the 78 dimensions of $%
{\Large e}_{6}$ redistributes accordingly to (\ref{linlevi}) namely%
\begin{equation}
\begin{tabular}{lll}
$78$ & $=$ & $35+43$ \\
$35$ & $=$ & $3+5+7+9+11$ \\
$43$ & $=$ & $\left( 1+21_{+}+21_{-}\right) $%
\end{tabular}%
\end{equation}%
with isospin j and conformal spin $s=j+1$ as follows:%
\begin{table}[H]
\centering
\begin{tabular}{ccccccccc}
\hline
$j$ & $=$ & $1$ & $2$ & $3$ & $4$ & $5$ & $;$ & $21$ \\ \hline
$2j+1$ & $=$ & $3$ & $5$ & $7$ & $9$ & $11$ & $;$ & $43$ \\ \hline
$s$ & $=$ & $2$ & $3$ & $4$ & $5$ & $6$ & $;$ & $22$ \\ \hline
\end{tabular}
\caption{Higher spin spectrum of the $E_{6}/A_{5}$ BTZ theory}
\label{AFs}
\end{table}%
Using this Levi-decomposition, the 78 generators of the exceptional ${\Large %
e}_{6}$ are splited in six $sl_{2}$ multiplets $W_{m_{s}}^{(s)}$ labelled by
the conformal spin s as follows%
\begin{table}[H]
\centering
\begin{tabular}{c|cccccccc}
$s$ &  & $2$ & $3$ & $4$ & $5$ & $6$ & $;$ & $22$ \\ \hline\hline
$W_{m_{s}}^{(s)}$ &  & $W_{m_{2}}^{(2)}$ & $W_{m_{3}}^{(3)}$ & $%
W_{m_{4}}^{(4)}$ & $W_{m_{5}}^{(5)}$ & $W_{m_{6}}^{(6)}$ & $;$ & $%
W_{m_{22}}^{(22)}$ \\
&  & $L_{m_{2}}$ & $F_{m_{3}}$ & $G_{m_{4}}$ & $H_{m_{5}}$ & $M_{m_{6}}$ & $;
$ & $N_{m_{22}}$ \\ \hline\hline
$|W_{m_{s}}^{(s)}|$ &  & $3$ & $5$ & $7$ & $9$ & $11$ & $;$ & $43$ \\ \hline
\end{tabular}%
\caption{Generators of the $E_{6}/A_{5}$ BTZ theory}
\end{table}%
with $m_{s}$ ranging like
\begin{equation}
\begin{tabular}{llllll}
$\left\vert m_{2}\right\vert \leq 1,$ & $\left\vert m_{3}\right\vert \leq 2,$
& $\left\vert m_{4}\right\vert \leq 3,$ & $\left\vert m_{5}\right\vert \leq
4,$ & $\left\vert m_{6}\right\vert \leq 5,$ & $\left\vert m_{22}\right\vert
\leq 21$%
\end{tabular}%
\end{equation}%
These generator multiplets $W_{m_{s}}^{(s)}$ are graphically collected in
the \textbf{Figure} \textbf{\ref{multii}};
their commutation relations defining the higher spin ${\Large e}_{6}$
algebra can be presented in a compact form like%
\begin{equation}
\left[ W_{m_{s}}^{(\tau )},W_{n_{\sigma }}^{(\sigma )}\right] =\sum_{s\in
J_{sl_{{\small 6}}}}c_{n_{\sigma },m_{\tau }|s}^{\tau ,\sigma }W_{m_{\tau
}+n_{\sigma }}^{(s)}  \label{we6}
\end{equation}%
where the $c_{n_{\sigma },m_{\tau }|s}^{\tau ,\sigma }$ are the constant
structures of the ${\Large e}_{6}\rightarrow sl_{6}$ algebra and where
higher spins $\tau ,\sigma \in J_{sl_{{\small 6}}}$ with interval $J_{sl_{%
{\small 6}}}=\{2,3,4,5,6,22\}.$ Explicitly, we can write\textrm{\ }(\ref{we6}%
) as follows%
\begin{equation}
\begin{tabular}{lll}
$\left[ W_{m_{s}}^{(\tau )},W_{n_{\sigma }}^{(\sigma )}\right] $ & $=$ & $%
c_{n_{\sigma },m_{\tau }|2}^{\tau ,\sigma }W_{m_{\tau }+n_{\sigma
}}^{(2)}+c_{n_{\sigma },m_{\tau }|3}^{\tau ,\sigma }W_{m_{\tau }+n_{\sigma
}}^{(3)}+c_{n_{\sigma },m_{\tau }|4}^{\tau ,\sigma }W_{m_{\tau }+n_{\sigma
}}^{(4)}+$ \\
&  & $c_{n_{\sigma },m_{\tau }|5}^{\tau ,\sigma }W_{m_{\tau }+n_{\sigma
}}^{(5)}+c_{n_{\sigma },m_{\tau }|6}^{\tau ,\sigma }W_{m_{\tau }+n_{\sigma
}}^{(6)}+c_{n_{\sigma },m_{\tau }|22}^{\tau ,\sigma }W_{m_{\tau }+n_{\sigma
}}^{(22)}$%
\end{tabular}%
\end{equation}%
Focusing on the left sector of the higher spin 3D gravity (similar
relationships hold for the twild right sector), we can expand the bulk gauge
field $A_{\mu }$ of the ${\Large e}_{6}$ theory as follows%
\begin{equation}
A_{\mu }=\sum_{s\in J_{sl_{{\small 6}}}}\dsum\limits_{m_{s}=1-s}^{s-1}A_{\mu
{\small (s)}}^{m_{s}}W_{m_{s}}^{(s)}  \label{bl}
\end{equation}%
with $s=2,3,4,5,6,22,$ and where $A_{\mu {\small (s)}}^{m_{s}}$ define the
gauge components along the 78 directions of the exceptional Lie algebra.
This expansion reads explicitly as follows%
\begin{equation}
\begin{tabular}{lll}
$A_{-}$ & $=$ & $A_{-{\small (2)}}^{m_{{\small 2}}}W_{m_{{\small 2}}}^{%
{\small (2)}}+A_{-{\small (3)}}^{m_{{\small 3}}}W_{m_{{\small 3}}}^{{\small %
(3)}}+A_{-{\small (4)}}^{m_{4}}W_{m_{{\small 4}}}^{{\small (4)}}+$ \\
&  & $A_{-{\small (5)}}^{m_{{\small 5}}}W_{m_{{\small 5}}}^{{\small (5)}%
}+A_{-{\small (6)}}^{m_{{\small 6}}}W_{m_{{\small 6}}}^{{\small (6)}}+A_{-%
{\small (22)}}^{m_{{\small 6}}}W_{m_{{\small 22}}}^{{\small (22)}}$%
\end{tabular}%
\end{equation}%
with repeated label $m_{s}$ taking values into the integral set $\left[
1-s,s-1\right] $. Its value on the 2D boundary of the 3D spacetime is given
by%
\begin{equation}
\mathfrak{A}_{\mu }=\sum_{s\in J_{sl_{{\small 6}}}}\dsum%
\limits_{m_{s}=1-s}^{s-1}\mathfrak{A}_{\mu {\small (s)}%
}^{m_{s}}W_{m_{s}}^{(s)}  \label{bn}
\end{equation}%
with vanishing radial component ($\mathfrak{A}_{\rho }=0).$ The boundary
potentials obey the two following constraints:

\begin{description}
\item[$\left( \mathbf{i}\right) $] the charges of the boundary component $%
\mathfrak{A}_{-}$ are taken in the DS gauge; i.e: the expansion of $%
\mathfrak{A}_{-}$ is restricted to $\mathfrak{A}_{-}^{{\small s-1}}W_{%
{\small s-1}}^{(s)}$ reading explicitly as follows%
\begin{equation}
\begin{tabular}{lll}
$\mathfrak{A}_{-}$ & $=$ & $L_{-}+\mathfrak{A}_{-}^{+}L_{+}+\mathcal{W}%
_{-}^{+2}W_{+2}+\mathcal{G}_{-}^{+3}G_{+3}+$ \\
&  & $\mathcal{H}_{-}^{+4}H_{+4}+\mathcal{M}_{-}^{+5}M_{+5}+\mathcal{N}%
_{-}^{+21}N_{+21}$%
\end{tabular}
\label{dds}
\end{equation}%
with $\mathfrak{A}_{-}^{-}=1.$ This boundary condition on the charges solves
the constraint relation $\left[ L_{+},\mathfrak{A}_{-}\right] =-2L_{0}$ due
to imposing\textrm{\ }the highest weight reduction $[L_{+},W_{s-1}^{(s)}]=0$
for higher spins $s.$ We also have the analogous for GR boundary conditions%
\begin{equation}
\mathfrak{A}_{-}^{{\small s-1}}=-\frac{2\pi }{\mathrm{k}}{\Large a}_{-}^{%
{\small s-1}}
\end{equation}

\item[$\left( \mathbf{ii}\right) $] The chemical potentials in $\mathfrak{A}%
_{+}=\sum \mathfrak{A}_{+}^{m_{s}}W_{m_{s}}^{(s)}$ are obtained by solving
the boundary field equation of motion given by
\begin{equation}
\partial _{-}\mathfrak{A}_{+}-\partial _{+}\mathfrak{A}_{-}+[\mathfrak{A}%
_{+},\mathfrak{A}_{-}]=0
\end{equation}%
By restricting to constant gauge field configurations $\partial _{-}%
\mathfrak{A}_{\pm }=\partial _{+}\mathfrak{A}_{\pm }=0,$ this field equation
reduces down to $[\mathfrak{A}_{+},\mathfrak{A}_{-}]=0$ which, by
substituting $\mathfrak{A}_{\pm }=\sum_{s}\sum_{m_{s}}\mathfrak{A}_{\pm
{\small (s)}}^{m_{s}}W_{m_{s}}^{(s)}$ and using (\ref{we6}), gives\footnote{%
As long as the chemical potentials on the boundary of AdS$_{3}$ are held
constant, i.e $\delta \mathfrak{A}_{+}=0,$ the variational principle remains
consistent as it is invariant on shell \cite{GR}.}
\begin{equation}
\sum_{\tau ,\sigma }\sum_{m_{\tau },n_{\sigma }}\mathfrak{A}_{+{\small (\tau
)}}^{m_{\tau }}\mathfrak{A}_{-{\small (\sigma )}}^{n_{\sigma }}[W_{m_{\tau
}}^{{\small (\tau )}},W_{n_{\sigma }}^{{\small (\sigma )}}]=0
\end{equation}%
expanding as follows%
\begin{equation}
\sum_{{\small s\in J}_{{\small sl}_{{\small 6}}}}\sum_{\tau ,\sigma
}\sum_{m_{\tau },n_{\sigma }}\mathfrak{A}_{+{\small (\tau )}}^{m_{\tau }}%
\mathfrak{A}_{-{\small (\sigma )}}^{n_{\sigma }}c_{n_{\sigma },m_{\tau
}|s}^{\tau ,\sigma }W_{m_{\tau }+n_{\sigma }}^{(s)}=0  \label{eqq}
\end{equation}
\end{description}

To write down an explicit expression of the fugacity components $\mathfrak{A}%
_{+}^{m_{\tau }},$ we have to solve the above equations of motion. Taking
into account the DS condition restricting $n_{\sigma }$ to its highest value
${\small \sigma -1}$ as in eq(\ref{dds}), the above (\ref{eqq}) becomes%
\begin{equation}
\sum_{{\small s\in J}_{{\small sl}_{{\small 6}}}}\sum_{\tau ,\sigma {\small %
\in J}_{{\small sl}_{{\small 6}}}\times {\small J}_{{\small sl}_{{\small 6}%
}}}\sum_{m_{\tau }=1-\tau }^{\tau -1}\mathfrak{A}_{+{\small (\tau )}%
}^{m_{\tau }}\mathfrak{A}_{-{\small (\sigma )}}^{{\small \sigma -1}}c_{%
{\small \sigma -1},m_{\tau }|s}^{\tau ,\sigma }W_{m_{\tau }+{\small \sigma -1%
}}^{(s)}=0
\end{equation}%
Because $W_{\mathrm{m}_{s}}^{(s)}$ is non vanishing only for $1-s\leq
\mathrm{m}_{s}{\small \leq s-1,}$ we can express $W_{m_{\tau }+{\small %
\sigma -1}}^{(s)}$ by using Kronecker function like $\delta _{\mathrm{m}%
_{s},m_{\tau }+{\small \sigma -1}}W_{\mathrm{m}_{s}}^{(s)}{\small ,}$ we can
put this relation into the form%
\begin{equation}
\sum_{{\small s\in J}_{{\small sl}_{{\small 6}}}}\left( \sum_{\tau ,\sigma
{\small \in J}_{{\small sl}_{{\small 6}}}\times {\small J}_{{\small sl}_{%
{\small 6}}}}\sum_{m_{\tau }=1-\tau }^{\tau -1}\mathfrak{A}_{+{\small (\tau )%
}}^{m_{\tau }}\mathfrak{A}_{-{\small (\sigma )}}^{{\small \sigma -1}}c_{%
{\small \sigma -1},m_{\tau }|s}^{\tau ,\sigma }\delta _{\mathrm{m}%
_{s},m_{\tau }+{\small \sigma -1}}\right) W_{\mathrm{m}_{s}}^{(s)}=0
\label{rel}
\end{equation}%
which can be presented like%
\begin{equation}
\sum_{{\small s}=2,3,4,5,6,22}\mathcal{U}_{(s)}^{\mathrm{m}_{s}}W_{\mathrm{m}%
_{s}}^{(s)}=0  \label{rela}
\end{equation}%
with summation under repeated indices and where we have set%
\begin{equation}
\mathcal{U}_{(s)}^{\mathrm{m}_{s}}=\sum_{\tau ,\sigma =2,3,4,5,6,22}\left(
\sum_{m_{\tau }=1-\tau }^{\tau -1}\mathfrak{A}_{+{\small (\tau )}}^{m_{\tau
}}\mathfrak{A}_{-{\small (\sigma )}}^{{\small \sigma -1}}c_{{\small \sigma -1%
},m_{\tau }|s}^{\tau ,\sigma }\delta _{\mathrm{m}_{s},m_{\tau }+{\small %
\sigma -1}}\right)
\end{equation}%
with $\left\vert \mathrm{m}_{s}\right\vert \leq {\small s-1.}$ For explicit
calculations, one can parameterise the expansion (\ref{rela}) like
\begin{equation}
\sum \mathcal{U}_{({\small 2)}}^{\mathrm{m}_{2}}W_{\mathrm{m}_{2}}^{(2)}+%
\mathcal{U}_{({\small 3)}}^{\mathrm{m}_{3}}W_{\mathrm{m}_{3}}^{(3)}+\mathcal{%
U}_{({\small 4)}}^{\mathrm{m}_{4}}W_{\mathrm{m}_{4}}^{(4)}+\mathcal{U}_{(%
{\small 5)}}^{\mathrm{m}_{5}}W_{\mathrm{m}_{5}}^{(5)}+\mathcal{U}_{({\small %
6)}}^{\mathrm{m}_{6}}W_{\mathrm{m}_{6}}^{(6)}+\mathcal{U}_{({\small 22)}}^{%
\mathrm{m}_{22}}W_{\mathrm{m}_{22}}^{(22)}=0
\end{equation}%
Moreover, the vanishing relations (\ref{rel}-\ref{rela}) require in turn the
vanishing of the $\mathcal{U}_{(s)}^{\mathrm{m}_{s}}$ coefficients of the
generator $W_{\mathrm{m}_{s}}^{(s)};$ i.e%
\begin{equation}
\mathcal{U}_{({\small 2)}}^{\mathrm{m}_{2}}=0,\quad \mathcal{U}_{({\small 3)}%
}^{\mathrm{m}_{3}}=0,\quad \mathcal{U}_{({\small 4)}}^{\mathrm{m}%
_{4}}=0,\quad \mathcal{U}_{({\small 5)}}^{\mathrm{m}_{5}}=0,\quad \mathcal{U}%
_{({\small 6)}}^{\mathrm{m}_{6}}=0,\quad \mathcal{U}_{({\small 22)}}^{%
\mathrm{m}_{22}}=0
\end{equation}%
For the example of the spin $s=2$, the corresponding three field equations
of the boundary gauge potential are given by $\mathcal{U}_{({\small 2)}}^{%
\mathrm{m}_{2}}=0$ with $\mathrm{m}_{2}=0,\pm 1$. To write down the explicit
expressions of these three relations, we first solve the constraint $\mathrm{%
m}_{s}=m_{\tau }+{\small \sigma -1}$ giving%
\begin{equation}
\begin{tabular}{lllll}
$m_{2}=1$ & : & $m_{\tau }+\sigma $ & $=$ & $2$ \\
$m_{2}=0$ & : & $m_{\tau }+\sigma $ & $=$ & $1$ \\
$m_{2}=-1$ & : & $m_{\tau }+\sigma $ & $=$ & $0$%
\end{tabular}%
\end{equation}%
Then, we compute%
\begin{equation}
\begin{tabular}{lllll}
$\mathcal{U}_{(2)}^{+}$ & : & $\dsum\limits_{\tau ,\sigma =2,3,4,5,6,22}%
\mathfrak{A}_{-{\small (\sigma )}}^{{\small \sigma -1}}\mathfrak{A}_{+%
{\small (\tau )}}^{2-\sigma }c_{{\small \sigma -1},2-\sigma |2}^{\tau
,\sigma }$ & $=$ & $0$ \\
$\mathcal{U}_{(2)}^{0}$ & : & $\dsum\limits_{\tau ,\sigma =2,3,4,5,6,22}%
\mathfrak{A}_{-{\small (\sigma )}}^{{\small \sigma -1}}\mathfrak{A}_{+%
{\small (\tau )}}^{1-\sigma }c_{{\small \sigma -1},1-\sigma |2}^{\tau
,\sigma }$ & $=$ & $0$ \\
$\mathcal{U}_{(2)}^{-}$ & : & $\dsum\limits_{\tau ,\sigma =2,3,4,5,6,22}%
\mathfrak{A}_{-{\small (\sigma )}}^{{\small \sigma -1}}\mathfrak{A}_{+%
{\small (\tau )}}^{-\sigma }c_{{\small \sigma -1},-\sigma |2}^{\tau ,\sigma
} $ & $=$ & $0$%
\end{tabular}%
\end{equation}%
The solution of these relations depend on the knowledge of the structure
constants $c_{{\small \sigma -1},m_{\tau }|s}^{\tau ,\sigma }$ of the E$_{6}$
gauge symmetry.\newline
For the example of the higher spin $s=3$, the five field equations of the
boundary gauge potential are given by $\mathcal{U}_{({\small 3)}}^{\mathrm{m}%
_{3}}=0$ with $\mathrm{m}_{3}=0,\pm 1,\pm 2$. To write down their
expressions, we first solve the constraint $\mathrm{m}_{3}=m_{\tau }+{\small %
\sigma -1}$ giving%
\begin{equation}
\begin{tabular}{lllll}
$m_{3}=2$ & : & $m_{\tau }+\sigma $ & $=$ & $3$ \\
$m_{3}=1$ & : & $m_{\tau }+\sigma $ & $=$ & $2$ \\
$m_{3}=0$ & : & $m_{\tau }+\sigma $ & $=$ & $1$ \\
$m_{3}=-1$ & : & $m_{\tau }+\sigma $ & $=$ & $0$ \\
$m_{3}=-2$ & : & $m_{\tau }+\sigma $ & $=$ & $-1$%
\end{tabular}%
\end{equation}%
Then, we get%
\begin{equation}
\begin{tabular}{lllll}
$\mathcal{U}_{(3)}^{++}$ & : & $\dsum\limits_{\tau ,\sigma =2,3,4,5,6,22}%
\mathfrak{A}_{-{\small (\sigma )}}^{{\small \sigma -1}}\mathfrak{A}_{+%
{\small (\tau )}}^{3-\sigma }c_{{\small \sigma -1},3-\sigma |3}^{\tau
,\sigma }$ & $=$ & $0$ \\
$\mathcal{U}_{(3)}^{+}$ & : & $\dsum\limits_{\tau ,\sigma =2,3,4,5,6,22}%
\mathfrak{A}_{-{\small (\sigma )}}^{{\small \sigma -1}}\mathfrak{A}_{+%
{\small (\tau )}}^{2-\sigma }c_{{\small \sigma -1},2-\sigma |3}^{\tau
,\sigma }$ & $=$ & $0$ \\
$\mathcal{U}_{(3)}^{0}$ & : & $\dsum\limits_{\tau ,\sigma =2,3,4,5,6,22}%
\mathfrak{A}_{-{\small (\sigma )}}^{{\small \sigma -1}}\mathfrak{A}_{+%
{\small (\tau )}}^{1-\sigma }c_{{\small \sigma -1},1-\sigma |3}^{\tau
,\sigma }$ & $=$ & $0$ \\
$\mathcal{U}_{(3)}^{-}$ & : & $\dsum\limits_{\tau ,\sigma =2,3,4,5,6,22}%
\mathfrak{A}_{-{\small (\sigma )}}^{{\small \sigma -1}}\mathfrak{A}_{+%
{\small (\tau )}}^{-\sigma }c_{{\small \sigma -1},-\sigma |3}^{\tau ,\sigma
} $ & $=$ & $0$ \\
$\mathcal{U}_{(3)}^{--}$ & : & $\dsum\limits_{\tau ,\sigma =2,3,4,5,6,22}%
\mathfrak{A}_{-{\small (\sigma )}}^{{\small \sigma -1}}\mathfrak{A}_{+%
{\small (\tau )}}^{--\sigma }c_{{\small \sigma -1},-1-\sigma |3}^{\tau
,\sigma }$ & $=$ & $0$%
\end{tabular}%
\end{equation}%
Similar relations hold for the other values of the higher spins namely $%
s=4,5,6,22.$

\subsection{Higher spins in the exceptional $E_{6}/SO_{5,5}$ BTZ theory}

\begin{figure}[ph]
\begin{center}
\includegraphics[width=12cm]{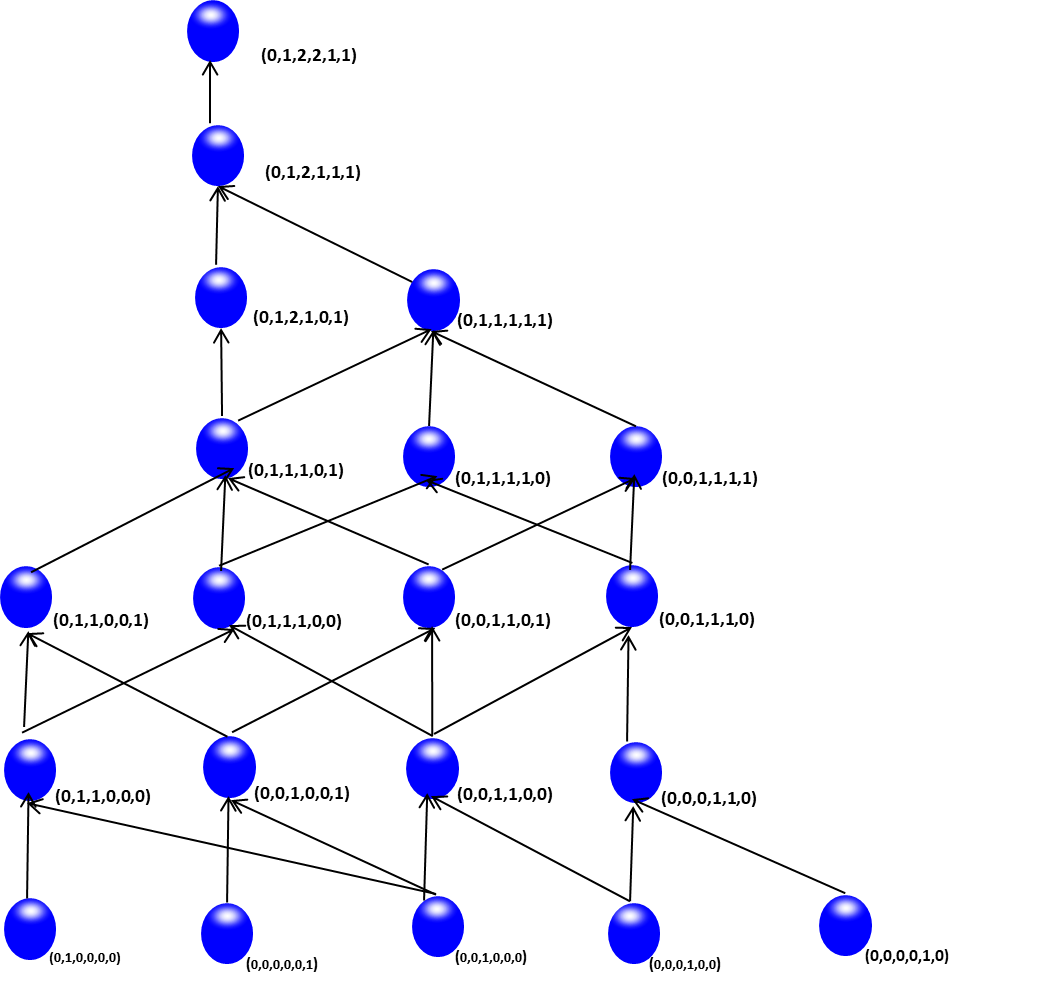}
\end{center}
\par
\vspace{-0.5cm}
\caption{Hasse diagram of the orthogonal so(5,5). This is the
Levi-subalgebra of the exceptional model $e_{6}$ obtained by cutting the
node $\protect\alpha _{1}$.}
\label{so10}
\end{figure}

\begin{figure}[ph]
\begin{center}
\includegraphics[width=14cm]{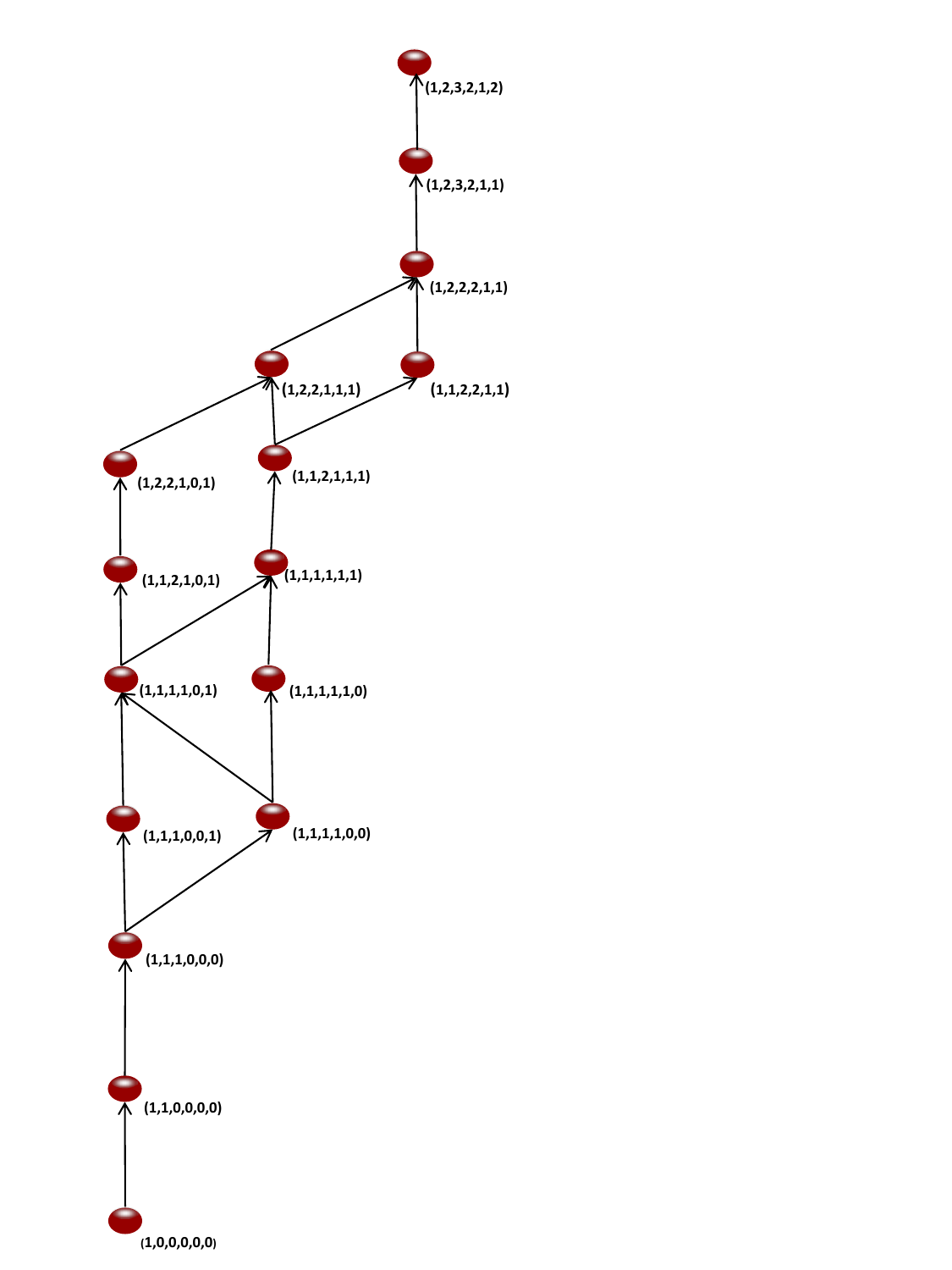}
\end{center}
\par
\vspace{-0.5cm}
\caption{the Hasse diagram of the 16-dimensional spinorial representation of
so(5,5) within the exceptional model $e_{6}$. The root in the bottom
corresponds to the cutted root $\protect\alpha _{1}$ in the
Levi-decomposition.}
\label{16}
\end{figure}

\begin{figure}[ph]
\begin{center}
\includegraphics[width=13cm]{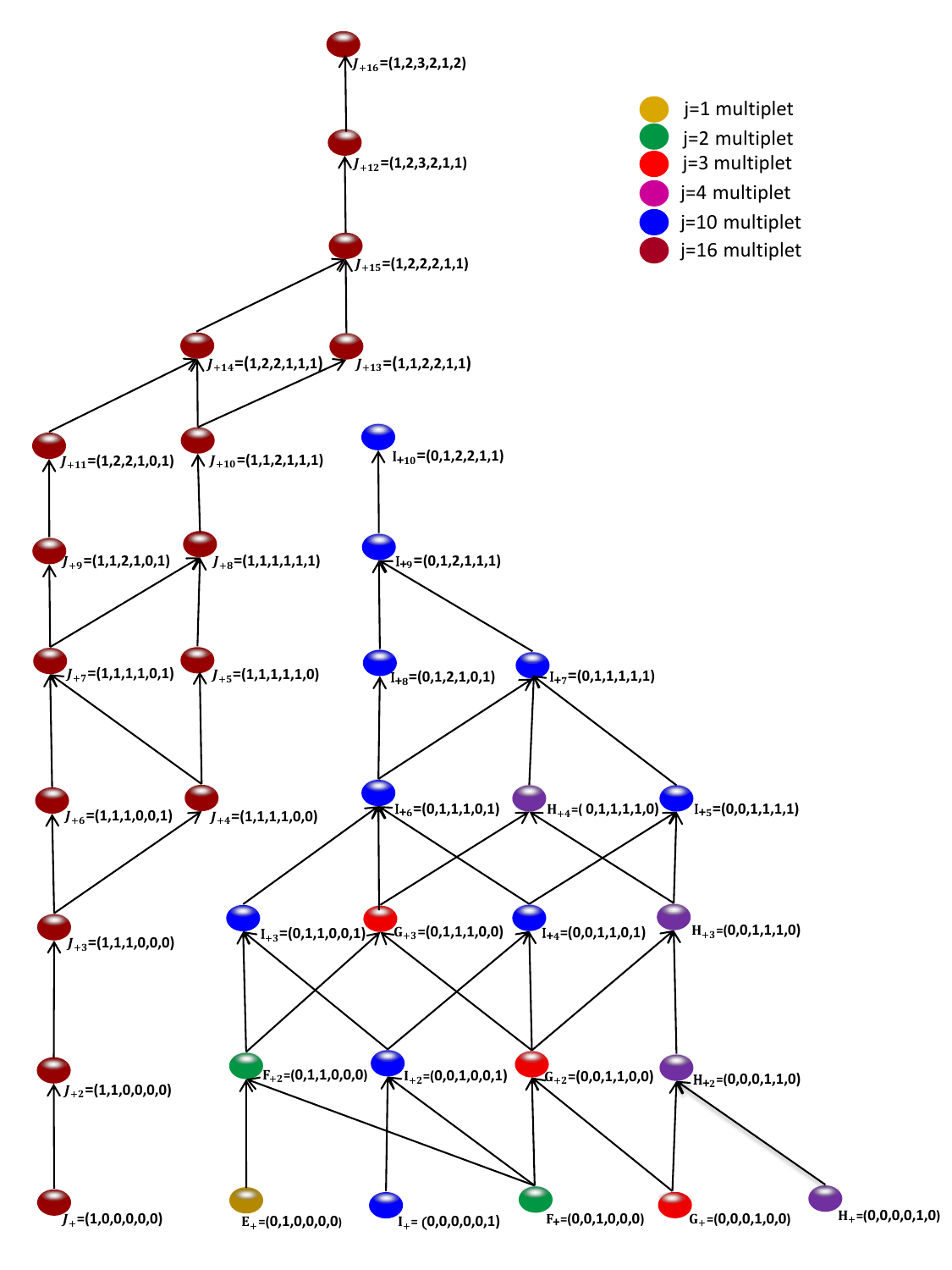}
\end{center}
\par
\vspace{-0.5cm}
\caption{The six isomultiplets making E$_{6}$ for the orthogonal based
Levi-decomposition. }
\label{multi}
\end{figure}

This investigation concerns the Levi-decomposition given by the cutting
pattern $E_{6}\rightarrow SO_{5,5}$ in the Tits Satake diagram \textbf{EI }%
of the \textbf{Figure }\ref{F1}. As, we showed before in (\ref{414}), this
higher spin decomposition has six isospin j multiplets and higher spins as
follows%
\begin{equation}
\begin{tabular}{lll}
$j\in I_{{\small so}_{{\small 10}}}$ & $=$ & $\{1,2,3,4,10,16\}$ \\
$s\in J_{{\small so}_{{\small 10}}}$ & $=$ & $\{2,3,4,5,11,17\}$%
\end{tabular}
\label{splitd5}
\end{equation}%
This follows from the red node cutting corresponding to the root $\alpha
_{1},$ and the resulting Tits-Satake diagram reveals the sub-Tits-Satake
diagram of the split real form of $SO_{5,5}$, see \textbf{Figure} \ref{F2}.
The associated Levi-subalgebra is given by $\boldsymbol{l}_{\mu
_{1}}=gl\left( 1,\mathbb{R}\right) \oplus SO_{5,5}$ where the 40 roots of $%
\Phi _{SO_{5,5}}$ form a subset of the root system $\Phi _{e_{6}}$ and $\mu
_{1}$ the coweight dual to $\alpha _{1}.$ Notice that the 40 roots of $\Phi
_{SO_{5,5}}$ are generated only by the five elementary roots $\alpha
_{2},\alpha _{3},\alpha _{4},\alpha _{5},\alpha _{6}$ and the Levi-
decomposition ${\Large e}_{6}=\boldsymbol{l}_{\mu }\oplus \boldsymbol{n}%
_{+}\oplus \boldsymbol{n}_{-}$ reads as follows%
\begin{equation}
{\Large e}_{6}\rightarrow \boldsymbol{l}_{\mu _{1}}\oplus \boldsymbol{16}%
_{+}\oplus \boldsymbol{16}_{-}
\end{equation}%
Notice that the 36+36 step operators $\mathcal{E}_{\pm \alpha }$ of ${\Large %
e}_{6}$ are split in the Levi- decomposition as: $\left( \mathbf{i}\right) $
20+20 step operators $\mathcal{E}_{\pm \beta }^{so_{{\small 5,5}}}$
generating $so\left( 5,5\right) $ given by the \textbf{Figure} \ref{so10}.
\textbf{\ }and 32 $\left( \mathbf{ii}\right) $ generating the spinor
representations; 16 step operators $\mathcal{E}_{+\gamma }^{\boldsymbol{16}}$
generating the nilpotent subalgebra $\boldsymbol{16}_{+}$, and other 16
operators $\mathcal{E}_{-\gamma }^{\overline{\boldsymbol{16}}}$ generating
the $\boldsymbol{16}_{-};$ they are given by the \textbf{Figure} \ref{16}.
This splitting distributes the 78 dimensions of ${\Large e}_{6}$ like
\begin{equation}
78=45+1+16_{+}+16_{-}
\end{equation}%
where%
\begin{equation}
45=3+5+7+9+21
\end{equation}%
corresponding to the Lorentz spin $j$ in (\ref{splitd5}). The 78 generators
of the exceptional ${\Large e}_{6}$ are splited in six $sl_{2}$ multiplets $%
W_{m_{s}}^{(s)}$ labelled by the conformal spin s as follows%
\begin{table}[H]
\centering
\begin{tabular}{c|ccccccc}
$s$ & $2$ & $3$ & $4$ & $5$ & $11$ & $;$ & $17$ \\ \hline
$W_{m_{s}}^{(s)}$ & $W_{m_{2}}^{(2)}$ & $W_{m_{3}}^{(3)}$ & $W_{m_{4}}^{(4)}$
& $W_{m_{5}}^{(5)}$ & $W_{m_{11}}^{(11)}$ & $;$ & $W_{m_{17}}^{(17)}$ \\
& $L_{m_{2}}$ & $F_{m_{3}}$ & $G_{m_{4}}$ & $H_{m_{5}}$ & $I_{m_{11}}$ & $;$
& $J_{m_{17}}$ \\ \hline
$|W_{m_{s}}^{(s)}|$ & $3$ & $5$ & $7$ & $9$ & $21$ & $;$ & $33$ \\ \hline
\end{tabular}%
\caption{Generators of the $E_{6}/SO_{5,5}$ BTZ theory}
\end{table}%
with $m_{s}$ ranging like
\begin{equation}
\begin{tabular}{llllll}
$\left\vert m_{2}\right\vert \leq 1,$ & $\left\vert m_{3}\right\vert \leq 2,$
& $\left\vert m_{4}\right\vert \leq 3,$ & $\left\vert m_{5}\right\vert \leq
4,$ & $\left\vert m_{11}\right\vert \leq 10,$ & $\left\vert
m_{17}\right\vert \leq 17$%
\end{tabular}%
\end{equation}%
The so(5,5) multiplets in addition to the 16$_{+}$ nilpotent subalgebra
multiplets as well as their associated generators are collected in the next
\textbf{Figure} \ref{multi}

\end{document}